\documentclass{aa}
\usepackage{graphicx}
\usepackage{psfig}
\begin{document}
   \title{Luminosity Functions beyond the spectroscopic limit}

   \subtitle{I. Method and near-infrared LFs in the HDF-N and HDF-S}

   \author{Micol Bolzonella\inst{1,2}, Roser Pell\'o\inst{2}, 
           Dario Maccagni\inst{1}}

   \offprints{Micol Bolzonella}

   \institute{IASF-MI, via Bassini 15, I-20133 Milano, Italy \\
              \email{micol,dario@mi.iasf.cnr.it}
         \and Observatoire Midi-Pyr\'en\'ees, UMR 5572,
              14 Avenue E. Belin, F-31400 Toulouse, France \\
             \email{roser@ast.obs-mip.fr}
             }

   \date{Received , 2001; accepted , 2002}

\abstract{We have developed a Monte Carlo method to compute the
luminosity function of galaxies, based on photometric redshifts, which
takes into account the non-gaussianity of the probability functions,
and the presence of degenerate solutions in redshift.  In this paper
we describe the method and the mock tests performed to check its
reliability.  The NIR luminosity functions and the redshift
distributions are determined for near infrared subsamples on the HDF-N
and HDF-S.  The results on the evolution of the NIR LF, the stellar
mass function, and the luminosity density, are presented and discussed
in view of the implications for the galaxy formation models.  The main
results are the lack of substantial evolution of the bright end of the
NIR LF and the absence of decline of the luminosity density up to a
redshift $z \sim 2$, implying that most of the stellar population in
massive galaxies was already in place at such redshift.
 }

\titlerunning{LFs beyond spectroscopic limit. I. Near IR}
\authorrunning{M. Bolzonella et al.}

\maketitle


\section{Introduction}

The study of galaxy formation and evolution implies the availability
of statistical samples at large look-back times. At large redshifts,
though, only star forming galaxies will be entering the samples
obtained in the visible bands and, to be able to probe their stellar
masses, observations at longer wavelengths are needed.  In the last
decade, deep photometric galaxy samples have become available, namely
through the observations by HST of the HDF-N (Williams et al.\
\cite{williams}) and HDF-S (Casertano et al.\ \cite{casertano}), which
have been coordinated with complementary observations from the ground
at near-infrared (NIR) wavelengths (Dickinson et al.\cite{dick}; da
Costa et al.\ \cite{dacosta}).  At the same time, reliable photometric
redshift techniques have been developed, allowing estimates of the
distances of faint galaxies for which no spectroscopic redshifts can
be obtained nowadays, even with the most powerful telescopes (e.g.\
Connolly et al. \ \cite{con}; Wang et al.\ \cite{wan}; Giallongo et
al.\ \cite{gia}; Fern\'andez-Soto et al.\ \cite{fsoto}; Arnouts et
al.\ \cite{arnouts}; Furusawa et al. \cite{furu}; Rodighiero et al.\
\cite{rodighiero}; Le Borgne \& Rocca-Volmerange \cite{leborgne};
Bolzonella et al.\ \cite{hyperz} and the references therein).

One of the main issues for photometric redshifts is to study the
evolution of galaxies beyond the spectroscopic limits. The relatively
high number of objects accessible to photometry per redshift bin
allows to enlarge the spectroscopic samples towards the faintest
magnitudes, thus increasing the number of objects accessible to
statistical studies per redshift bin. Such slicing procedure can be
adopted to derive, for instance, redshift distributions, luminosity
functions in different bands, or rest-frame colours as a function of
absolute magnitudes, among the relevant quantities to compare with the
predictions derived from the different models of galaxy formation and
evolution. This approach has been recently used to infer the star
formation history at high redshift from the UV luminosity density, to
analyse the stellar population and the evolutionary properties of
distant galaxies (e.g.\ SubbaRao et al.\ \cite{subba}; Gwyn \&
Hartwick \cite{gwyn}; Sawicki et al.\ \cite{sawicki}; Connolly et al.\
\cite{con}; Pascarelle et al.\ \cite{pascarelle}; Giallongo et al.\
\cite{gia}; Fern\'andez-Soto et al.\ \cite{fsoto}; Poli et al.\
\cite{poli}), or to derive the evolution of the clustering properties
(Arnouts et al.\ \cite{arnouts}; Magliocchetti \& Maddox
\cite{maglio}, Arnouts et al.\ \cite{arnouts2}).

We have developed a method to compute luminosity functions (hereafter
LFs), based on our public code \emph{hyperz} to determine photometric
redshifts (Bolzonella et al.\ \cite{hyperz}).  This original method is
a Monte Carlo approach, different from the ones proposed by SubbaRao
et al. (\cite{subba}) and Dye et al. (\cite{dye}) in the way of
accounting for the non-gaussianity of the probability functions, and
specially to include degenerate solutions in redshift.  In this paper
we present the method and the tests performed on mock catalogues, and
we apply it specifically to derive NIR LFs and their evolution on the
HDF-N and HDF-S.

The NIR luminosity is directly linked to the total stellar mass, and
barely affected by the presence of dust extinction or starbursts.
According to Kauffmann \& Charlot (\cite{kauff}), the NIR LF and its
evolution constitute a powerful test to discriminate between the
different scenarios of galaxy formation, i.e.\ if galaxies were
assembled early, according to a monolithic scenario, or recently from
mergers.  The theoretical NIR LFs derived by Kauffmann \& Charlot
exhibit a sharp difference between the two models at redshifts $z>1$.
The PLE models foresee a constant bright-end for the LF, whereas
hierarchical models are expected to undergo a shift towards faint
magnitudes with increasing redshift.  In this paper we compare the
theoretical predictions with the observations on the HDFs, in order to
extend the analysis performed from spectroscopic surveys (Glazebrook
et al.\ \cite{glaze}, Songaila et al.\ \cite{songaila}, Cowie et al.\
\cite{cowie2}) up to $z \sim 2$.  The comparison between the present
NIR LF results and a similar study in the optical-UV bands, obtained
with the same photometric redshift approach (Bolzonella et al.\ Paper
II, in preparation), could provide with new insights on the galaxy
formation scenario.

The plan of this paper is the following. Section~\ref{catalogue} gives
a brief description of the photometric catalogues used to select the
samples, their properties being discussed in Sect.~\ref{sample}.  In
Sect.~\ref{lf} we describe the technique conceived to compute the
luminosity functions using the \emph{hyperz} photometric redshift
outputs, and the test of the method through mock catalogues. The
results obtained on the HDFs near-infrared LF are presented and
discussed in Section~\ref{resu}, together with the Luminosity Density
and the Mass Function derived from them. The implications of the
present results on the galaxy formation models are discussed in
Sect.~\ref{discuss}, and the main results are summarized in
Sect.~\ref{sum}.

Throughout this paper we adopt the cosmological parameters $\Omega_0 =
1$, $\Omega_\Lambda = 0$, when not differently specified.  Magnitudes
are given in the AB system (Oke \cite{oke}).  Throughout the paper,
the Hubble Space Telescope filters F300W, F450W, F606W and F814W are
named $U_{300}$, $B_{450}$, $V_{606}$ and $I_{814}$ respectively.


\section{Photometric catalogues}
\label{catalogue}

The HDF-N (Williams et al.~\cite{williams}) and HDF-S (Casertano et
al.~\cite{casertano}) are the best data sets to which the photometric
redshift techniques can be applied, because of the wavelength coverage
extending from the $U_{300}$ to the $K_s$ bands through a combination
of space and ground based data, and of the accurate photometry
available.  Table \ref{filters} gives the characteristics of the
filters involved.

For the HDF-N we adopted the photometric catalogue provided by
Fern\'andez-Soto et al.\ (\cite{fsoto}).  These authors used the
SExtractor (Bertin \& Arnouts \cite{bertin}) package and detected
objects in the $I_{814}$ image.  The final catalogue consists of a
total of $1067$ galaxies (after exclusion of known stars) in
$5.31$\,arcmin$^2$.  The reference aperture used to measure the fluxes
in the different bands is defined by the threshold isophote on the
$I_{814}$ image.  This catalogue is divided in 2 zones: the deepest
one (Z1) consists of 946 objects in $3.92$ arcmin$^2$ and it has been
limited to $I_{814}(AB) \le 28$ (in ``best'' magnitudes, see
SExtractor user manual), the shallowest one (Z2, including field edges
and the planetary camera) contains 121 objects to $I_{814}(AB) \le
26$, in an area of $1.39$ arcmin$^2$.

For the HDF-S we adopted the photometric catalogue prepared by
Vanzella et al.\ \cite{vanzella} and available on the
web\footnote{http://www.stecf.org/hstprogrammes/ISAAC/HDFSdata.html}.
The catalogue contains $1611$ objects, detected in the
$V_{606}+I_{814}$ image and with magnitudes measured inside a variable
aperture suited to compute photometric redshifts.  The NIR imaging has
been carried out with the near-infrared spectrometer/imager VLT-ISAAC
and the photometry has been measured taking into account the different
PSF of NIR images.

\begin{table}
\caption{Characteristics of photometric data used in this paper:
effective wavelength, bandpass and AB conversion (conv$_{\rm AB} =
m_{\rm AB} - m_{\rm Vega}$) for the different filters, and limiting
magnitudes in the HDF-N and HDF-S (see text).}

\begin{tabular}{cccccc}
Filter & $\lambda_{\rm eff}$ & $\Delta \lambda$ & conv$_{\rm AB}$
& $m_{\rm lim}^{\rm HDF-N}$  & $m_{\rm lim}^{\rm HDF-S}$  \\
       &    \AA              &    \AA           & mag
& mag & mag  \\

\hline
 $U_{300}$ &  3010 &  854 &  1.381 & 29.0 & 28.3 \\
 $B_{450}$ &  4575 &  878 & -0.064 & 28.5 & 28.5 \\
 $V_{606}$ &  6039 & 1882 &  0.132 & 28.5 & 28.5 \\
 $I_{814}$ &  8010 & 1451 &  0.439 & 28.0 & 28.0 \\
 $J$       & 12532 & 2651 &  0.937 & 25.5 & 25.8 \\
 $H$       & 16515 & 2903 &  1.407 & 25.0 & 25.0 \\
 $K_s$     & 21638 & 2724 &  1.871 & 25.0 & 25.3 
\end{tabular}
\label{filters} 
\end{table}


\section{Sample properties}
\label{sample}

\subsection{Photometric Redshifts}

\begin{figure*}
{\centering \leavevmode
\psfig{file=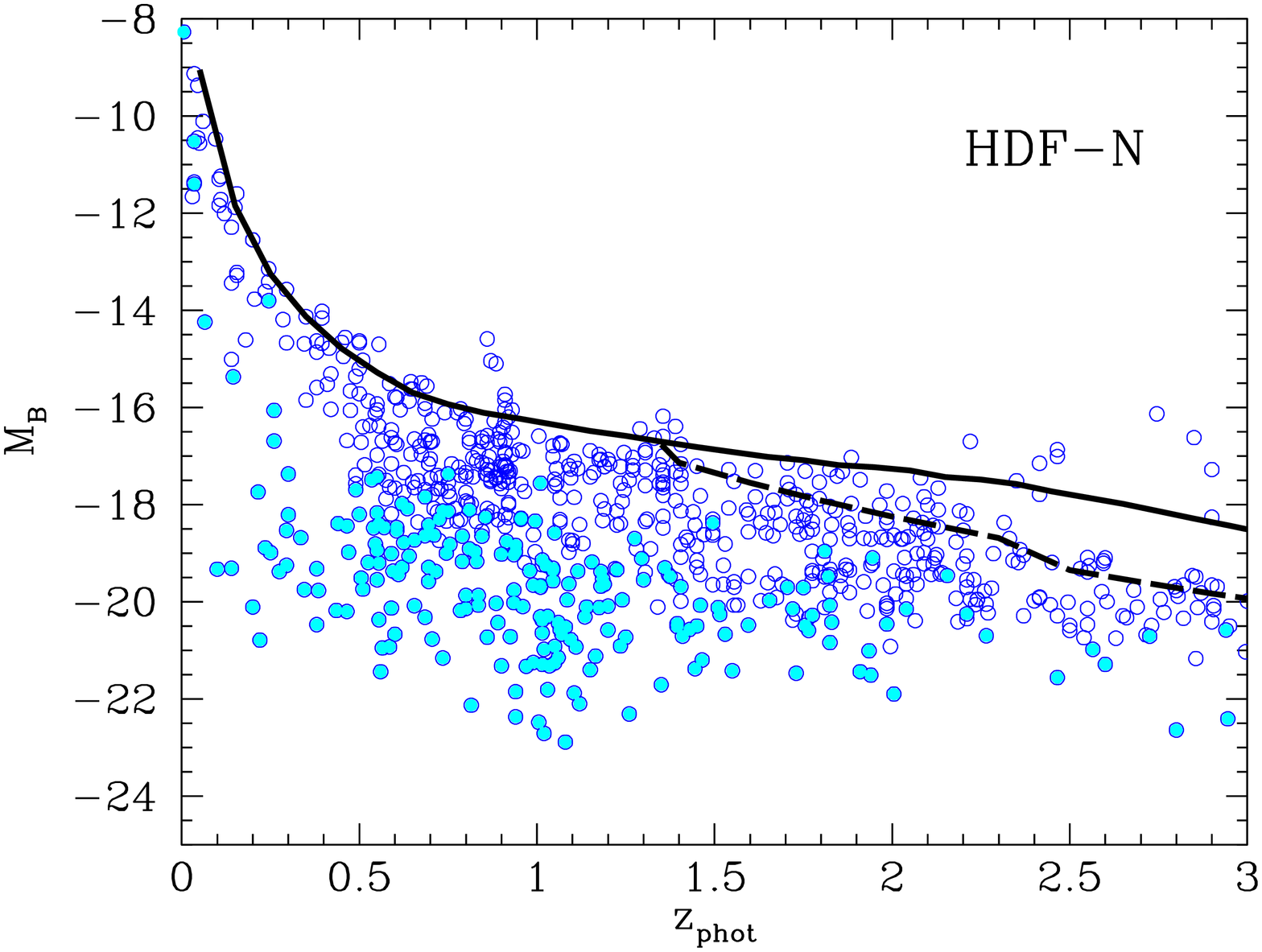,width=.49\textwidth,angle=270} \hfil
\psfig{file=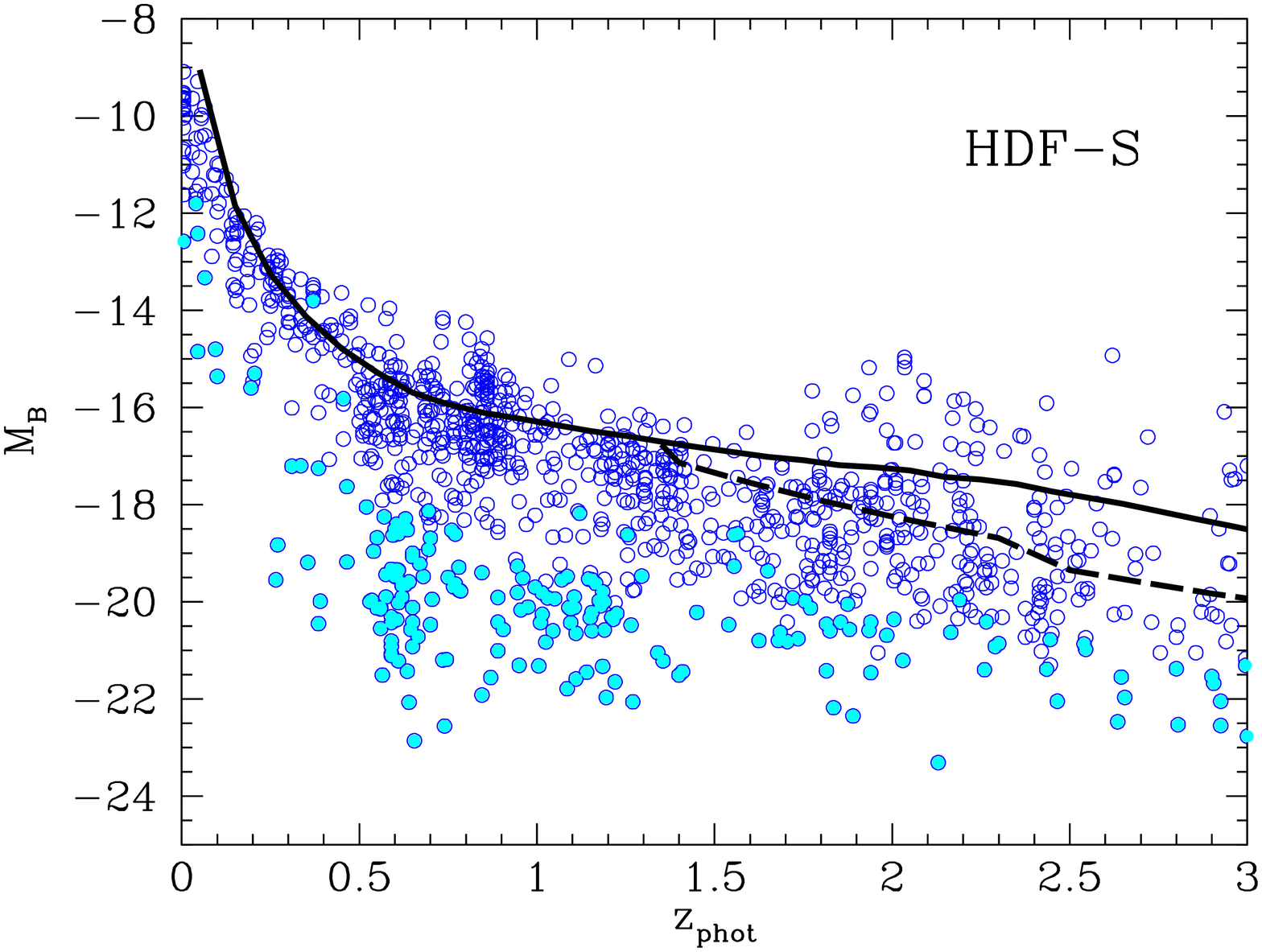,width=.49\textwidth,angle=270} } 
\caption{Absolute magnitude in the $B$-band as a function of
photometric redshift in the HDF-N and HDF-S for objects with
$B_{450} \le 28.5$.  The filled dots represent the subsample with
$K_s \le 24$, used to estimate the luminosity functions in Section
\ref{lf}.  The solid lines mark the limiting absolute magnitude
computed using the apparent limiting magnitude of the sample in the
$B_{450}$ band, and the mean $k$-correction over a mix of spectral
types.  The dashed lines display the same result when using the
limiting magnitudes for the NIR filters.}
\label{zmabs}
\end{figure*}

To compute photometric redshifts, we converted the fluxes $f_\nu$ into
AB magnitudes, assigning the magnitude value of $99$ (corresponding to
an undetected object in our photometric redshift scheme) in case of
negative fluxes, negative error fluxes, or $S/N = f_\nu/\Delta f_\nu <
1$. Moreover, we added in quadrature a photometric error of
$0.02$ magnitudes in all filters, to account for systematics in the
zeropoints.

Then, we applied the public code
\emph{hyperz}\footnote{http://webast.obs-mip.fr/hyperz/ and mirror
sites} to compute the photometric redshifts of galaxies in the HDF-N
and HDF-S.  A full description of the technique and the analysis of
the performances on the HDFs can be found in Bolzonella et al.\
(\cite{hyperz}).  Here we only recall that an interesting
characteristic of the \emph{hyperz} code is the possibility of
computing probability functions in the redshift space.  We will make
full use of this facility to derive the LFs.

A full description of the different parameters used by \emph{hyperz}
can be found in the User's Manual of the code, available on the web pages.
The most relevant parameters used here are the following: \\ 
-- The limiting magnitudes applied in case of undetected objects are
given in Table \ref{filters}.  We choose $P(<m)=0.8$ in the cumulative
histograms to set the adopted limiting magnitudes, which roughly
correspond to objects with a $S/N  \sim 1$. \\
-- We used Calzetti's (\cite{calzetti}) law to produce reddened
templates, with $A_V$ ranging from $0$ to $1.2$ magnitudes.  \\
-- The Lyman forest blanketing is modelled according to the standard
prescriptions of Madau (\cite{madau}), without variations along the
different lines of sight. \\
-- As galaxy template set we selected the $5$ ``standard'' SEDs built
using the GISSEL98 library by Bruzual \& Charlot (\cite{bruzual}),
with solar metallicity; the CWW SEDs were not considered because they
are redundant. \\
-- The range of absolute magnitudes $M_B$ constraining the allowed
solutions is $[-28,-9]$. \\

\begin{figure*}
{\centering \leavevmode
\psfig{file=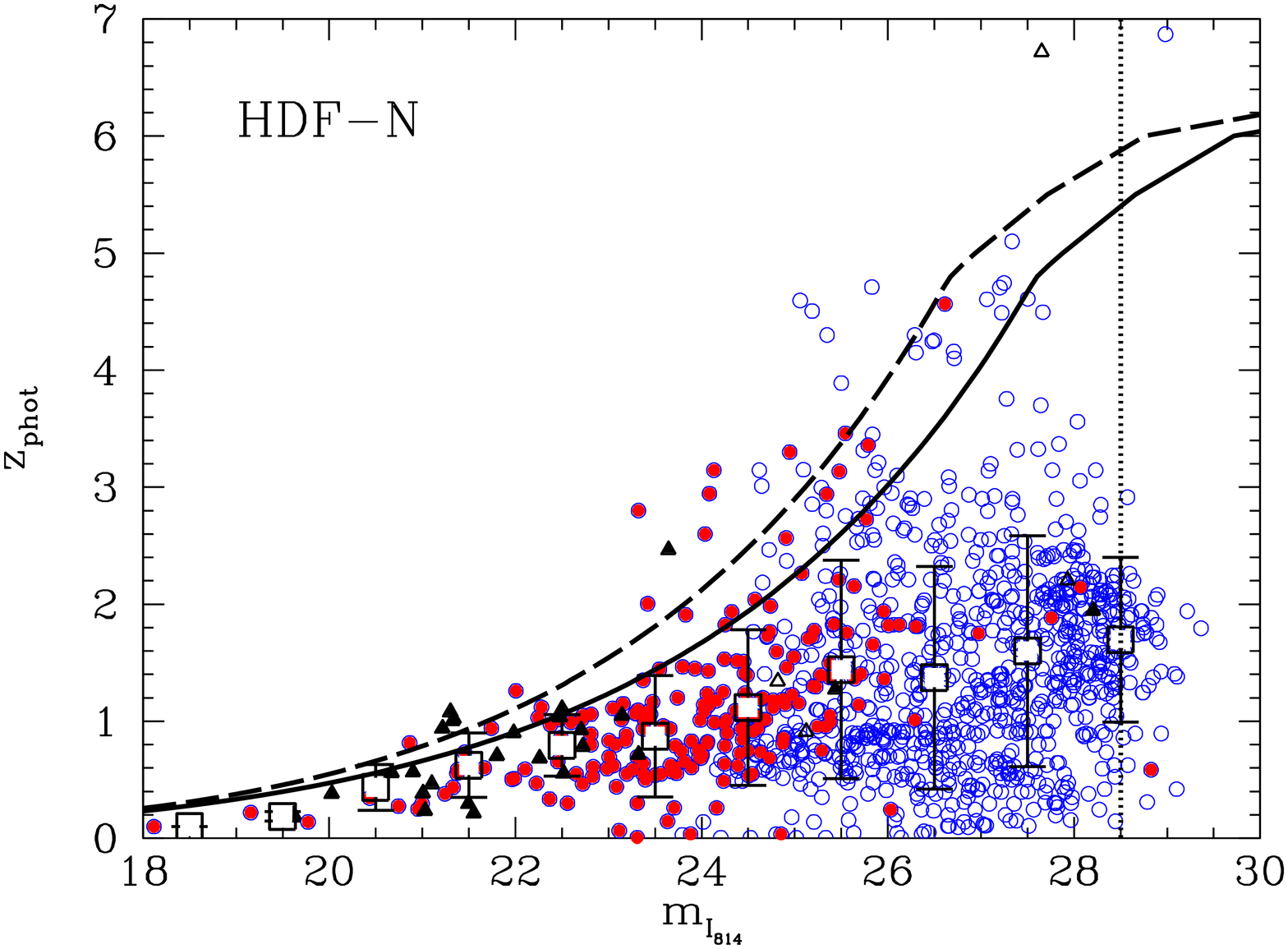,width=.49\textwidth,angle=270} \hfil
\psfig{file=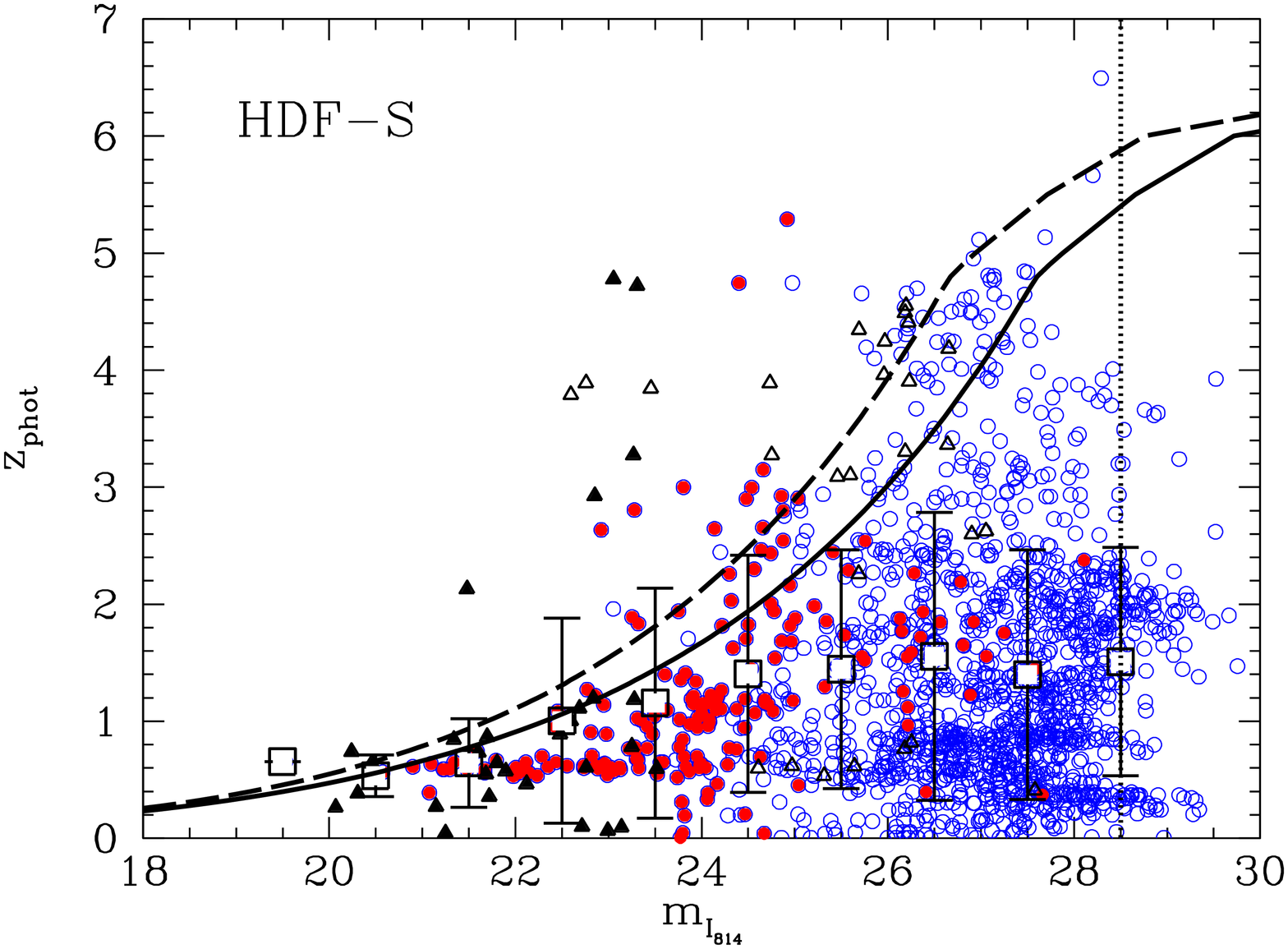,width=.49\textwidth,angle=270} } 
\caption{Hubble diagram for objects in the HDF-N and HDF-S samples.
Filled symbols represent the subsample with $K_s \le 24$.  Triangles
show the objects for which $P(z_{\rm phot})=0$ everywhere.  Dashed and
solid lines display the location on this diagram of a local $L^*$
galaxy, with cosmological parameters ($\Omega_0=1$, $\Omega_\Lambda =
0$) and ($\Omega_0=0.3$, $\Omega_\Lambda = 0.7$) respectively, with
$H_0 = 50\,{\rm km\,s^{-1}\,Mpc^{-1}}$.  The vertical dotted line
marks the $I_{814}=28.5$ magnitude at which the signal-to-noise ratio
is $S/N=3$.  The mean redshift and the $1\sigma$ dispersion have also
been plotted. }
\label{zmapp}
\end{figure*}

\begin{figure*}
{\centering \leavevmode
\psfig{file=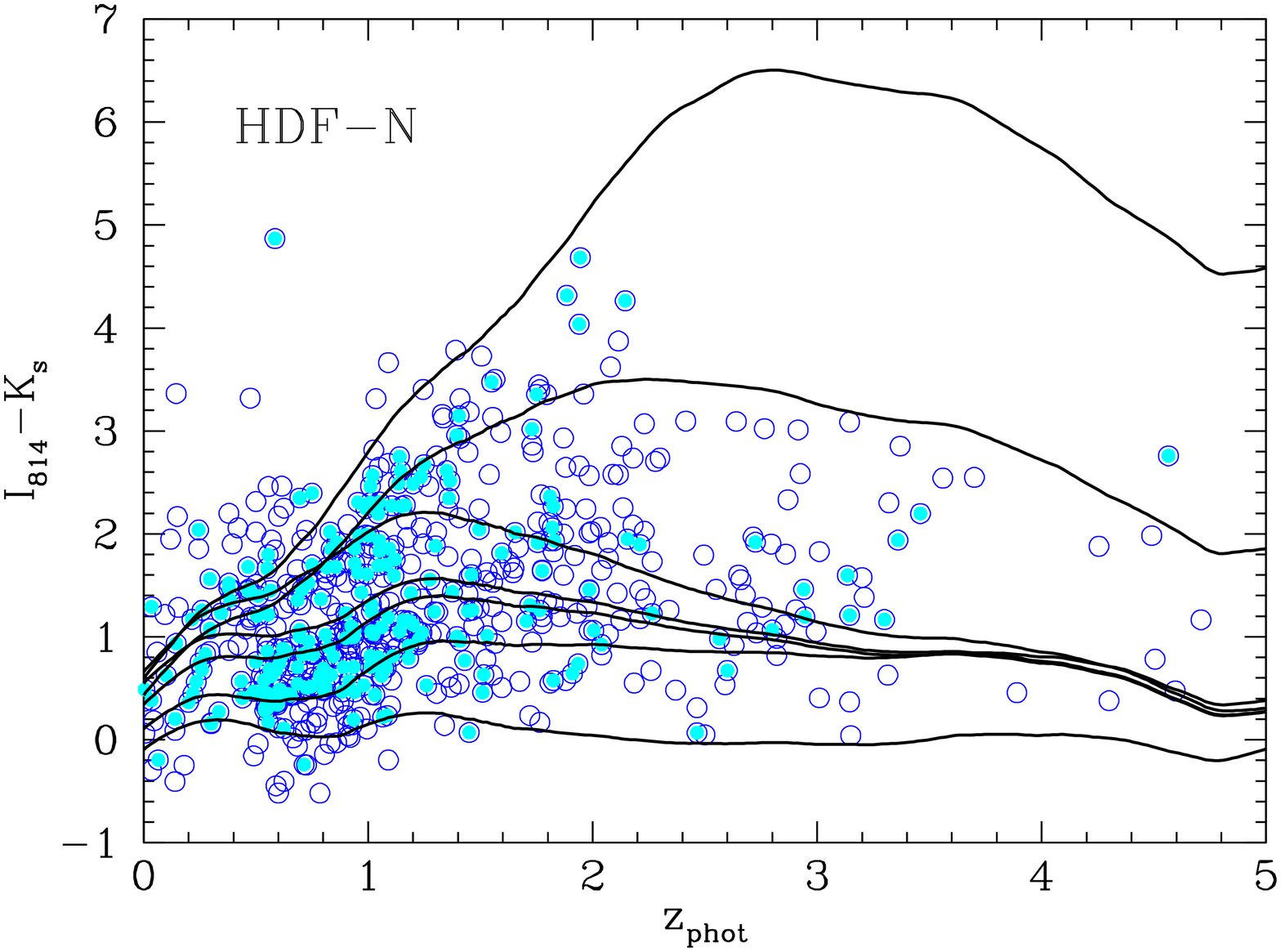,width=.49\textwidth,angle=270} \hfil
\psfig{file=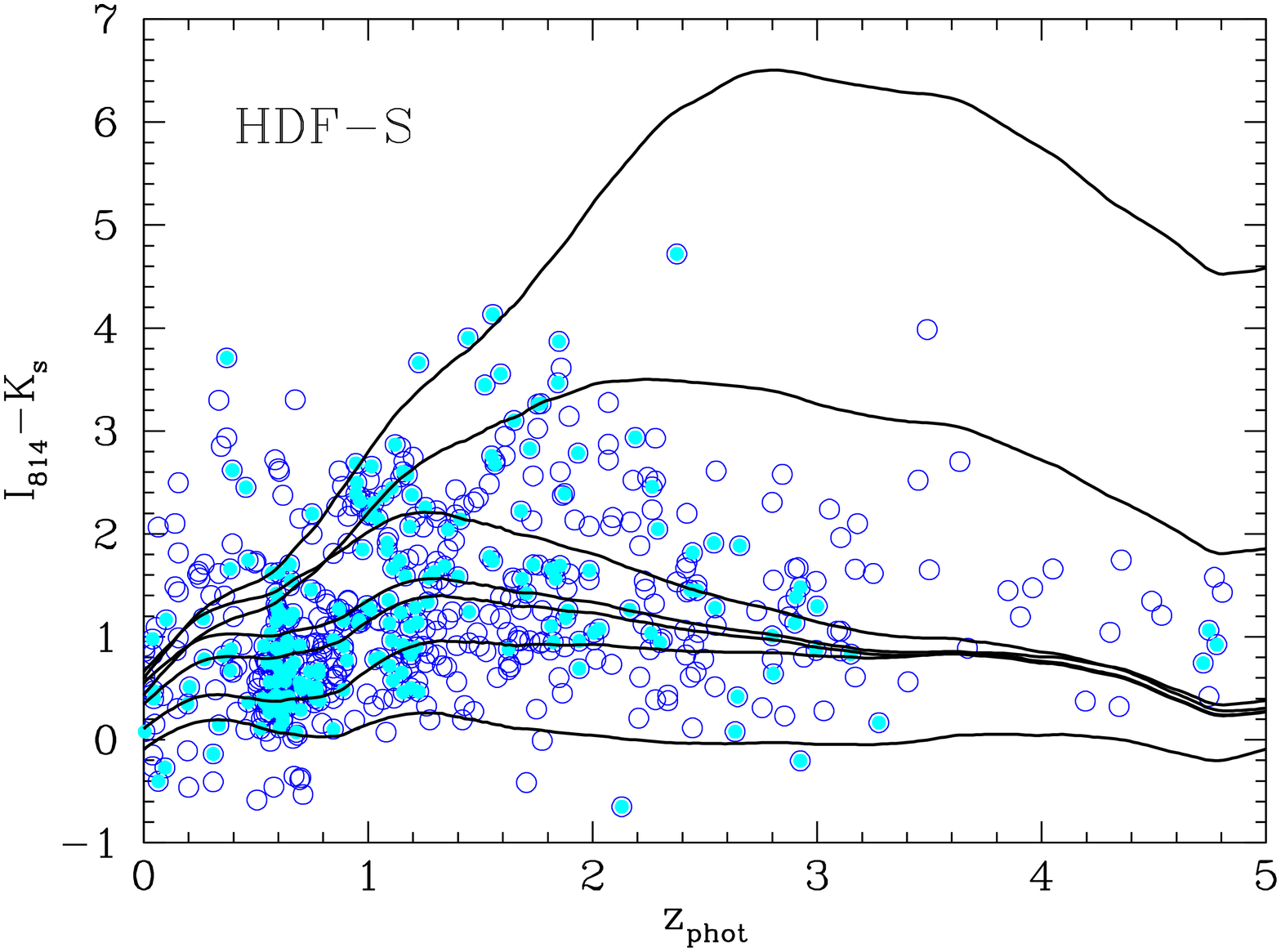,width=.49\textwidth,angle=270} } 
\caption{Colour $I_{814}-K_s$ as a function of photometric redshift,
compared to the predictions for different synthetic SEDs.  From top to
bottom: Elliptical galaxy of fixed $10$\ and $5$ Gyr age, evolving E,
Sa, Im, and Im of fixed $1$ and $0.1$\,Gyr age. Empty dots illustrate
the sample with $K_s \le 26$ in the HDF-N and HDF-S samples,
corresponding to $S/N=1$, whereas filled dots represent objects with
$K_s \le 24$, i.e. with $S/N \ge 3$.  The observed colours are
enclosed in the limits of colours foreseen from the synthetic SEDs.}
\label{z_ik}
\end{figure*}

A comparison between the photometric and spectroscopic redshifts has
been shown in Bolzonella et al.\ (\cite{hyperz}).  Here we assume that
galaxies in the photometric sample will follow the same behaviour,
i.e.\ the same dispersion around the true values.  Beyond $I_{814}=25$
our spectroscopic knowledge falls dramatically.  Very few objects are
available to calibrate the photometric redshifts in this domain, and
so larger discrepancies are always possible.  Nonetheless, these
objects will remain beyond the spectroscopic limits till the arrival
of larger telescopes in the future, and photometric redshifts are
presently the only available tool to study them.

We used the results of the photometric redshift estimate to examine
some properties of the sample and to give an estimate of the
evolution in the samples. 

Figure~\ref{zmabs} displays the absolute magnitude in the $B$-band,
obtained as standard output of \emph{hyperz}, as a function of the
photometric redshift for objects with $B_{450} \le 28.5$.  The thick
solid line represents the limiting absolute magnitude computed for a
$B = 28.5$ object and a mean $k$-correction computed over a mix of
spectral types, with the method explained below in Sect.~\ref{kcorr}.
For comparison, the thick dashed lines show the same result when using
the limiting magnitudes given in Table \ref{filters} for the NIR
filters.  It is worth to remark the sequence of theoretical absolute
magnitudes, in agreement with the observational limits, obtained
without imposing stringent constraints in $M_B$.

In Fig.~\ref{zmapp} we plot the observed $I_{814}$ magnitudes versus
the photometric redshift, viz the Hubble diagram.  The mean redshift
per magnitude bin is also shown, with its $1\sigma$ dispersion.  The
faintest objects are also the farthest ones, as we expect for an
expanding universe.  The position on this Hubble diagram of a local
$L^*$ galaxy is also shown for comparison, for the two sets of
cosmological parameters used in this paper.  In the HDF-S there
are some objects well above the $L^*$ lines: most of them are objects
with $P(z_{\rm phot})=0$, shown as triangles.  Nearly half of these
objects are classified by SExtractor as stars in the Vanzella et
al. (\cite{vanzella}) catalogue.  When we fit their SEDs with stellar
templates (Pickles \cite{pickles}) and quasar spectra (according to
the method described by Hatziminaoglou et al.~\cite{evanthia}), only
one of these objects has colors fully consistent with a highly
reddened star.  In all the other cases, these objects are not well
fitted neither with the standard SEDs for galaxies nor with the
stellar or quasar templates.  Thus, we have no reason to exclude them
when computing LFs. 

The inspection of these two figures indicates that the statistical
properties of the photometric redshift samples display the expected
behaviour.  According to our previous simulations, aiming to reproduce
the properties of the HDFs (Bolzonella et al.\ \cite{hyperz}), the
redshift distribution beyond the spectroscopic limits can be
considered as reliable.

In Fig.~\ref{z_ik} we plot the apparent colour $I_{814}-K_s$ versus
the photometric redshift, as well as the colours computed from the
synthetic SEDs of different types: from top to bottom, an Elliptical
galaxy at fixed ages of $10$\,Gyr and $5$\,Gyr, evolving E, Sa, Im, Im
of fixed $1$ Gyr and $0.1$\,Gyr age.  A few objects redder than old
ellipticals seem to be present at $z<1$ in both fields.  Also, a group
of objects with $I_{814}-K_s > 2.6$ and $1 \la z \la 2$ is detected:
these objects are EROs according to the usual criteria (e.g.\ Cimatti
et al.\ \cite{cimatti}; Scodeggio \& Silva \cite{scodeggio}; Moriondo
et al.\ \cite{mori}).


\subsection{Redshift distribution}

\begin{figure*}
{\centering \leavevmode
\psfig{file=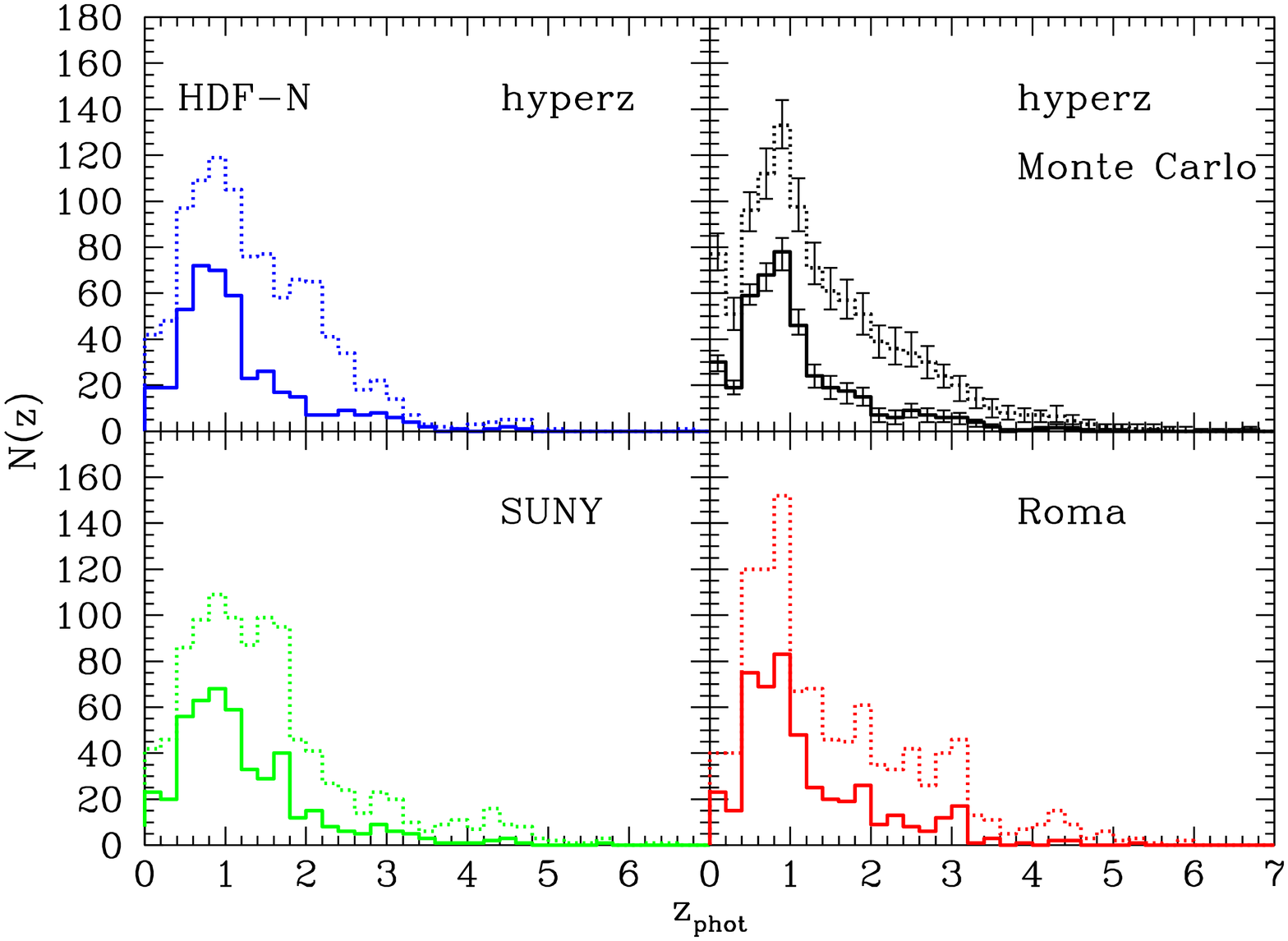,width=.49\textwidth,angle=270} \hfil
\psfig{file=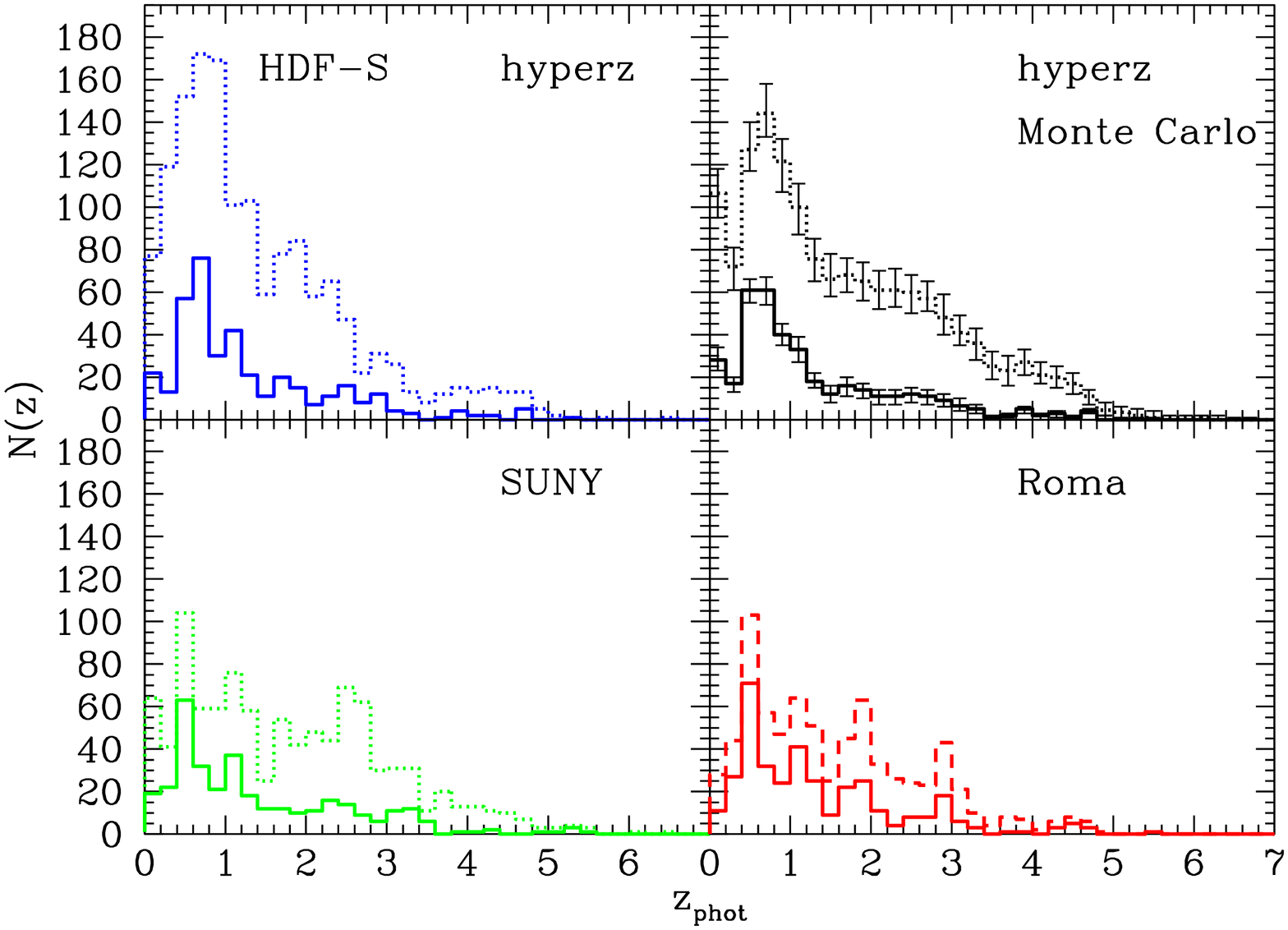,width=.49\textwidth,angle=270} } 
\caption{Redshift distributions obtained for the $I_{814}$ selected
samples in the HDF-N and HDF-S by different authors.  \emph{Left
panel:} $N(z)$ in the HDF-N obtained using the SUNY catalogue by
hyperz, hyperz with a Monte Carlo method, Fern\'andez-Soto et al.\
(\cite{fsoto}) and Fontana et al.\ (\cite{fonta1}), with limiting
magnitudes of $I_{814} = 28.5$ (dotted line) and $I_{814} = 26$ (solid
line).  \emph{Right panel:} $N(z)$ in the HDF-S, with the same two
limiting magnitudes.  Fontana et al.\ (\cite{fonta1}) used the SUNY
catalogue, limiting their analysis to $I_{814} = 27.5$ (dashed
line). }
\label{nzall}
\end{figure*}

\begin{figure*}
{\centering \leavevmode
\psfig{file=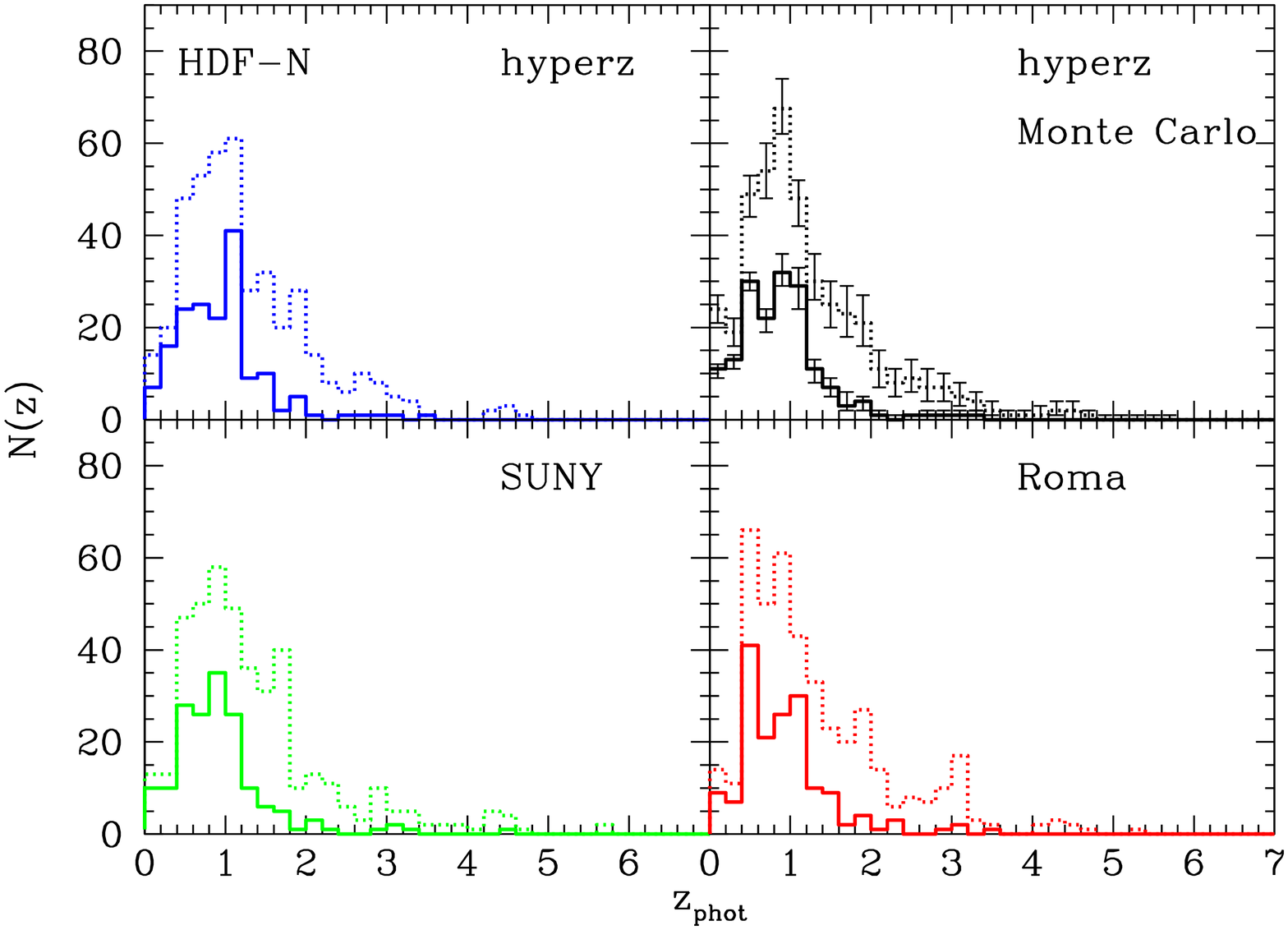,width=.49\textwidth,angle=270} \hfil
\psfig{file=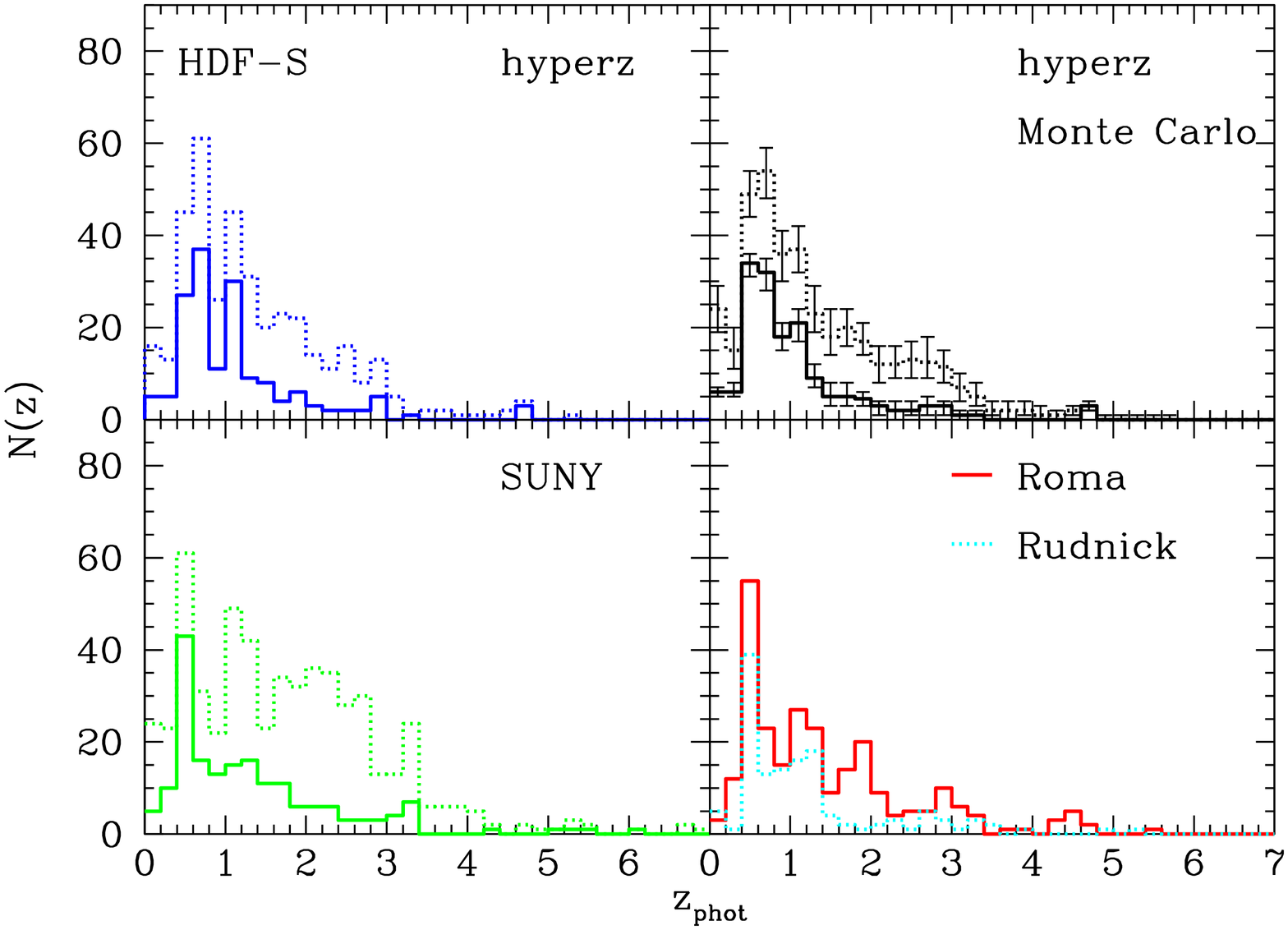,width=.49\textwidth,angle=270} } 
\caption{Redshift distributions obtained for the $K_s$ selected
subsamples in the HDF-N and HDF-S by different authors.  \emph{Left
panel:} $N(z)$ in the HDF-N obtained using the SUNY catalogue by
hyperz, hyperz with a Monte Carlo method, Fern\'andez-Soto et al.\
(\cite{fsoto}) and Fontana et al.\ (\cite{fonta1}) with two limiting
magnitudes: $K_s \le 25$ (dotted line) and $K_s \le 23.5$ (solid
line).  \emph{Right panel:} the same in the HDF-S.  In the lower right
panel the solid line represents the $N(z)$ at $K_s \le 23.5$ obtained
by Fontana et al. (\cite{fonta1}) and the dotted line the $N(z)$ at
the same limit by Rudnick et al. (\cite{rudnick}).}
\label{nzksel}
\end{figure*}

Figure~\ref{nzall} shows the redshift distributions, as obtained from
the photometric redshift computation, for the HDFs catalogues.  We
have compared our result on the HDF-N with the redshift distributions
obtained by Fern\'andez-Soto et al.\ (\cite{fsoto}, SUNY group) and
Fontana et al.\ (\cite{fonta1}, Roma group), using the same catalog
(see Section~\ref{catalogue}).  In the HDF-S, we compared our redshift
distribution, obtained with the catalogue by Vanzella et
al. (\cite{vanzella}) with the $N(z)$ computed using the SUNY
catalogue by Fern\'andez-Soto et al. \ (\cite{fsoto}) and Fontana et
al.\ (\cite{fonta1}).  We have also computed a median redshift
distribution, obtained with the Monte Carlo method we applied to
estimate the luminosity function, and detailed in Section
\ref{montecarlo}.  This redshift distribution takes into account the
probability functions of individual objects, in such a way that
objects will be scattered in their allowed range of photometric
redshifts, according to their $P(z_{\rm phot})$.  Objects with flat
$P(z_{\rm phot})$ will spread over a wide range of $z_{\rm phot}$,
whereas objects with narrow $P(z_{\rm phot})$ will be always assigned
to a redshift close to the best fit one.  We show the result of 100
iterations, represented by the median and the error bars, computed as
10\% and 90\% of the values distribution in each redshift bin.

Considering the fainter sample of the HDF-N presented in the
left panel of Figure \ref{nzall}, the Kolmogorov-Smirnov test shows
that the hyperz and SUNY distributions are compatible, whereas the
hyperz and Roma distributions are not.  We reject the hypothesis that
two distributions were issued from the same parent distributions when
the significance level is lower than a conservative value of 1\%.
Using the same criterium, the median $N(z)$ computed with the
hyperz-Monte Carlo method is compatible with both the SUNY and the
Roma distributions.  At $I_{814} \le 26$, the KS test shows that
each $N(z)$ is consistent with the others.

\begin{figure*}
{\centering \leavevmode
\psfig{file=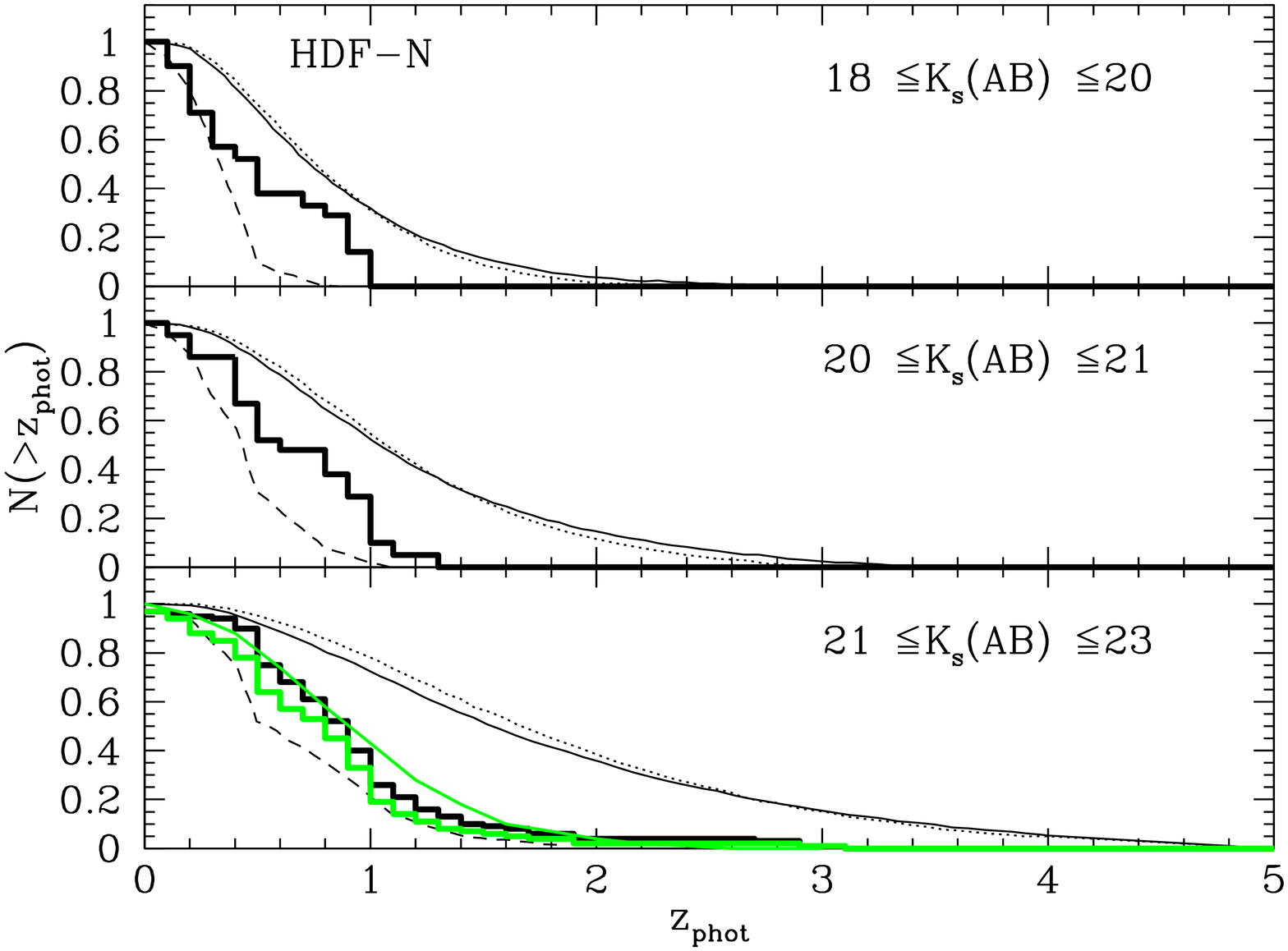,width=.49\textwidth,angle=270} \hfil
\psfig{file=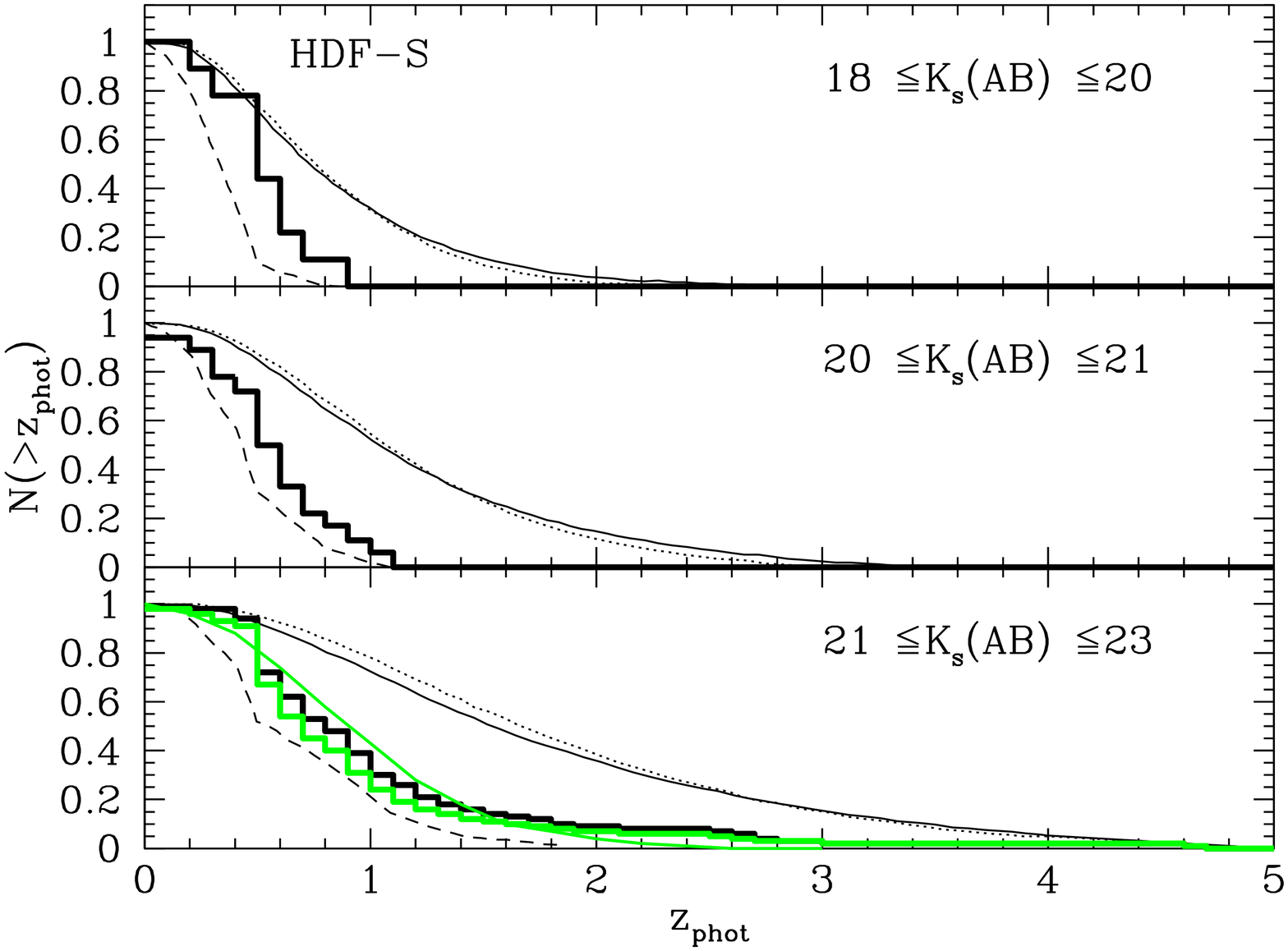,width=.49\textwidth,angle=270} } 
\caption{Cumulative redshift distribution in the $K_s$ band selected
samples in HDF-N and HDF-S obtained with the Monte Carlo method (thick
solid lines), compared to the theoretical expectations by Kauffmann \&
Charlot (\cite{kauff}): solid and dotted lines correspond to
$\Omega_0=1$ and $\Omega_0=0.2$ PLE models with null cosmological
constant, whereas dashed lines correspond to their hierarchical galaxy
formation model based on a $\Omega_0=1$ cosmology.  The $3$
apparent $K_s$ magnitude bins roughly correspond to the original ones
used by Kauffmann \& Charlot (\cite{kauff}).  The light gray
cumulative histogram in the lower panels represents $N(>z)$ for the
sample with $K_s<23$, whereas the light gray continuous line is the
theoretical expectation at the same limiting magnitude computed by
Fontana et al.\ (\cite{fontana}) for a hierarchical model based on a
$\Lambda$CDM cosmology.}
\label{nzcum}
\end{figure*}

In the HDF-S, we show the comparison of $N(z)$ for the same two
limiting magnitudes, $I_{814} \le 28.5$ and $I_{814} \le 26$.  Fontana
et al.\ (\cite{fonta1}) selected their subsample imposing $I_{814} \le
27.5$ (dashed line).  The KS test for the subsample with $I_{814} \le
28.5$ shows that the distributions are not drawn from the same parent
distribution, but considering the bright subsample with $I_{814} \le
26$ the distributions are fully consistent each other, even if they
are obtained from different photometric catalogues.

In Figure \ref{nzksel} we show the same comparison carried out for the
$K_s$-band selected subsample we used in the following of this paper.
We have analysed the distributions obtained when selecting objects
with $K_s \le 25$ and $K_s \le 23.5$.  In the HDF-N the agreement
among the different $N(z)$ distributions is remarkable: by means of
the Kolmogorov-Smirnov test we obtain that we can reject the
possibility that the distributions were drawn from different parent
distributions with a confidence level ranging from $23$\% (minimum)
and $70$\% (maximum) at $K_s \le 25$, and a probability ranging from
$41$\% to $100$\% at $K_s \le 23.5$.

In the HDF-S the same comparison shows that the hyperz and SUNY
distributions are not compatible, even choosing the conservative value
of $1$\% for the confidence level, whereas the hyperz and Roma results
are marginally consistent.  On the other hand, the redshift
distributions of the bright subsample show a better agreement, with
probabilities in each case larger than the chosen limit, even if the
catalogues are different, in particular in the NIR dataset.  In the
lower right panel we also show the redshift distribution obtained by
Rudnick et al.\ (\cite{rudnick}) for a sample limited to $K_s \le
23.5$ and a catalogue built by these authors using ISAAC-VLT images.
Also in this case we found that the redshift distributions are
compatible.  Even considering the subsample with $K_s \le 24$,
i.e. the limit chosen computing LFs, the redshift distributions are in
general consistent.

In conclusion, some small differences are observed in the redshift
distribution when using different methods, mostly affecting the
faintest samples, whereas the results are compatible for the bright
ones.  At faint limits in magnitude, the redshift distribution
obtained with the hyperz-Monte Carlo method avoids introducing
spurious features, and it makes the relevant $N(z)$ distributions
compatible. At brighter limits, by means of the comparison with the
other photometric redshift estimates, we have shown that the results
are consistent and characterized by a high quality in the redshift
determination.  The differences among different authors become very
smooth when the Monte Carlo approach is used; this procedure allow us
to compute reliable LFs in Sect.~\ref{resu}.

\subsection{Cumulative redshift distribution}

Kauffmann \& Charlot (\cite{kauff}) argued that the cumulative
redshift distribution in the $K$-band can be used as a test for the
scenarios of galaxy formation.  In Fig.~\ref{nzcum} we have compared
our data with the theoretical expectations given by Kauffmann \&
Charlot (\cite{kauff}), in the case of monolithic (PLE) and
hierarchical galaxy formation scenarios.  The original $K$-band
magnitude bins have been shifted by $\sim 2$ magnitudes to match the
AB magnitudes used here.  The fraction of galaxies at high redshifts
is much larger in the case of the PLE models than in the hierarchical
scenario.  The observed HDFs cumulative redshift distribution lie
between the two models and, in any case, it is always below the
predictions of the PLE models.  This result was already pointed out by
Kauffmann \& Charlot (\cite{kauff}), when comparing their predictions
with the results found with the Songaila et al.\ (\cite{songaila}) and
Cowie et al.\ (\cite{cowie2}) samples, in the two brightest magnitude
bins. The present results on the HDFs extend this trend towards the
faintest magnitude bins. However, this results must be considered with
caution, mainly because of the small size of the surveyed fields and
uncertainties affecting the models.  For instance, Kauffmann \&
Charlot derived their PLE model from the $B$-band LF.  The
expectations derived from monolithic and hierarchical scenarios should
be more compatible in the forthcoming new generation of models
(Pozzetti, private communication).  On the other hand, expectations
derived by Fontana et al.\ (\cite{fontana}) for a hierarchical model
in the framework of a $\Lambda$CDM cosmology seem to be in good
agreement with the HDFs data.  Due to the models uncertainties
affecting the comparison of cumulative redshift distributions, we
applied the more powerful test on the $K$-band LF proposed by
Kauffmann \& Charlot (\cite{kauff}), whose results are shown in
Sect.~\ref{resu}.


\section{Luminosity Functions}
\label{lf}

The Luminosity Function represents the number of objects per unit
volume with luminosities in the range $[L,L+dL]$. Differently from
galaxy counts, distances are involved in the LF computation, where the
intrinsic luminosities are considered rather then the apparent ones
(except the LFs of objects belonging to a unique structure, like
galaxy clusters).  This characteristic makes the LF an important
cosmological test, containing much more information than galaxy
counts.

The LFs are of crucial importance in the description of sample
statistical properties in observational cosmology. The LF can measure
the amount of luminous matter in the universe, depending on the
cosmological parameters. Moreover, the analysis of its characteristics
and its evolution provides fundamental insights on the galaxy
evolution mechanisms and can constrain the formation epoch.  Studying
the LFs in a wide range of absolute magnitudes is a necessity to
understand the galaxy formation process.  To attain this goal, more
and more fainter apparent magnitudes must be reached.


\subsection{LF estimators}
\label{lf_estim}

We applied three different methods to estimate the LF: the $1/V_{\rm
max}$, the $C^-$ and the STY methods. Willmer (\cite{willmer}) and
Takeuchi et al.\ (\cite{take}) have recently reviewed and compared these
estimators.

The $1/V_{\rm max}$ is the so-called \emph{classical method}, first
published by Schmidt (\cite{schmidt}), and detailed later by Felten
(\cite{felten}).  It was conceived for quasars, as many of the other
LF estimators, but it is extensively applied in the galaxy LF
computation:
\begin{equation}
V_{\rm max} = \int_{\min(z_2,z_{\rm max})}^{\max(z_1,z_{\rm min})}
\frac{dV}{dz} \,dz
\label{vmax}
\end{equation}
where $V_{\rm max}$ is the minimum comoving volume in the survey in
which the galaxy $i$ with absolute magnitude $M_i$ remains observable,
given $z_1$ and $z_2$, the redshift range of the survey.  It is a
non-parametric method, i.e.\ it does not assume a shape for the
LF.  One of the advantages of this method is its simplicity. Moreover,
it is a non-parametric estimator and thus it gives the shape and the
normalization at the same time.  Also, it is an unbiased estimator of
the source density. Of course, there are also some shortcomings. First
of all, the $1/V_{\rm max}$ estimator is very sensitive to density
fluctuations, especially in pencil beam surveys.  Furthermore, one
loses information on where in the magnitude bin a galaxy is located,
but for small bins ($dM$) it is a reasonable estimate.


The $C^-$ method was introduced by Lynden-Bell (\cite{lynden}); the
original technique was simplified and developed by Cho{\l}oniewski
(\cite{cholo}) in such a way as to compute simultaneously the shape of
the LF and the density. We adopted the Cho{\l}oniewski approach: the
observed distribution of galaxies is assumed to be separable in its
dependences on the absolute magnitude $M$ and the redshift $z$.
As the $1/V_{\rm max}$ estimator, this method is non parametric, but
it has the advantage of being insensitive to density inhomogeneities.
SubbaRao et al.\ (\cite{subba}) realized a modified version of this
method, to take into account a continuum distribution of redshifts
arising from the photometric redshift computation.  We discuss this
case later.


We also applied the STY method proposed by Sandage, Tammann \& Yahil
(\cite{sty}).  This estimator uses a maximum likelihood technique to
find the most probable parameters of an analytical LF $\phi(M)$, in
general assumed to be the Schechter function:
\begin{equation}
\phi(L)\, dL  =  \phi^* \left(\frac{L}{L^*}\right)^\alpha 
\exp \left(-\frac{L}{L^*}\right)\, d\!\left(\frac{L}{L^*}\right) \: ,
\label{phi_l}
\end{equation}
that, transformed into magnitudes, becomes:
\begin{eqnarray}
\phi(M)\, dM &  = & 0.4 \ln(10) \phi^* 10^{-0.4(M-M^*)(\alpha+1)} \times  
\nonumber \\
\            &    & \exp [-10^{-0.4(M-M^*)}] \,dM \:.
\label{phi_m}
\end{eqnarray}
Because the data are not binned, the method takes advantage of all the
information in the sample.  Imposing a functional shape, we cannot
test if the assumed $\phi(M)$ is a good representation of data.  Then
the non-parametric methods allow to plot the data points and the STY
estimator can be used to search for the better parametrization.  The
unknown parameters of equations \ref{phi_l} and \ref{phi_m} are the
normalization $\phi^*$, the faint end slope $\alpha$, that assumes
negative values, and the characteristic luminosity $L^*$ or magnitude
$M^*$, that marks the separation between the exponential law
prevailing in the bright part of the LF and the power law with index
$\alpha$ dominant in the faint end.  All these parameters can depend
on the morphological type of galaxies under consideration.  Several
methods have been conceived to compute in an unbiased way the mean
galaxy density and thus the parameter $\phi^*$.  A detailed discussion
of the density estimators can be found in Davis \& Huchra
(\cite{davis}) or in Willmer (\cite{willmer}).

Davis \& Huchra (\cite{davis}) derived a \emph{minimum variance
estimator}:
\begin{equation}
\bar{n} = \frac{\sum_{i=1}^N N_i(z_i)\, w(z_i)}
{\int_{z_1}^{z_2} s(z)\, w(z)\, \frac{dV}{dz}\, dz}
\label{n_mve}
\end{equation}
where $N_i$ is the number of galaxies at $z=z_i$ (in general $N_i=1$,
but it can take different values if galaxies are divided in redshift
bins), whereas $w(z_i)$ are the weights.  The estimator called $n_3$
by Davis \& Huchra (\cite{davis}) can be derived from equation
\ref{n_mve} by taking $w(z_i)=1$:
\begin{equation}
n_3 = \frac{N}{\int_{z_1}^{z_2} s(z)\, \frac{dV}{dz}\, dz} \: .
\label{n_3}
\end{equation}
Here all the observed galaxies are equally weighed and then the
estimator is quite stable, but it is affected by large scale
inhomogeneities: if $\left<V/V_{\rm max}\right>>0.5$, the deduced
value of $n_3$ can be overestimated.


\subsection{Beyond the spectroscopic limit: the Monte Carlo approach}
\label{montecarlo}

One of the problems in the study of LFs is the availability of
extended samples, in terms of the number of galaxies involved in such
samples, in the range of magnitudes attained (to study in detail the
faint-end behaviour) and in large redshift domains (to study the
evolution beyond $z\simeq 1$).  However, such a sample is not easy to
acquire in spectroscopic surveys, and has not yet been obtained.

A viable alternative solution is the use of the deeper and faster
photometric surveys, which allow to use photometric redshifts instead
of spectroscopic ones.  A spectroscopic subsample is anyhow
recommended, to calibrate and check photometric redshifts.  A suitable
survey according to these requirements is, once more, the HDFs.
Different groups put a particular effort in the attempt to compute the
LF of HDF-N (Gwyn \& Hartwick \cite{gwyn}; Sawicki et al.\
\cite{sawicki}; Mobasher et al.\ \cite{moba2}; Takeuchi et al.\
\cite{take}).

Up to now, the approaches used to compute the Luminosity Functions by
means of the photometric redshift technique, in the HDF-N or other
fields, can be summarized as follows:
\begin{itemize}

\item 
the photometric redshift is assumed to be the true redshift of the
galaxy, i.e.\ $z_{\rm phot} = z_{\rm true}$ (Gwyn \& Hartwick
\cite{gwyn}; Sawicki et al.\ \cite{sawicki}; Mobasher et
al.\ \cite{moba2}; Takeuchi et al.\ \cite{take});

\item 
for each object, a smooth distribution of redshifts around $z_{\rm
phot}$ is considered, with a Gaussian probability function $P(z)$. The
rms of the Gaussian distribution can be either a fixed value or a
value depending on the apparent magnitude of the galaxy under
consideration. In this case, the rms is derived from the expectations
obtained from the analysis of the spectroscopic subsample.  The
smoothed redshift can be used in different ways: SubbaRao et
al.\ (\cite{subba}) conceived a modified version of the $C^-$ method to
account for a continuous distribution of redshifts around $z_{\rm
phot}$.  Liu et al.\ (\cite{liu}) obtained an estimate of the LF using
the $1/V_{\rm max}$ method and attributing a series of fractional
contributions to the luminosity distribution in the redshift space
around $z_{\rm phot}$. Dye et al.\ \cite{dye} used the Gaussian
distribution to realize a series of Monte Carlo iterations, where the
value of each $z_{\rm phot}$ is chosen randomly according to the
Gaussian distribution.

\end{itemize}

These approaches are subject to some shortcomings, due to the nature
of the photometric redshift technique itself.  First, the photometric
redshift is only an estimate of the true redshift, and then we can
only suppose $z_{\rm phot} \approx z_{\rm true}$.  Second, in most
cases the probability distribution as a function of redshift, $P(z)$,
cannot be assimilated to a Gaussian function, because degeneracies and
uncertainties produce complex distribution functions, often with the
presence of several distinct peaks. These features of the photometric
redshift estimate lead to different distances $d_L$, absolute
magnitudes $M$, volumes $V_{\rm max}$ and also to different
$k$-corrections, because in general the spectral type of the best fit
at a fixed redshift can be different from the best fit type of the
surrounding redshift steps.  Then, all these details about the
photometric redshift technique must be taken into account when
computing the luminosity functions of galaxies belonging to a
photometric survey.

\begin{figure*}
{\centering \leavevmode
\psfig{file=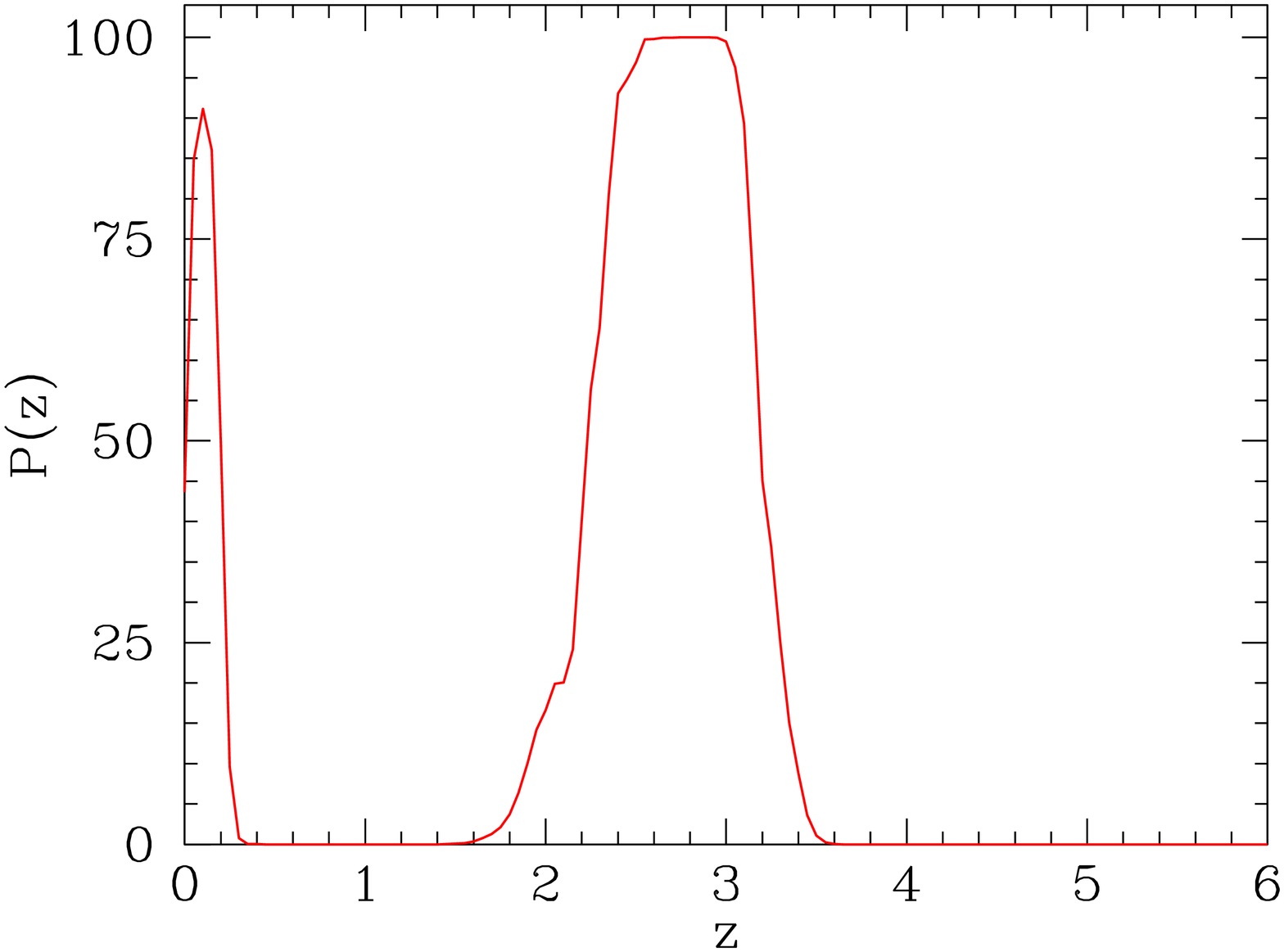,width=.49\textwidth,angle=270} \hfil
\psfig{file=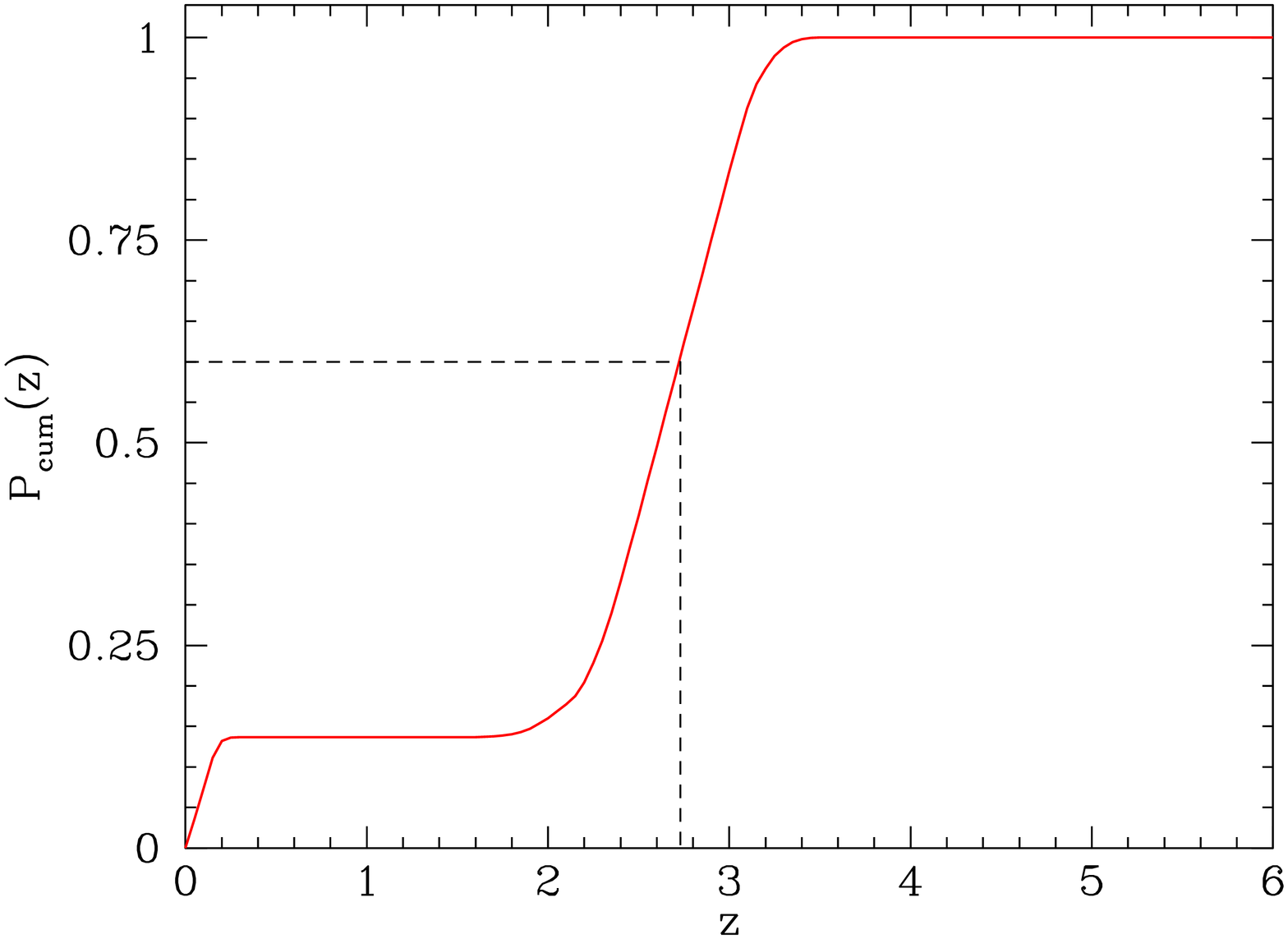,width=.49\textwidth,angle=270} } 
\caption{An example of the procedure to select the photometric
redshift randomly and according to the probability distribution of the
object.  \emph{Left:} The function $P(z)$ as computed by
\emph{hyperz}, renormalized to the maximum value.  \emph{Right:} The
corresponding cumulative probability function.  A random number
between $0$ and $1$ is drawn. The value of $z_{prob}$ is computed by
linear interpolation between the two redshift steps bracketing the
value of the cumulative function equal to the random number. }
\label{zrand}
\end{figure*}

To compute the LFs in the HDF-N and HDF-S, we adopted a Monte Carlo
approach, different from the Dye et al.\ (\cite{dye}) method in the
way of accounting for the non-gaussianity of the photometric redshift
errors.  Specifically, to assign the photometric redshift used in each
iteration of the LF estimate, we build the cumulative function $P_{\rm
cum}(z)$ from the $P(z)$, as described in Fig.~\ref{zrand}.  During
each Monte Carlo iteration, for every galaxy we randomly select a
number between $0$ and $1$, corresponding to a value of the redshift
$z_{\rm prob}$ that we assign to the considered galaxy.  The
probability of obtaining a given value of $z_{\rm prob}$ is related to
the $P(z)$ distribution: when the $P_{\rm cum}(z)$ remains horizontal,
it means that the $\chi^2$ probability relative to those redshifts is
near to $0$, and then it is almost impossible to select them by
choosing a number in the interval $[0,1]$.  Viceversa, vertical
regions of the $P_{\rm cum}(z)$ curve correspond to very likely values
of the photometric redshift.

Proceeding in this way, we use all the informations contained in the
{\tt .log\_phot} file produced as output by \emph{hyperz}, we take
into account the existence of multiple solutions, and we are able to
compute the correct $k$-corrections, knowing all the characteristics
of the best fit SED at $z=z_{\rm prob}$, i.e.\ the associated spectral
type, age and $A_V$.

The final data points of the LF or its fitting parameters will be
evaluated by means of the median over many Monte Carlo realizations.  
The errors will be immediately
found after sorting, by locating the values $x_k$ whose indexes $k$
correspond to the assigned probability.  In this way we can compute the
confidence intervals at different levels.

Another possibility to take into account the uncertainties and the
information contained in the photometric redshift procedure, is the
method described by Arnouts et al.\ (\cite{arnouts}): they performed
Monte Carlo simulations to test the effect of the photometric errors
on redshift estimates.  They assigned a random magnitude according to
the photometric rms and verified if the changes of redshift could affect
the statistics inside the redshift slices they used to divide the
sample.  We chose not to follow this approach, because the degeneracy
among different parameters can lead to a smoothed $P(z)$, even when
the photometric errors are small.  In the procedure followed by
Arnouts et al.\ (\cite{arnouts}), the presence of secondary and
significant peaks in $P(z)$ is not taken into account.

In the following, we adopt a cosmological model with parameters
$\Omega_0=1$, $\Omega_\Lambda = 0$ to facilitate the comparison of our
LF results with other surveys.  We compute the LFs also for the
fashionable cosmological model $\Omega_0=0.3$, $\Omega_\Lambda = 0.7$.


\subsection{$k$-corrections}
\label{kcorr}

Usually, $k$-corrections in redshift surveys are computed from spectra
at $z=0$, after attributing a spectro-morphological type to each
object (e.g. Lilly et al.\ \cite{lilly}; Loveday et al.\
\cite{loveday2}).  When it is impossible to separate galaxy types, a
statistical $k$-correction can be computed considering a mix of
morphological types (e.g.\ Zucca et al.\ \cite{zucca}).  The SED
fitting technique used to determine photometric redshifts allows to
compute a well suited $k$-correction for each object, obtained
directly from the best fit SED.

The absolute magnitude of an object at redshift $z$, in a given filter
$M_{\lambda}$ is:
\begin{equation}
M_{\lambda}  =  m_{\lambda} - 5\log d_L - 25 - k_{\lambda}(z) \:,
\end{equation}
where $d_L$ is the luminosity distance in Mpc, $m_{\lambda}$ is the
apparent magnitude, and $k_{\lambda}(z)$, the $k$-correction, is
defined by:
\begin{eqnarray}
k_{\lambda}(z) & = & M_{\lambda}(z,t_0)-M_{\lambda}(0,t_0) \nonumber \\
               & = & 
2.5 \log \frac{\int f(\lambda,t_0) R(\lambda)\,d\lambda}
{\int f(\frac{\lambda}{1+z},t_0) R(\lambda)\,d\lambda} 
+ 2.5\log(1+z) \:.
\label{kcorrdef}
\end{eqnarray}
Here $t_0$ is the time corresponding to the present epoch ($z=0$);
therefore, the $k$-correction is a pure geometrical correction which
transforms the observed magnitude into the magnitude in the rest-frame
of the observed galaxy, assuming no spectral evolution.  This approach
is straightforward at low redshifts, where the differences induced by
spectral evolution are negligible, and then the empirical SED
$f(\lambda,t_0)$ of a galaxy at $t_0$ ($z \sim 0$) can be used to
derive the $k$-correction of a galaxy seen at redshift $z$.  For high
redshift galaxies, the uncertainties on the evolution of the SEDs from
$f(\lambda,t_0)$ to $f(\lambda,t_z)$, that is the evolutionary
correction, make the previous approach not reliable in the context of
this paper.  The most straightforward way to compute the
$k$-correction is to apply equation \ref{kcorrdef} on the best fit SED
at redshift $z$.

In practice, to minimize the assumption on the best fit SED, we choose
the apparent magnitude in the filter $i$ which is closest to the
rest-frame filter selected for the LF, that we call $k$, and we
compute the absolute magnitude in the AB photometric system through
the equation:
\begin{eqnarray}
M_k & = & m_i - 2.5 \log 
\frac{\int f_{\rm SED}\left(\frac{\lambda}{1+z}\right) R_k(\lambda)\,d\lambda}
{\int f_{\rm SED}\left(\frac{\lambda}{1+z}\right) R_i(\lambda)\,d\lambda} 
\nonumber \\
    &   & + 2.5 \log \frac{\int f_{\rm Vega}(\lambda) R_k(\lambda)\,d\lambda}
{\int f_{\rm Vega}(\lambda) R_i(\lambda)\,d\lambda} \nonumber \\
    &   & - 5\log d_L - 25 
- 2.5 \log \frac{\int f_{\rm SED}(\lambda) R_k(\lambda)\,d\lambda \,(1+z)}
{\int f_{\rm SED}\left(\frac{\lambda}{1+z}\right) R_k(\lambda)\,d\lambda} 
\nonumber \\
    &   & - {\rm conv}_{{\rm AB},i} + {\rm conv}_{{\rm AB},k}
\label{mabs}
\end{eqnarray}
where the first two lines give the correction to obtain the apparent
magnitude in the filter $k$, the third line corresponds to the passage
from apparent to absolute magnitudes and the forth line contains the
corrections to transform the input AB magnitudes in the Vega system
for the $i$ filter and viceversa for the $k$ filter, to obtain the
final AB magnitude.

Thus, we do not introduce the evolutionary correction explicitly, but
we use the most reliable $k$-corrected magnitudes, based on the best
fit SEDs.  In this way, the evolution of the galaxy population can be
directly compared with the expectations from the different models of
galaxy formation and evolution.

\begin{figure}
\centerline{\psfig{file=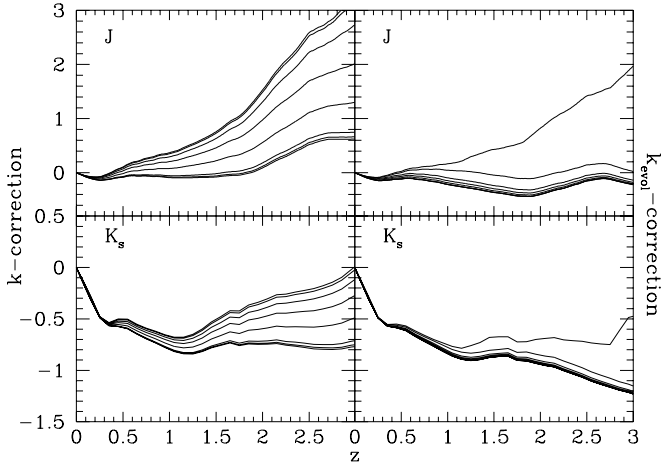,width=0.49\textwidth,angle=270}}
\caption{$k$-corrections in $J$ and $K_s$ bands, for the same spectral
types contained in the \emph{hyperz} package. Lines from top to bottom
represent $k$-correction from early to late type galaxies.
\emph{Left:} $k$-correction in magnitudes, computed from the different
$t_0$ SEDs.  \emph{Right:} $k$-correction in magnitudes, computed from
the evolving SEDs.}
\label{kcorrfig}
\end{figure}

One of the advantages of working on NIR wavebands is that the
$k$-corrections are small and nearly independent on spectral type,
thus minimizing the uncertainties in the estimate of the absolute
magnitudes.  In particular, shallow redshift surveys use $-2.5z$ for
all galaxy types in the $K$ band (Loveday \cite{loveday}; Glazebrook
et al.\ \cite{glaze}).  This represents a good approximation for
galaxies with $z<0.30$, but not in our case, because we are dealing
with higher redshift objects.  Therefore, we adopted a consistent
technique, using the SED corresponding to the best fit at the selected
$z_{\rm phot}$ from the Monte Carlo procedure.  In Fig.~\ref{kcorrfig}
we show the $k$-corrections in the $J$ and $K_s$ bands used in the LF
computation, for both the usual correction based on the $t_0$ SED,
and the $k_{\rm evol}$-correction, computed from the evolved SED at
$t_z$, assuming that galaxies form at $z=10$.  The lines represent,
from top to bottom, the $k$-corrections from early to late type
galaxies.  The $k_{\rm evol}$-correction may not represent the
correction actually applied, because galaxies can have best fit SEDs
with younger ages than $t_z$, but the plot is valid to show the
overall trend.

In particular for the $K_s$-band, we can remark that up to redshift
$z=3$ the $k$ and $k_{\rm evol}$-corrections are nearly independent on
the spectral type and small.  For these reasons we considered reliable
the extrapolation of absolute magnitudes up to at least redshift $\sim
2$ in the $K_s$ filter and, with some caution, up to $z \sim 3$.


\subsection{Incompleteness}
\label{incompl}

The advantage of using a photometric catalogue is that it is less
subject to incompleteness then spectroscopic redshift surveys.  The
incompleteness in redshift surveys, that can affect the LF estimate,
can be not only magnitude-dependent, but it can also arise as a
function of galaxy type or redshift, and in some cases it is
impossible to take it into account.

\begin{figure}
\centerline{\psfig{file=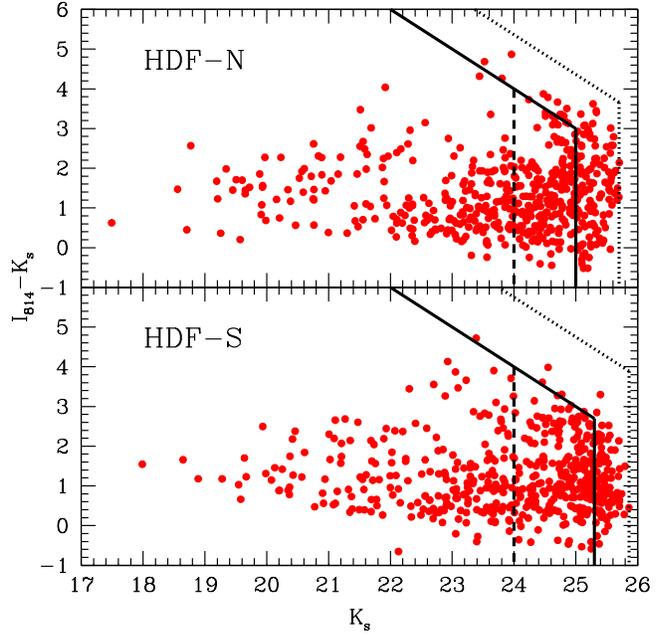,width=0.49\textwidth}}
\caption{$I_{814}- K_s$ CM diagrams for the $K_s$-band selected
subsamples.  Thick solid lines illustrate the limiting colours
computed according to Table \ref{filters}, corresponding to $80$\% of
the cumulative histogram of apparent magnitudes, whereas thick dotted
lines correspond to the very faint limit of magnitudes in the whole
catalogue. The dashed lines represent the actual limits used to
compute the LFs. }
\label{kik}
\end{figure}

Obviously, also photometric surveys are affected by incompleteness.
The SUNY catalogue in the HDF-N is, by construction, an $I_{814}$ band
selected sample (Fernandez-Soto et al. \cite{fsoto}), whereas objects
in the ISAAC HDF-S catalog are detected on the $V_{606}+I_{814}$
image.  Even though we use a $K_s$ limited sample in the subsequent
calculations, it is worth to check on the possible color-selection
effects which could affect the two fields in different ways.  Figures
\ref{z_ik} and \ref{kik} display the color-redshift and
color-magnitude diagrams.  As shown in Fig.~\ref{kik}, very red
objects in $I_{814}- K_s$, with faint $K_s \ge 24.5$ magnitudes, could
be missed due to selection criteria.  To avoid such effects of colour
selection, we adopted a limiting magnitude $K_s=24$, corresponding to
S/N $\sim$ 3. Moreover, this selection allows us to use the whole area
of $5.31$\,arcmin$^2$ in the HDF-N, combination of the Z1 and Z2
zones, because at this limit the colour distributions of objects with
redshifts between $0$ and $2$ have a similar average:
$\left<I_{814}-K_s\right> = 1.28$ in Z1 and $0.95$ in Z2, with a large
dispersion in both zones.  In the Z2 zone, even for the faintest
objects in $K_s$, the colours up to $2$ -- $2.5$ are allowed: in the
Z1 zone most objects have colours below this value, thus we do not
introduce any bias by combining the data belonging to the two zones.

At the selected limit in magnitude, blue galaxies have about the same
chance to be observed than the reddest ones in the $K_s$-band selected
subsamples in both fields.

Another type of incompleteness could arise from the surface brightness
effect: when objects are detected at bright surface brightness limit,
then the LF estimate could be affected, with $M^*$ becoming fainter,
$\phi^*$ smaller and $\alpha$ slightly flatter (Cross \&
Driver~\cite{cross} and references therein).  In our case, the
detection up to faint surface brightness used by the authors of the
catalogues ($\mu_{\rm lim} (I_814) \simeq 26$\,arcsec$^{-2}$ in both
the HDF-N and HDF-S) will not induce significant effects on the LF
estimates.  In particular, the bright end of the LF, on which we base
our conclusions, will not suffer strongly from the mentioned effect.
The same inference can be demonstrated for other types of
incompleteness, such as the detection and measurement algorithm, or
the cosmological dimming of surface brightness, discussed by Yoshii
(\cite{yoshii}) and Totani \& Yoshii (\cite{totani}), affecting the
very faint part of the sample at the limit of the selection and then
unable to invalidate our conclusions.

The quantity $V_{\rm max}$ used in the $1/V_{\rm max}$ method to
compute LFs can also be used to test the completeness of the sample:
if the set of observed galaxies is complete, we expect that they
populate uniformly the volume of the survey, i.e.\ that the galaxies
are randomly distributed inside their $V_{\rm max}$ volume.  This
corresponds to the condition $\left<V/V_{\rm max}\right> = 0.5$, where
$V$ is the volume characteristic of each galaxy, given its redshift
and the limiting magnitude of the survey.  However, this line of
reasoning is valid only if the population does not evolve in
luminosity and it is spatially homogeneous.  Larger or smaller values
can have different origins.  When the sample is subject to magnitude
incompleteness to the limiting magnitude (the more distant galaxies
become undetectable), the volume $V_{\rm max}$ becomes too big and we
have $\left<V/V_{\rm max}\right> < 0.5$.  The same effect can be the
result of luminosity evolution, if the nearest objects are also the
intrinsically brightest ones.  A value $\left<V/V_{\rm max}\right> >
0.5$ could be the effect of luminosity evolution, with the brightest
objects being the most distant ones.  In Sect.~\ref{resu} we list
the values of $\left<V/V_{\rm max}\right>$ averaged over $100$ Monte
Carlo realizations, which actually range between $0.43$ and $0.54$,
thus very close to the theoretical completeness value.


\subsection{Test of the method through mock catalogues}
\label{test}

The reliability of the method has been tested by means of mock
catalogues.  To this aim, we used template galaxies belonging to four
spectral types, built from the GISSEL98 library (Bruzual \& Charlot
\cite{bruzual}), corresponding to a star formation in a single burst,
star formations with timescales $\tau = 3$\,Gyr and $\tau = 15$\,Gyr,
and a continuous star formation, each one with Scalo (\cite{scalo})
IMF.  The four spectral types match the coulours of Elliptical, Sa, Sc
and Irregular galaxies.  A single, nonevolving luminosity function is
used, with a fraction of $\phi^*$ assigned to each type following the
mix of morphological types in the local universe used by Pozzetti et
al. (\cite{pozz}) to build their PLE models.  In particular, we
assigned a fraction of $0.28$, $0.47$, $0.22$ and $0.03$ to the four
types, from early to late, assuming that these fractions remain valid
beyond the original limit of $b_J \le 16.5$. 
 
\begin{figure*}
\vspace{-1.cm}
\centerline{\psfig{file=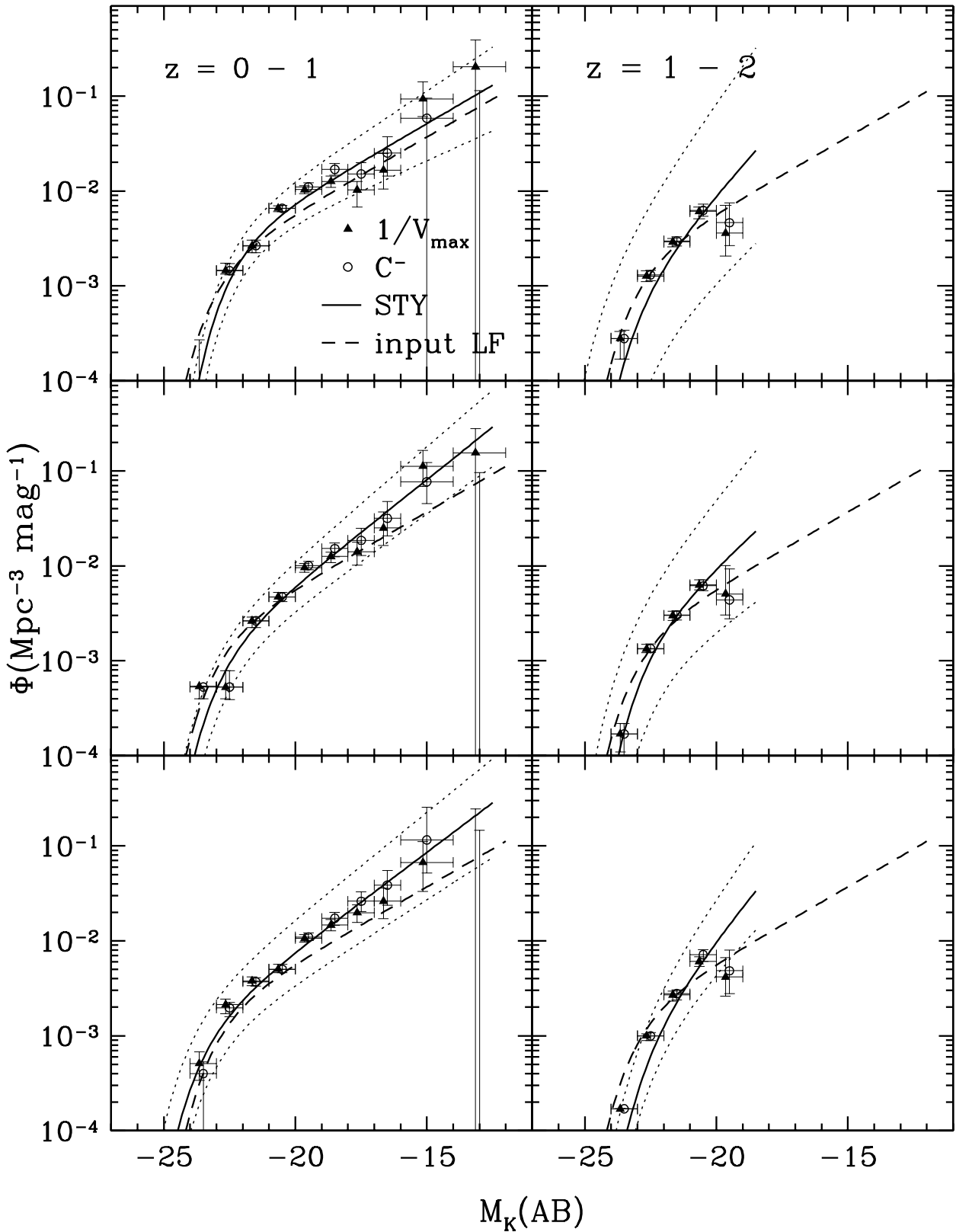,width=0.97\textwidth}}
\caption{The input and recovered LFs for $3$ mock catalogues in two
redshift ranges, selecting objects with $K \le 24$, i.e. $S/N \ge 3$.
Triangles and circles represent the binned LFs recovered with the
$1/V_{\rm max}$ and the $C^-$ methods respectively.  The points
relative to the $1/V_{\rm max}$ method have been shifted by $-0.15$
magnitudes to avoid superpositions.  Error bars represent $10$ and
$90$\% of the distribution of values obtained with the Monte Carlo
method.  The solid line is the Schechter function recovered using the
STY method, and the dotted lines are the $10$ and $90$\% STY
functions.  The thick dashed line is the input LF. }
\label{lfKsimul}
\end{figure*}

The number of galaxies in a redshift slice $[z,z+\Delta z]$ and in a
range of absolute magnitude $[M,M+\Delta M]$ has been computed by the
following integrals:
\begin{eqnarray}
\lefteqn{N(z,z+\Delta z; M,M+\Delta M) =} \nonumber \\
&& \qquad \qquad \omega \int_z^{z+\Delta z} \int_M^{M+\Delta M} 
\phi_{\rm input}(M)\,dM \frac{dV}{dz}\,dz \: ,
\end{eqnarray}
where $\omega$ is the solid angle covered by the survey, $\phi_{\rm
input}(M)$ is the input Luminosity Function and $dV$ is the volume,
dependent on the cosmology.  Apparent magnitudes in all the
filters are computed from the absolute magnitude $M$, the template
spectra and the redshift $z$.  We decided to compute apparent
magnitudes using the inversion of equation \ref{mabs}, viz applying
only the $k$-correction on the evolved SED, because here we are
interested only in testing the reliability of the output compared with
the input data, using the same procedure adopted for real data.  In a
realistic case the evolutionary correction becomes important:
recovering absolute magnitudes with equation \ref{mabs} should produce
an evolution of the measured LF even if the input LF is nonevolving,
as discussed in Sect.~\ref{kcorr}.

For the input LF we imposed a Schechter functional form in the $K$
band, with parameters $\alpha=-1.40$, $M^*_{K}=-23.14$ in AB
magnitudes and a normalization $\phi^*=0.002\,{\rm Mpc}^{-3}$, setting
$H_0 = 50\,{\rm km\, s^{-1}\, Mpc^{-1}}$.  We used the same cosmology
to build mock catalogues and to recover the LF.  We discuss the
influence of the world models in this kind of calculation in
Sect.~\ref{cosmo}.

In our simulated catalogues, we reproduce the same observational
effects affecting the real catalogues of the HDFs.  In particular, the
signal-to-noise ratio behaviour in each filter band is set to be
consistent with the data.  Galaxies included in the mock catalogues
have magnitudes brighter than the limiting magnitude, otherwise their
magnitude is set equal to $99$, corresponding to a non detected object
in the syntax of \emph{hyperz}.  The limiting magnitude corresponds to
a signal-to-noise ratio $S/N=1$, for consistency with the real HDFs
catalogues, but we considered only objects with $S/N \ge 3$ to
estimate LFs.  The photometric error is computed as a function of
magnitude for each filter, after specifying a signal-to-noise ratio
reached at a given magnitude, matching the same ratios computed for
the HDFs catalogues and using a similar procedure as in Bolzonella et
al.\ (\cite{hyperz}).  A random reddening with $A_V$ ranging from $0$
to $1$ is also applied to SEDs.  To allow a photometric redshift
estimate, we included in the catalogues only objects detected in at
least $3$ filters.  The number of galaxies included in a mock
catalogue has been computed using a surface similar to the HDFs, i.e.\
$5$\,arcmin$^2$.  In this way we obtained $N_{\rm obj}$ objects, to
which we added a fluctuation due to Poissonian statistics, $\pm {\rm
random}[0,\sqrt{N_{\rm obj}}]$.  Objects have been randomly selected
from a larger field, to allow the selection of bright objects at low
redshift.  The final catalogues have roughly the same number of
objects as the HDFs, with the same observational characteristics.
  
\begin{figure}
\centerline{\psfig{file=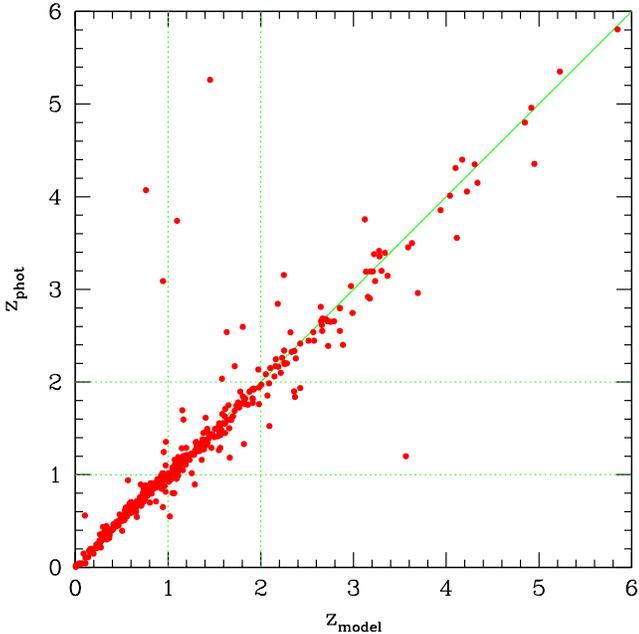,width=0.49\textwidth}}
\caption{Comparison between the input redshift $z_{\rm model}$ and the
photometric redshift best fit $z_{\rm phot}$ for the objects with $K
\le 24$} in the simulated catalogue used to recover the input LF in the
upper panel of Figure~\ref{lfKsimul}.  Dotted lines identify the two
considered redshift bins.
\label{zzsimul}
\end{figure}

Next, we computed photometric redshifts for these simulated galaxies,
using all the available SEDs, and we used the \emph{hyperz} outputs
(the probability function of redshift, the SED parameters) to estimate
absolute magnitudes and then the Luminosity Functions by means of the
Monte Carlo method described in the previous section, selecting
objects in the $K_s$ and with magnitudes $K_s \le 24$, i.e. $S/N \ge
3$.

The recovered LFs for $3$ random catalogues are shown in
Fig.~\ref{lfKsimul} for two redshift ranges.  The mean parameters
obtained averaging over a set of $10$ random catalogues are
$\overline{\phi^*} = 0.00267\pm 0.00141\,{\rm Mpc}^{-3}$,
$\overline{M^*_K} = -22.958 \pm 0.771$ and $\overline{\alpha} =
-1.441\pm 0.083$ in the redshift range $z=0$ -- $1$.  The estimated
errors in these values correspond to the standard deviation
($1\sigma$) of the distribution of values used to compute the
arithmetic mean and do not take into account the errors inferred from
each Monte Carlo realization.  The agreement of these values with the
input ones is remarkable.  In the redshift range $z=1$ -- $2$ the
recovered parameters are $\overline{\phi^*} = 0.00259 \pm
0.00207\,{\rm Mpc}^{-3}$, $\overline{M^*_K} = -22.611 \pm 0.489$,
$\overline{\alpha} = -1.681 \pm 0.208$.  In this case the
normalization is well recovered, even if the large error reflects the
large scatter in the values obtained in different realizations; the
value of $M^*_K$ is consistent with the input one, whereas the
recovered $\alpha$ is slightly overestimated even considering the
$1\sigma$ error.

The comparison between the input value of the redshift and the
photometric redshift best fit is shown in Fig.~\ref{zzsimul} for the
objects with $S/N \ge 3$ in $K$ magnitudes: for these objects the
photometric redshift is a very good estimate of the input one and we
do not see the degeneracy between different redshift ranges,
characteristic of the faintest objects with lower signal-to-noise
ratio (see Bolzonella et al.\ \cite{hyperz}).  Only few objects are
erroneously attributed to high redshifts: these objects are very faint
galaxies, non detected in the $U_{300}$ and $V_{450}$ filters,
determining a confusion in the location of the Lyman break.

\begin{figure*}
{\centering \leavevmode 
\psfig{file=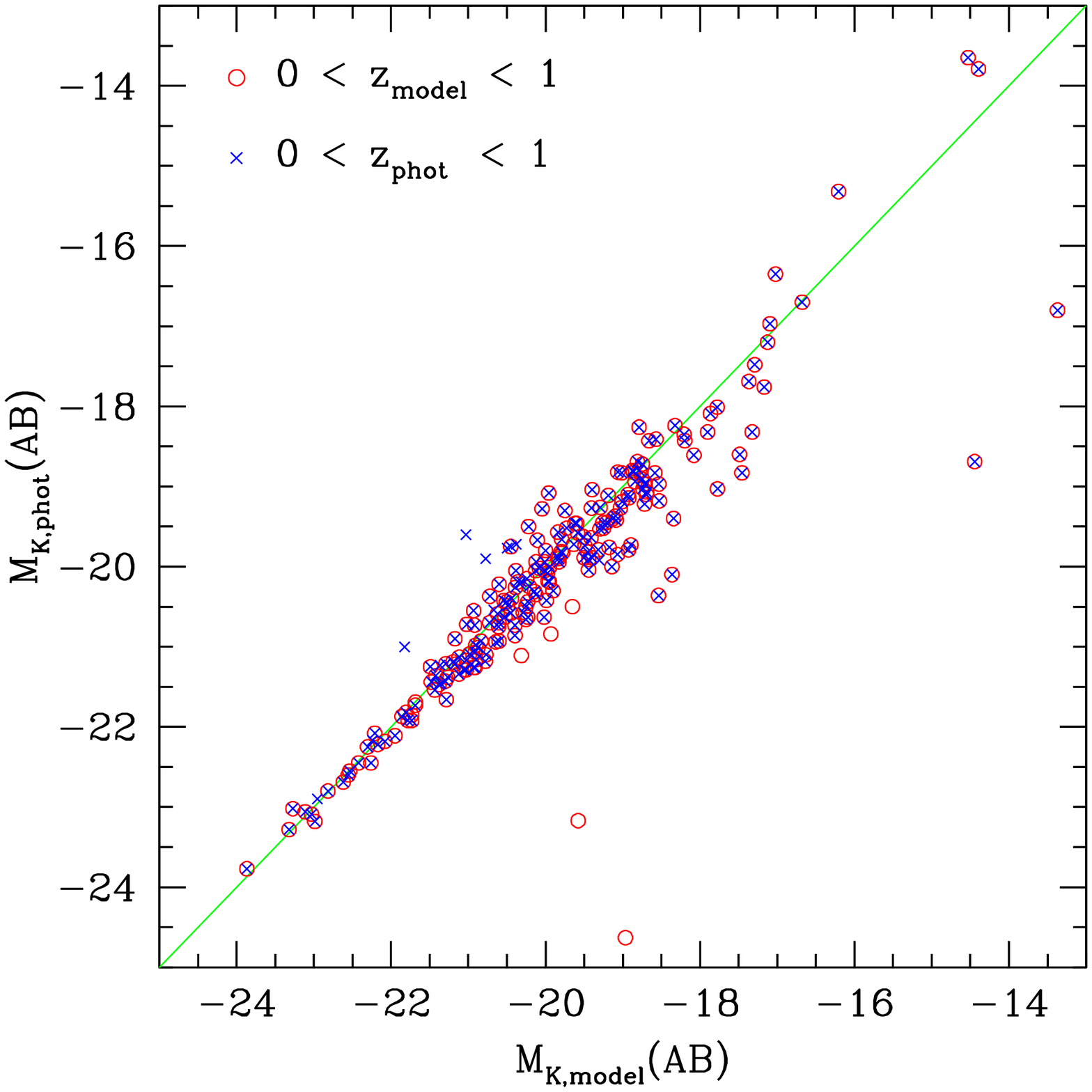,width=.49\textwidth} \hfil 
\psfig{file=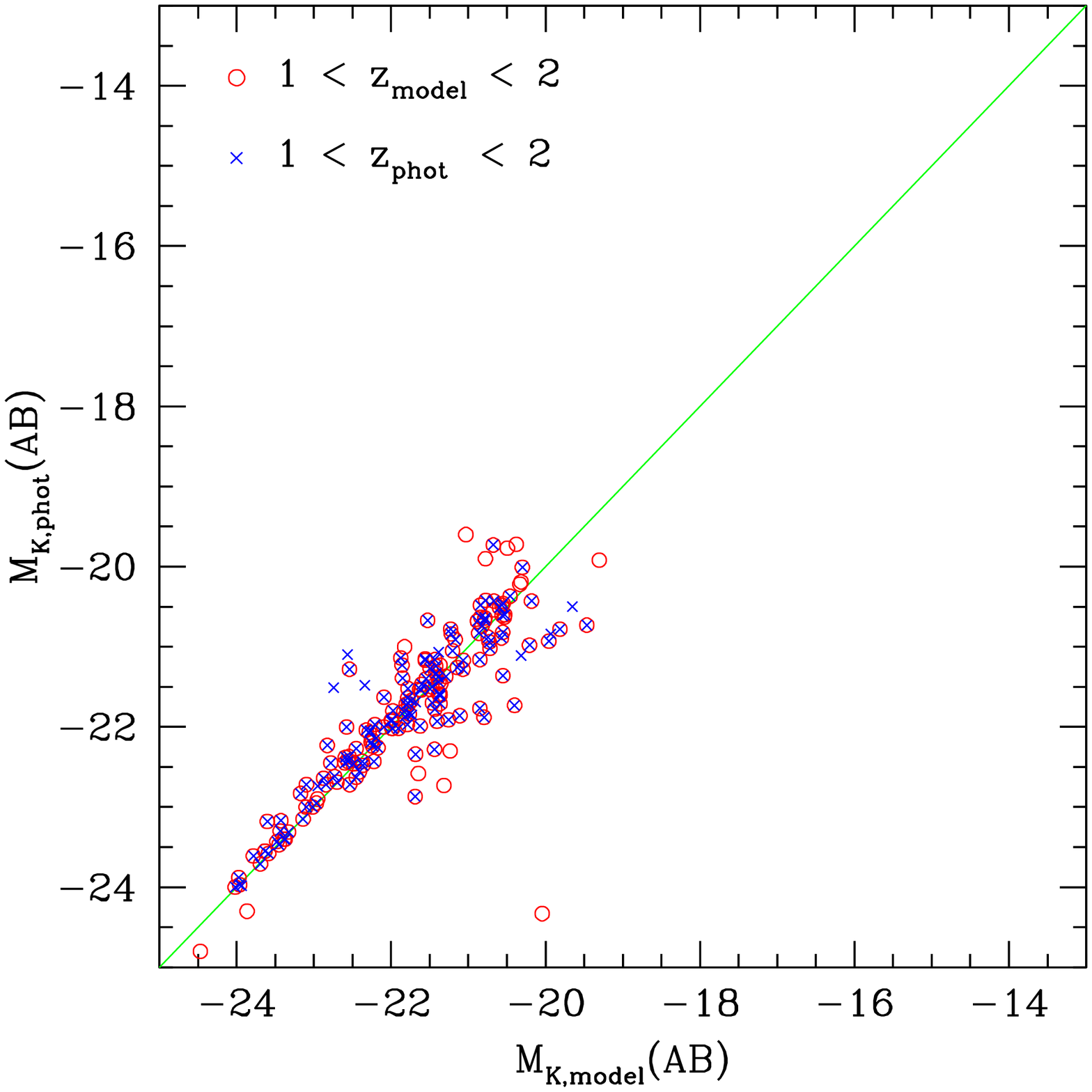,width=.49\textwidth}} 
\caption{Comparison between the input absolute magnitude $M_{K,{\rm
model}}$ and the absolute magnitude computed using the photometric
redshift best fit $M_{K,{\rm phot}}$, for the mock catalogue used in
the upper panel of Fig.~\ref{lfKsimul} and objects with $K \le 24$.
\emph{Left:} redshift range $z=0$ -- $1$.  \emph{Right:} $z=1$ -- $2$.
Circles and crosses represent the points selected following the input
redshift and the photometric redshift respectively. }
\label{mmsimul}
\end{figure*}

In Fig.~\ref{mmsimul} the input absolute magnitude is compared to the
absolute magnitude computed by \emph{hyperz}, using the redshift and
the spectral type best fit.  The objects represented in these two
panels have been selected using their photometric and their input
redshifts, in the two redshift bins shown in Fig.~\ref{lfKsimul}.  We
considered only objects with $K \le 24$, i.e. the same objects used in
the LF estimate.  The agreement is very good, with very few outliers.
In the left panel, we can notice the rapid decrease of galaxies with
$M_K \ga -16$, producing large error bars in the binned estimate of
the LF.  In the range between $z=1$ and $2$, the lack of faint objects
due to observational limits prevents a good estimate of the faint-end
slope $\alpha$ using the STY method. Nevertheless, the bright end is
well reproduced by the binned methods.

Similar analysis comparing methods for LF estimate through simulations
have been carried out by other authors.  Willmer (\cite{willmer})
found that the STY method tends to slightly underestimate the
faint-end compared to the input value.  In our case, we found a small
overestimate of $\alpha$, but the large error bars are consistent with
the input value.  Concerning the non parametric methods, the study of
Takeuchi et al.\ (\cite{take}) demonstrated that for large and
spatially homogeneous samples the LF estimate is not biased, whereas
the faint-end is subject to large fluctuations when the sample is
small.  The overestimate of the low redshift LF on the faintest bins
is similar to ours, according to their figures, and still consistent
with the input LF due to the large error bars.  On the contrary,
at higher redshift we see a decline in the faintest bins, that is also
been shown by Liu et al.\ (\cite{liu}).  

In summary, the procedure used to recover the LF in the range $z=0$ --
$1$ can be considered as reliable, and we did not try to take into
account the small systematic effects mentioned above. In the range
$z=1$ -- $2$, the results of the STY method have to be taken with
care, but the non parametric estimate of the LF still provides a good
fit of the bright-end.


\section{Near infrared luminosity functions in the HDFs}
\label{resu}

\begin{figure*}
{\centering \leavevmode
\psfig{file=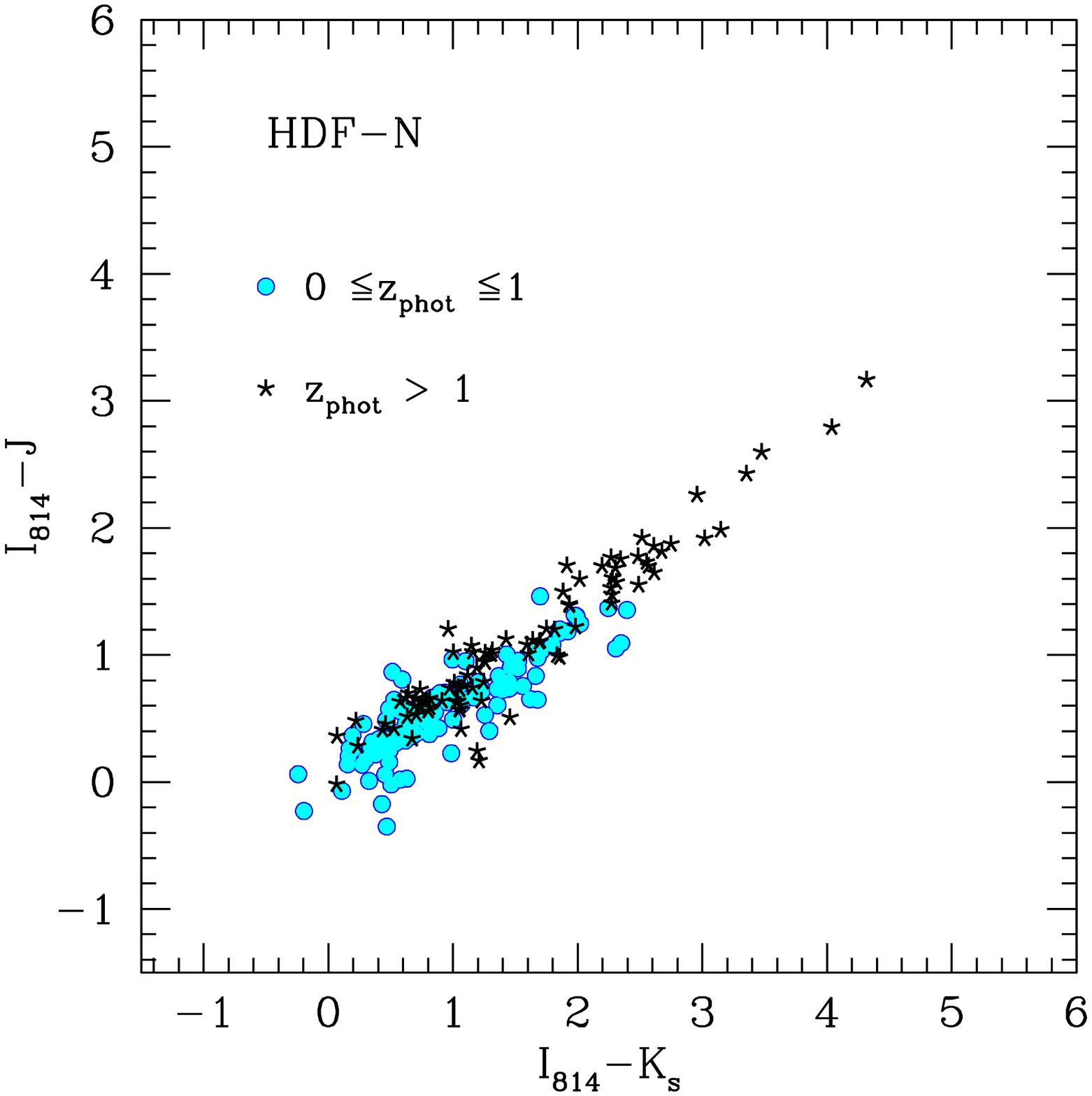,width=.49\textwidth} \hfil
\psfig{file=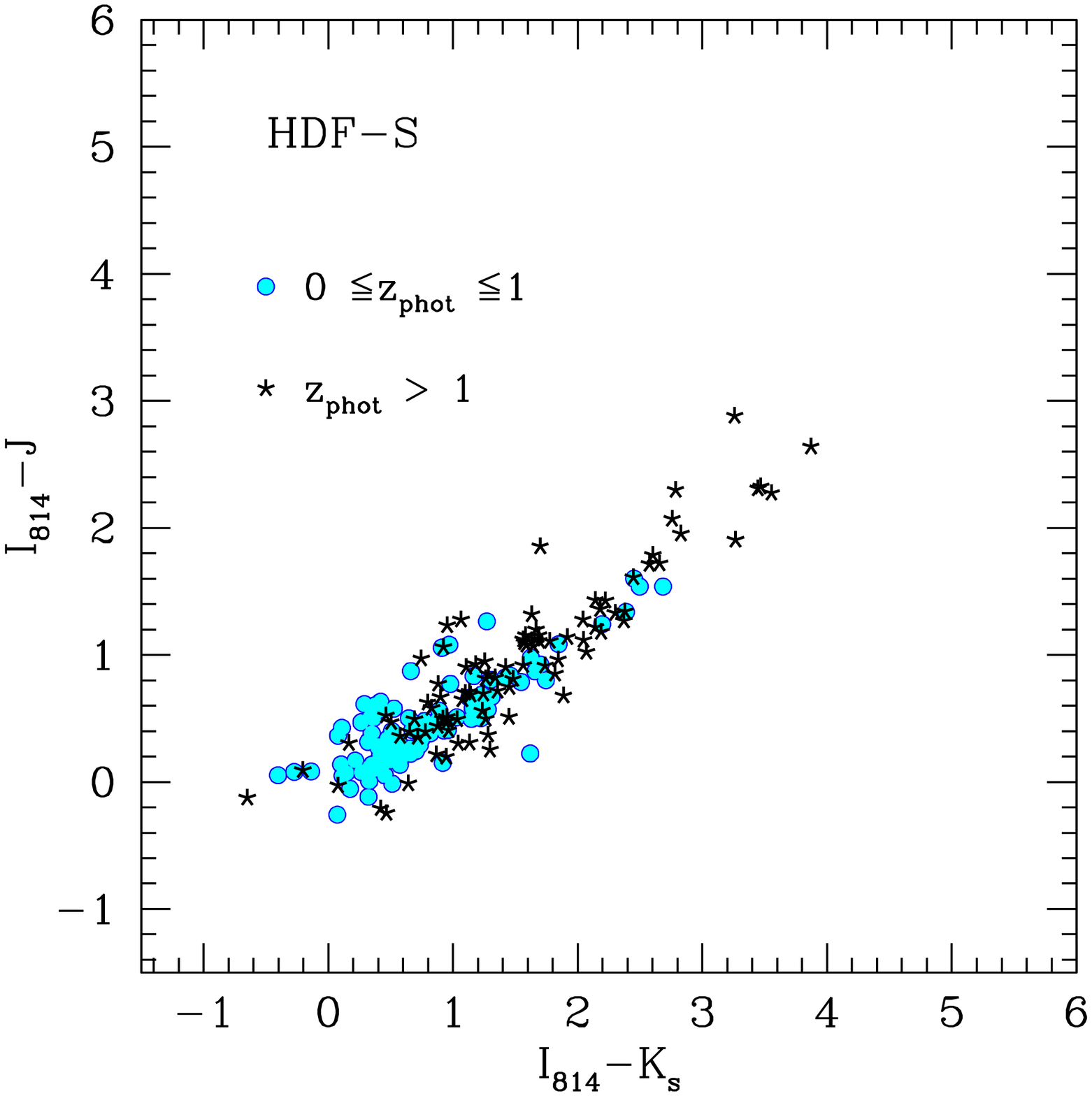,width=.49\textwidth}} 
\caption{NIR colour-colour plot: $I_{814}-K_s$ vs $I_{814}-J$ for the
HDF-N and HDF-S.  Limiting magnitudes are $I_{814}=28.5$, $J=24.6$ and
$K_s=24.0$, corresponding to $S/N=3$. }
\label{ikij}
\end{figure*}

We compute the LFs using the three methods described in
Sect.~\ref{lf_estim}.  In particular, for the non-parametric methods
we adopt a binning of $1$ magnitude or $2$ magnitudes at the faint end
to have a conspicuous number of objects in each bin.  To compute the
luminosity functions we divide the sample in redshift slices, larger
than the typical errors of photometric redshifts, to minimize the
change of redshift bin and to study the redshift evolution of the LFs.
The adopted limiting magnitude is $K_s = 24$ in both the HDF-N and
HDF-S.  Following the procedure described above, we iterate the
computation of the binned or parametrized LF.  We choose to realize
$100$ iterations, sufficient to estimate the effect of random
selection of redshifts.  During the photometric redshift calculation
we impose a range of absolute $B$ magnitudes: solutions with $M_B$
outside the range $[-28,-9]$ are considered forbidden also when
estimating the LF.

Because the $K_s$-band is the reddest one, the absolute magnitudes
have been computed using always the $K_s$ apparent magnitudes.
Studying the $K$-band at redshifts $z\sim 2$ means to map the
rest-frame $I$-band emission: in Fig.~\ref{ikij} we show that these
magnitudes are strictly correlated and thus they basically map the
same stellar population.  This behaviour can be explained considering
the emitting stellar population: even at $z \sim 2$ the luminosity at
the wavelengths covered by the $K_s$ filter is always produced by the
old star population, the $4000$\,\AA\ break still being inside the
$J$ filter.  Furthermore, $K_s$ magnitudes are not affected by recent
bursts of star formation.  For this reason, and because of the
characteristics of the $k$-correction discussed in
Sect.~\ref{kcorr}, we considered that we can safely compute
$K_s$-band absolute magnitudes at least up to a redshift of $2$.

We have also estimated the $J$-band LF in the redshift range $z=[0,1]$
from the $J$-selected subsample, and in the redshift range $[1,2]$ by
selecting the objects in the $K_s$-band sample that better approximate
the $J$ filter rest frame.

We will study the LF in the optical bands and in the UV in a
forthcoming paper (Bolzonella et al. in preparation).


\subsection{$K_s$-band LF}

Up to now, the $K_s$-band LF has been estimated only for data in
redshift surveys, and for this reason the accessible range of
redshifts to be explored was limited.  In Table~\ref{ktable} we report
previous estimates, mainly retrieved from the table by Loveday
(\cite{loveday}), completed with a guess of the magnitude and redshift
ranges of the samples.  Magnitudes have been transformed into the AB
system by means of the conversion provided in Table~\ref{filters} and
the dependence from the Hubble constant has been explicited using the
Hubble parameter $h=H_0/(100\,{\rm km\, s^{-1}\, Mpc^{-1}})$.

\tabcolsep 0.1cm
\begin{table*}
\caption{Compilation of other $K_s$-band LFs for field galaxies in
redshift surveys, in general obtained with cosmological parameters
$\Omega_0 =1$; the data are transformed into AB magnitudes in the
filter $K$.  Errors are in general given at $1\sigma$ of confidence
likelihood contours.  Modified and updated from Loveday's
(\cite{loveday}) table.  }
\begin{tabular}{lr@{$\pm$}lr@{$\pm$}lr@{$\pm$}lccl}
Author(s)   & \multicolumn{2}{c}{$\phi^*$ [$h^{3}\,$Mpc$^{-3}$]}
            & \multicolumn{2}{c}{$M^*_K-5\log h$}
            & \multicolumn{2}{c}{$\alpha$}
            & $M_K-5\log h$ range & $z$ range & other remarks\\
\hline
Mobasher et al.\ (\cite{mobasher})   & $0.0112$ & $0.0016$ 
                          & $-21.51$ & $0.3^1$ 
                          & $-1.0$   & $0.3$ 
                          & $[-23.5,-19.0]$ & $0.0\le z \la 0.1$ 
                          & AARS \\
Glazebrook et al.\ (\cite{glaze}) & $0.029$  & $0.007$ 
                          & $-21.15$ & $0.23^2$ 
                          & $-1.04$  & $0.31$
                          & $[-22.5,-18.5]$ & $0.0<z<0.2$ 
                          & \\
Glazebrook et al.\ (\cite{glaze}) & $0.019$  & $ 0.002$ 
                          & $-21.44$ & $ 0.11^2$ 
                          & $-1.04$  & $0.31$ 
                          & $[-24.0,-18.5]$ & $0.0<z<0.8$ 
                          & \\
Cowie et al.\ (\cite{cowie2})      & \multicolumn{2}{c}{$0.016$} 
                          & \multicolumn{2}{l}{$-21.63$} 
                          & \multicolumn{2}{l}{$-1.25$} 
                          & $[-23.5,-16.5]$ & $0.0<z \la 1.6$ 
                          & $K$-band selected\\
Gardner et al.\ (\cite{gardner})$^3$& \multicolumn{2}{c}{$0.0182$} 
                          & $-21.50$ & $0.17$ 
                          & $-1.03$  & $0.24$ 
                          & $[-23.5,-18.0]$ & $\left<z\right>=0.14$ 
                          & $K$-band selected \\
Szokoly et al.\ (\cite{szokoly})    & $0.012$ & $ 0.004$ 
                          &$-21.72$ & $ 0.3$ 
                          & $-1.27$ & $ 0.2$
                          & $[-23.5,-18.5]$ & $0.0<z \la 0.5$ 
                          & $K$-band selected \\
Loveday (\cite{loveday})            & $0.012$ & $ 0.008$ 
                          & $-21.71$ & $ 0.42$ 
                          & $-1.16$ & $ 0.19$
                          & $[-24.0,-14.0]$ & $\left<z\right>=0.05$ 
                          & Stromlo-APM \\
Kochanek et al.\ (\cite{kocha})$^4$   & $0.0116$ & $ 0.001$ 
                          & $-21.52$ & $ 0.05$ 
                          & $-1.09$ & $ 0.06$
                          & $[-24.0,-18.5]$ & $0.01 \la z \la 0.05$ 
                          & 2MASS \\
Cole et al.\ (\cite{cole1})      & $0.0108$ & $ 0.0016$ 
                          & $-21.57$ & $ 0.03$ 
                          & $-0.96$ & $ 0.05$
                          & $[-24.0,-17.0]$ & $0 < z \la 0.2$ 
                          & 2dF - $\Lambda$CDM \\
Balogh et al.\ (\cite{balogh})$^5$ & \multicolumn{2}{c}{---} 
                          & $-21.61$ & $ 0.08$ 
                          & $-1.10$ & $ 0.14$
                          & $[-24.0,-18.0]$ & $0 < z \la 0.18$ 
                          & \\ 
\hline
\end{tabular}

$^1$ {\small From Loveday (\cite{loveday}): added $0.22$ magnitudes due
to different method for $k$-corrections.}\\ 
$^2$ {\small From Loveday (\cite{loveday}): aperture correction of 
$-0.30$ magnitudes.}\\ 
$^3$ {\small The value adopted is relative to the $q_0=0.5$ model with 
only $k$-correction. }\\ 
$^4$ {\small Isophotal magnitude selection $7<K_{20}<11.25$.}\\ 
$^5$ {\small Normalized to the weighted number of galaxies brighter
than $K_{\rm Vega}=-21.5$.}
\label{ktable}
\end{table*}

Using the HDF-N and HDF-S catalogues we can reach unprecedented
depths: we estimated for the first time the LFs in the $K_s$ band for
objects with $z_{\rm phot} \in [0,1]$ and $[1,2]$.  

We selected the subsamples in each redshift range after the
randomization procedure of redshifts.  The values of $\left<V/V_{\rm
max}\right>$ averaged over the Monte Carlo realizations are $0.54,
0.44$ in the redshift ranges $z_{\rm phot} \in [0,1], [1,2]$ for the
HDF-N and $0.51,0.47$ for the HDF-S in the same redshift ranges.

\begin{figure*}
\centerline{\psfig{file=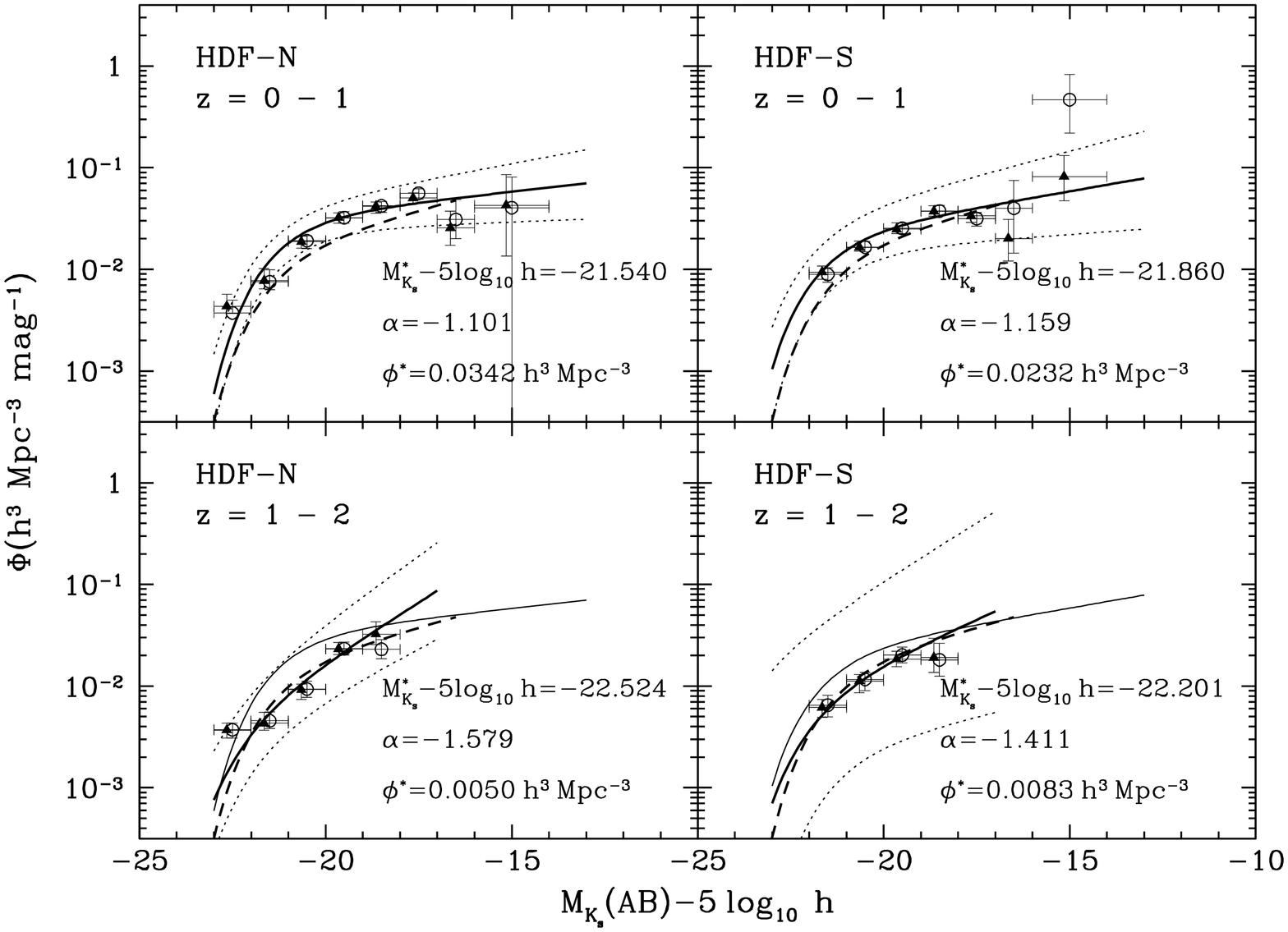,width=0.99\textwidth,angle=270}}
\caption{$K_s$-band LFs for galaxies in the HDF-N and HDF-S in two
redshift bins, assuming a limiting magnitude $K_s=24.0$ and a
cosmological model with $\Omega_0=1$.  Triangles and circles represent
the median of the results obtained with the $1/V_{\rm max}$ and the
$C^-$ methods respectively.  The upper panels represent the LF for
galaxies with $z_{\rm phot}$ in $[0,1]$, the lower in $[1,2]$. The
points relative to the $1/V_{\rm max}$ method have been shifted of
$-0.15$ magnitudes for clarity.  The error bars corresponding to $10$
and $90$ percentiles of the distribution are also shown.  The solid
line shows the LF obtained with the STY method.  The LF obtained by
Cowie et al.\ (\protect\cite{cowie2}) (dashed line) is shown as
reference LF.  In the lower panels, the $z_{\rm phot} = [0,1]$ LF
(thin solid line) is also displayed for comparison.}
\label{lf_k_hdfns}
\end{figure*}

Figure~\ref{lf_k_hdfns} illustrates our estimate of the $K_s$ band LFs
for the HDF-N and the HDF-S.  There is a good agreement between the
local $K$-band LF computed by Cowie et al.\ (\cite{cowie,cowie2}) and
the $z_{\rm phot} \in [0,1]$ sample, especially in the HDF-S. In the
case of the $z_{\rm phot} \in [1,2]$ sample, our results are still
compatible with the local values.  A slight negative evolution is
observed between the $[0,1]$ and $[1,2]$ bins when comparing the
non-parametric estimates for galaxies fainter than $M_K=-21$.  This
trend is hardly significant.  The fact that the $1/V_{\rm max}$ and
$C^-$ estimates are very close means that there are no artifacts due
to clustering.

\tabcolsep 0.15cm
\renewcommand{\baselinestretch}{1.3}
\begin{table*}
\caption{Parameters of the $K_s$-band LF for galaxies in the HDF-N and
HDF-S obtained with the STY method. }
\begin{center}
\begin{tabular}{c|l|c|cccc}
$z$ range & Cosmology  & Field 
        & $\phi^*$ [$h^{3}\,$Mpc$^{-3}$]
        & $M^*_K({\rm AB})-5\log h$ & $\alpha$ 
        & $M_K-5\log h$ range \\
\hline
$0 \le z < 1$ & $\Omega_0=1$, $\Omega_\Lambda=0$ &  HDF-N 
              & $0.0342^{+0.0065}_{-0.0084}$ 
              & $-21.540^{+0.164}_{-0.247}$ 
              & $-1.101^{+0.066}_{-0.070}$ 
              & $[-23.0,-12.0]$ \\
              &                                  &  HDF-S 
              & $0.0232^{+0.0087}_{-0.0073}$ 
              & $-21.860^{+0.286}_{-0.279}$ 
              & $-1.159^{+0.093}_{-0.084}$ 
              & $[-23.0,-12.0]$ \\
\cline{3-7}
              & $\Omega_0=0.3$, $\Omega_\Lambda=0.7$ &  HDF-N 
              & $0.0121^{+0.0029}_{-0.0036}$  
              & $-22.266^{+0.238}_{-0.501}$
              & $-1.164^{+0.064}_{-0.076}$
              & $[-24.0,-12.0]$ \\
              &                                      &  HDF-S 
              & $0.0100^{+0.0039}_{-0.0031}$ 
              & $-22.389^{+0.257}_{-0.297}$ 
              & $-1.170^{+0.089}_{-0.095}$ 
              & $[-23.0,-12.0]$ \\
\hline
$1 \le z < 2$ & $\Omega_0=1$, $\Omega_\Lambda=0$ &  HDF-N 
              & $0.0050^{+0.0029}_{-0.0017}$ 
              & $-22.524^{+0.373}_{-0.382}$ 
              & $-1.579^{+0.098}_{-0.079}$ 
              & $[-23.5,-17.0]$ \\
              &                                  & HDF-S 
              & $0.0083^{+0.0074}_{-0.0059}$ 
              & $-22.201^{+0.624}_{-1.64}$ 
              & $-1.411^{+0.080}_{-0.159}$ 
              & $[-22.5,-17.5]$ \\
\cline{3-7}
              & $\Omega_0=0.3$, $\Omega_\Lambda=0.7$ & HDF-N 
              & $0.0017^{+0.0008}_{-0.0005}$ 
              & $-23.200^{+0.259}_{-0.379}$ 
              & $-1.568^{+0.099}_{-0.088}$ 
              & $[-24.0,-17.5]$ \\
              &                                      & HDF-S 
              & $0.0024^{+0.0025}_{-0.0017}$ 
              & $-23.065^{+0.738}_{-1.541}$ 
              & $-1.424^{+0.222}_{-0.174}$
              & $[-23.0,-18.0]$ \\
\hline
\end{tabular}
\label{tab_lfk_sty}
\end{center}
\end{table*}
\renewcommand{\baselinestretch}{1}

In Table \ref{tab_lfk_sty} we list the parameters of the STY estimate
for the $[0,1]$ and $[1,2]$ samples.  The most impressive
results we can note from Fig.~\ref{lf_k_hdfns} are the very wide
range of absolute magnitudes covered by the data 
and the possibility of computing for the first time the NIR LFs at
redshifts in the range $[1,2]$, where many difficulties arise for the
traditional spectroscopy.  Moreover, this redshift range is of
paramount importance in the study of galaxy formation and evolution,
as we will discuss in Sect.~\ref{discuss}.  

\begin{figure*}
{\centering \leavevmode
\psfig{file=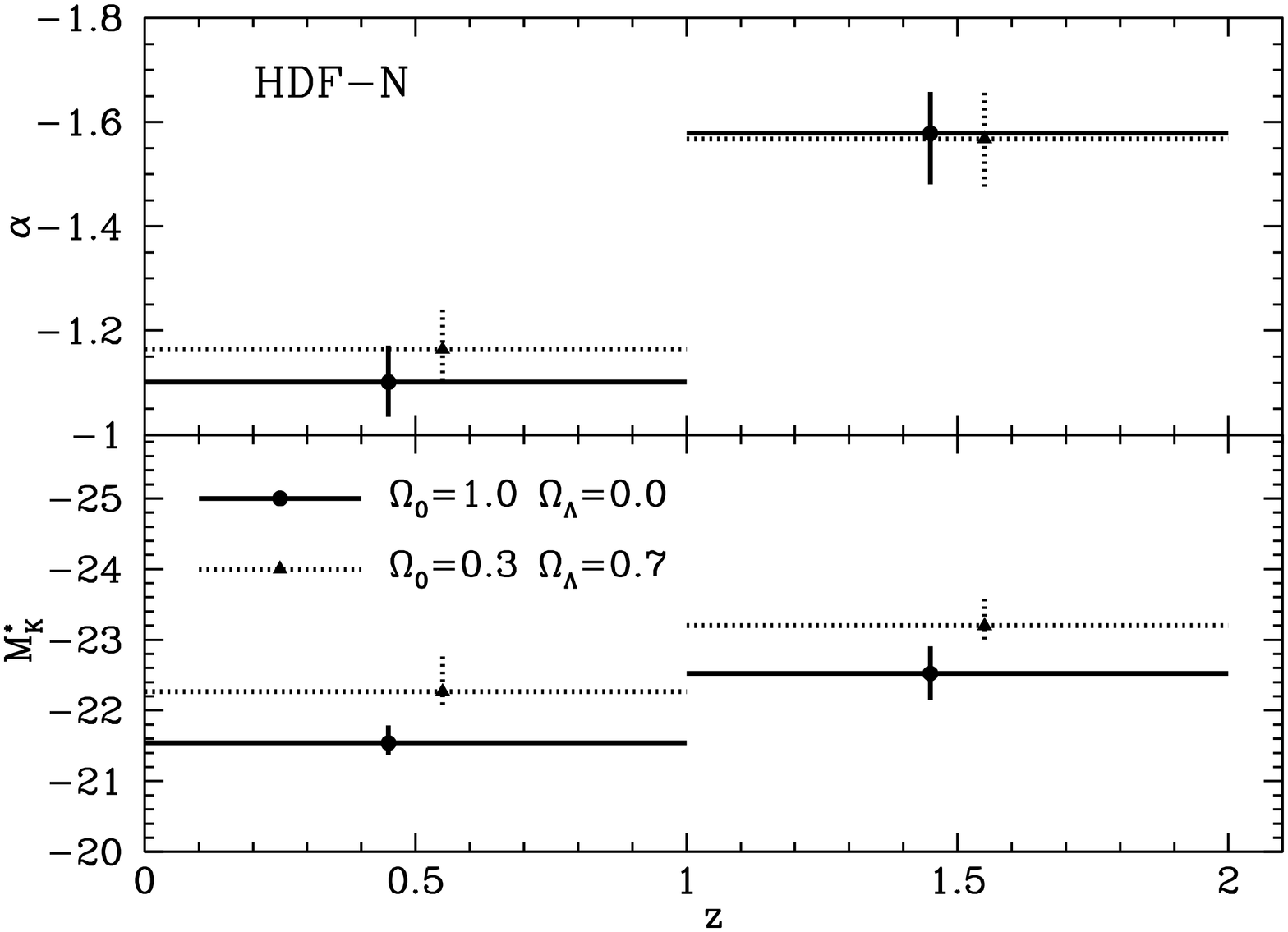,width=.49\textwidth,angle=270} \hfil
\psfig{file=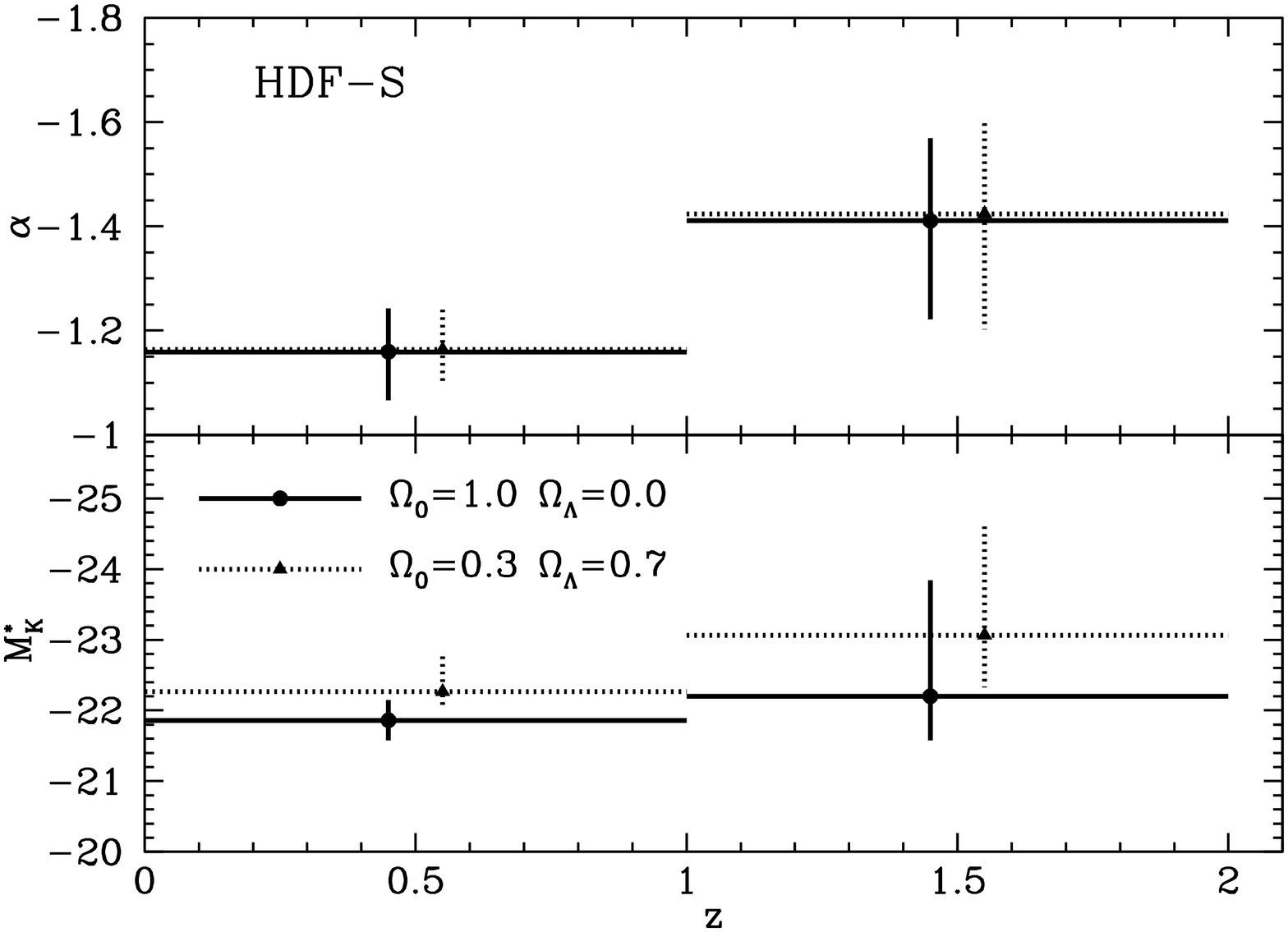,width=.49\textwidth,angle=270} } 
\caption{Parameters of the LFs as a function of the redshift bin for
the two cosmologies adopted in the paper.  Results obtained with the
$\Omega_0=1$ and $\Omega_0=0.3, \Omega_\Lambda=0.7$ cosmologies have
been shifted in redshift by $-0.05$ and $+0.05$ respectively.}
\label{zalpham}
\end{figure*}

In Fig.~\ref{zalpham} we summarize the results given in
Table~\ref{tab_lfk_sty}, showing the $\alpha$ and $M^*_K({\rm AB})$
parameters and their respective errors as a function of redshift,
derived for the two adopted cosmologies.  


\subsection{$J$-band LF}

We also computed LFs in the $J$-band. In this case, we selected
galaxies in the $J$ filter when considering the lowest redshift bin
[0,1], whereas we estimated the $J$-band in the highest redshift range
($z_{\rm phot} \in [1,2]$) using the $K_s$-band selected subsamples.
In this way we select the objects approximately in the $J$-band
rest-frame and we can check if the assumptions made for the $K_s$ band
LF computation were safe.  We selected objects in the HDF-N with $J
\le 24.6$ in the redshift range $[0,1]$ and with $K_s \le 24$ in
$z=[1,2]$, corresponding to objects with $S/N \ge 3$. At these limits
the colours in the HDF-N and HDF-S are very similar: at $J = 24.6$,
the mean $I_{814}-J$ in $[0,1]$ is $0.57$ in the HDF-N and $0.50$ in
the HDF-S; at $K_s = 24$ the mean $I_{814}-K$ in $z=[1,2]$ is $1.65$
in the HDF-N and it is $1.85$ in the HDF-S.

The values of $\left<V/V_{\rm max}\right>$ are $0.56, 0.43$ in the
redshift ranges $z_{\rm phot} \in [0,1]$ and $[1,2]$, respectively,
for the HDF-N, and $0.51, 0.47$ in the same redshift ranges for the
HDF-S.

\begin{figure*}
\centerline{\psfig{file=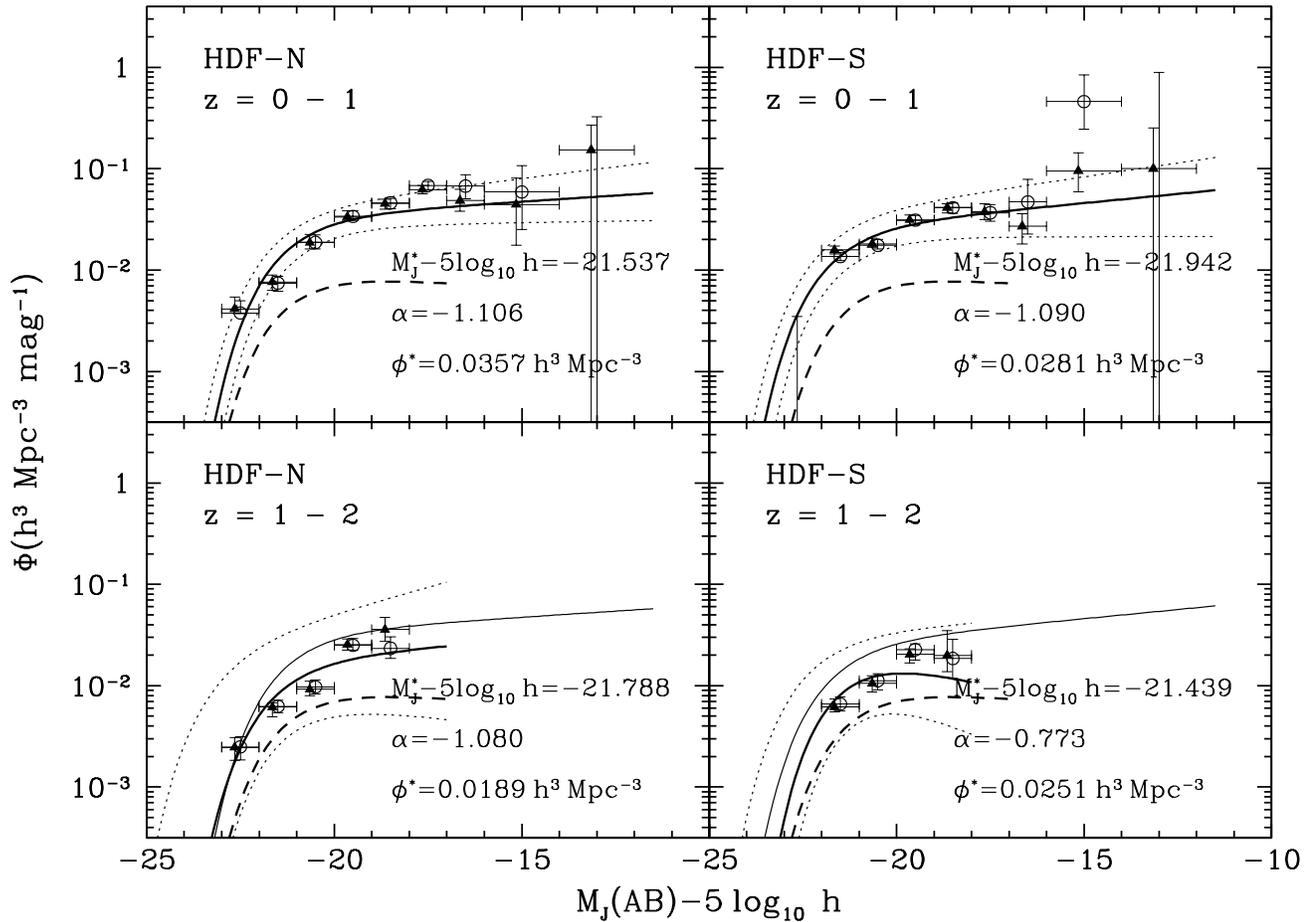,width=0.99\textwidth,angle=270}}
\caption{$J$-band LFs for galaxies in the HDF-N and HDF-S in two
redshift bins, assuming a limiting magnitude $J=24.6$ in the redshift
range $[0,1]$ and $K_s=24.0$ in $z=[1,2]$, and a cosmological model
with $\Omega_0=1$.  Same symbols as in Fig.~\ref{lf_k_hdfns}.  The LF
obtained by Cole et al.\ (\protect\cite{cole1}) (dashed line) assuming
the same cosmological model is also shown as reference LF.  In the
lower panels, the LF in the $z_{\rm phot} = [0,1]$ interval is also
displayed for comparison (thin solid line).}
\label{lf_j_hdfns}
\end{figure*}

\renewcommand{\baselinestretch}{1.3}
\begin{table*}
\caption{Parameters of the $J$-band LF for galaxies in the HDF-N and
HDF-S obtained with the STY method. }
\begin{center}
\begin{tabular}{c|l|c|cccc}
$z$ range & Cosmology  & Field 
        & $\phi^*$ [$h^{3}\,$Mpc$^{-3}$]
        & $M^*_J({\rm AB})-5\log h$ & $\alpha$
        & $M_J-5\log h$ range  \\
\hline
$0 \le z < 1$ & $\Omega_0=1$, $\Omega_\Lambda=0$ & HDF-N  
              & $0.0357^{+0.0064}_{-0.0064}$
              & $-21.537^{+0.225}_{-0.264}$
              & $-1.106^{+0.091}_{-0.055}$
              & $[-23.0,-11.0]$ \\
              &                                  & HDF-S  
              & $0.0281^{+0.0086}_{-0.0049}$
              & $-21.942^{+0.280}_{-0.253}$
              & $-1.090^{+0.089}_{-0.046}$
              & $[-23.0,-11.0]$ \\
\cline{3-7}
              & $\Omega_0=0.3$, $\Omega_\Lambda=0.7$ & HDF-N 
              & $0.0158^{+0.0022}_{-0.0027}$
              & $-22.077^{+0.185}_{-0.333}$ 
              & $-1.066^{+0.036}_{-0.044}$
              & $[-24.0,-11.5]$ \\
              &                                      & HDF-S 
              & $0.0129^{+0.0033}_{-0.0025}$
              & $-22.451^{+0.244}_{-0.230}$
              & $-1.082^{+0.075}_{-0.047}$
              & $[-23.5,-11.5]$ \\
\hline
$1 \le z < 2$ & $\Omega_0=1$, $\Omega_\Lambda=0$ & HDF-N 
              & $0.0189^{+0.0076}_{-0.0108}$
              & $-21.788^{+0.380}_{-1.423}$ 
              & $-1.080^{+0.193}_{-0.175}$
              & $[-23.5,-17.0]$ \\
              &                                  & HDF-S 
              & $0.0251^{+0.0084}_{-0.0126}$
              & $-21.439^{+0.400}_{-1.053}$
              & $-0.773^{+0.199}_{-0.301}$
              & $[-22.0,-17.0]$ \\
\cline{3-7}
              & $\Omega_0=0.3$, $\Omega_\Lambda=0.7$ & HDF-N 
              & $0.0061^{+0.0028}_{-0.0036}$
              & $-22.431^{+0.376}_{-1.652}$
              & $-1.047^{+0.175}_{-0.195}$
              & $[-24.0,-17.5]$ \\
              &                                       & HDF-S 
              & $0.0074^{+0.0026}_{-0.0040}$
              & $-22.357^{+0.605}_{-1.273}$
              & $-0.796^{+0.229}_{-0.317}$
              & $[-23.0,-17.5]$ \\
\hline
\end{tabular}
\label{tab_lfj_sty}
\end{center}
\end{table*}
\renewcommand{\baselinestretch}{1}
 
In Fig.~\ref{lf_j_hdfns} we plot the LFs obtained with the adopted
parametric and non parametric methods in the redshift ranges $[0,1]$
and $[1,2]$, as well as the Cole et al.\ (\cite{cole1}) local LF
estimated in the 2dFGRS, shown as a reference.  Our estimate and the
Cole et al.\ (\cite{cole1}) one, suitably transformed in AB magnitudes
($M^*_J=-21.40$, $\alpha=-0.93$, $\phi^*=0.0108$ computed with
cosmology $\Omega_0=1, \Omega_\Lambda=0$ and only $k$-correction to
match the same conditions we used), seem to be in disagreement, mainly
in the normalization.  However, the comparison between our LF estimate
obtained in the flat $\Lambda$-dominated cosmology and the analogous
one computed by Cole et al. \cite{cole1} partially mitigates the
difference. 

Table \ref{tab_lfj_sty} contains the values of the Schechter
parameters of the $J$-band LF obtained in the two redshift ranges for
the HDF-N and HDF-S.  We have also computed the LF in the $H$ and $I$
bands for the HDF-N and HDF-S catalogues.
In all cases, we found similar results for the two fields.


\subsection{Effect of cosmology}
\label{cosmo}

\begin{figure*}
{\centering \leavevmode
\psfig{file=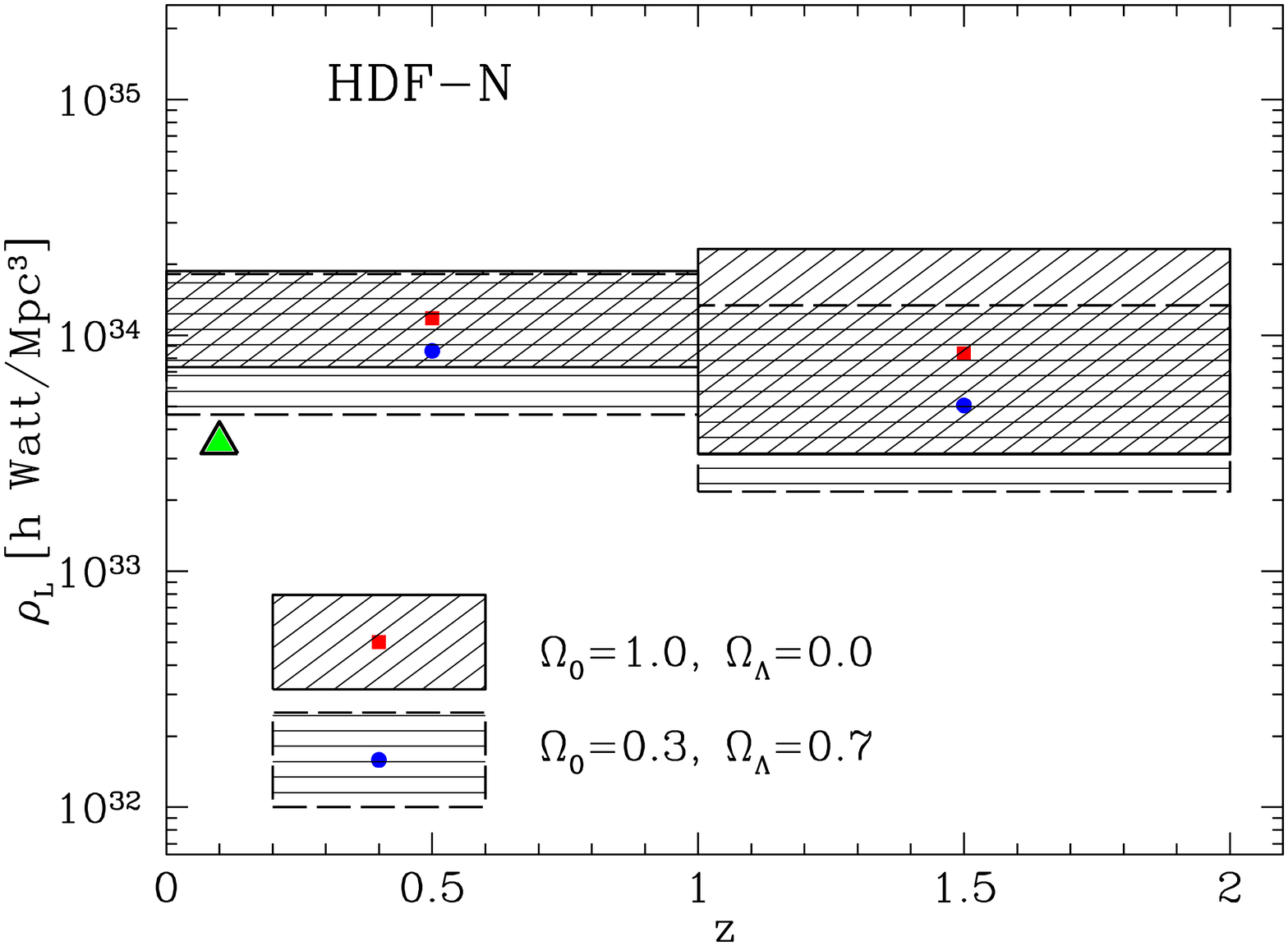,width=.49\textwidth,angle=270} \hfil
\psfig{file=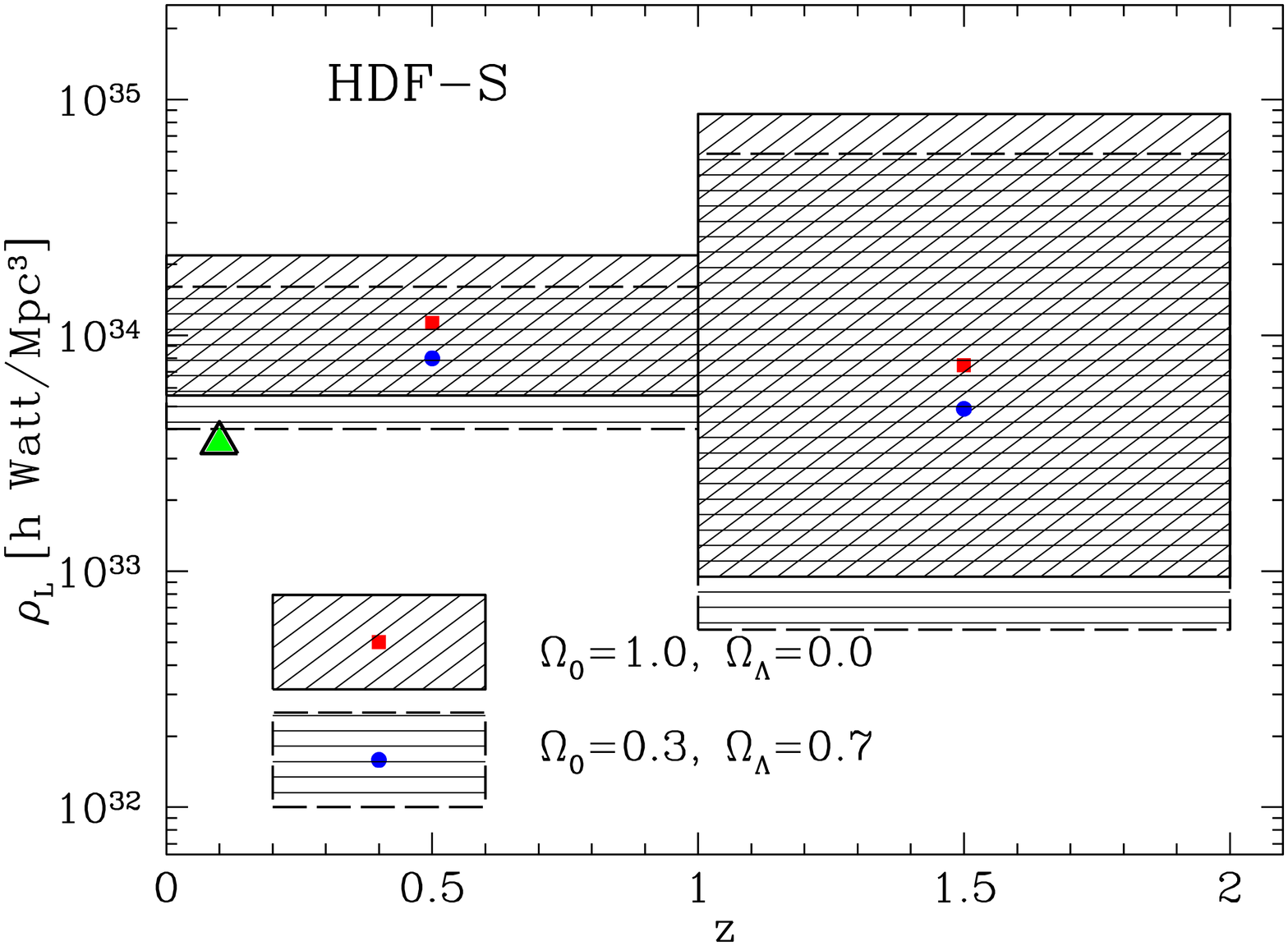,width=.49\textwidth,angle=270} }
\caption{The luminosity density in the $K_s$-band, as obtained from
the values listed in Table \ref{tab_lfk_sty}.
Squares: luminosity densities computed using the median values of the
STY estimate of the LF in the case of a $\Omega_0=1$ cosmological
model.  Circles: values of the luminosity density in a flat
cosmological constant dominated universe. The value obtained by Cole
et al.\ (\cite{cole1}) is also shown for comparison as a triangle. }
\label{lumdens}
\end{figure*}

We estimated LFs in two cosmologies: the ``old'' Standard CDM with
$\Omega_0=1$ and $\Omega_\Lambda=0$, and a flat cosmological constant
dominated model, that nowadays is the most accepted one, with
$\Omega_0=0.3$ and $\Omega_\Lambda=0.7$.  The influence of cosmology
in the photometric redshift computation is negligible, as discussed by
Bolzonella et al.\ (\cite{hyperz}), thus we used the same photometric
redshifts to estimate the LFs in different cosmologies.

When LFs are computed at low redshifts, we expect little or no
differences in the estimates, as in the case of the 2dFGRS by Cole et
al.\ (\cite{cole1}), where the difference in the value of $M^*_{K_s}$
between the two interesting models is $0.16$ magnitudes.  Therefore,
whereas local estimates are not affected by the change of cosmology,
when we are dealing in particular with galaxies at high redshifts the
variation of volume and distances become substantial.  As expected,
looking at Tables \ref{tab_lfk_sty} and \ref{tab_lfj_sty} we can
remark that the value of $M^*$ brightens when the lambda dominated
cosmology is assumed, whereas $\phi^*$ decreases because of the
increase of the comoving volume at a given redshift.  In fact, the
difference in distance modulus will affect absolute magnitudes,
whereas the difference in volumes affects the estimate of the $\phi^*$
parameter.


\subsection{NIR Luminosity Density}

Using the values of the Schechter parameters listed in Tables
\ref{tab_lfk_sty} and \ref{tab_lfj_sty} we computed the luminosity
density, given by
\begin{equation}
\rho_L = \int L \phi(L)\,dL = \phi^* L^* \Gamma(\alpha +2) \: .
\label{rho_l}
\end{equation}

Results are shown in Fig.~\ref{lumdens}, in units of $h\,{\rm
W\,Mpc^{-3}}$, derived from the usual solar luminosity values, after
integration and scaling through a stellar solar type SED (taken from
the library of Pickles \cite{pickles}).  Boxes represent the maximum
and minimum values obtained using the Schechter parameters in a given
redshift range.  We show the results for both the adopted cosmological
models: we can see that the luminosity density does not strongly
depend on cosmology, being the values relative to one model inside
the boxes of the second one.  The most remarkable characteristic
visible in Fig.~\ref{lumdens} is that the luminosity density is
consistent with non-evolution at least up to a redshift $z \sim 2$.
                                    
As expected from the comparison of our Schechter parameters with the
values in Table~\ref{ktable}, our estimates of $\rho_{K_s}$ are higher
than the luminosity densities computed for instance by Cole et al.\
(\cite{cole1}) for the 2dFGRS: their value in the $K_s$-band,
conveniently converted into $h\,{\rm W\,Mpc^{-3}}$ in the $K_s$
filter, is $\rho_{K_s}=3.5\times 10^{33}\,h\,{\rm W\,Mpc^{-3}}$.  It
is shown as a triangle in Fig.~\ref{lumdens}: it is marginally
consistent with our lower limit of the $z \in [0,1]$ estimate in the
HDF-N and HDF-S for the $\Lambda$-dominated model (the same model
adopted by Cole et al.\ \cite{cole1}).  In the $J$-band, our estimates
in the redshift range $[0,1]$ are $\rho_J = 3.53 \times 10^{34} (3.86
\times 10^{34})$ and $2.35\times 10^{34} (2.49 \times
10^{34})\,h\,{\rm W\,Mpc^{-3}}$ in the HDF-N (HDF-S), for the matter
and $\Lambda$-dominated models respectively, whereas the value found
by Cole et al.\ (\cite{cole1}) is $6.8\times 10^{33}\,h\,{\rm
W\,Mpc^{-3}}$.
 
In a recent paper, Wright (\cite{wright}) already claimed that the
value obtained by Cole et al.\ (\cite{cole1}) does not match the
extrapolation to the NIR from the optical luminosity
densities obtained in the SDSS by Blanton et al.\ (\cite{blanton}). The
discrepancy is about a factor of $2.3$. On the contrary, the
present results are in much better agreement with the SDSS optical
luminosity densities. This result has to be considered with caution
because of the small size of the HDFs.


\subsection{Mass Function}

\begin{figure*}
{\centering \leavevmode
\psfig{file=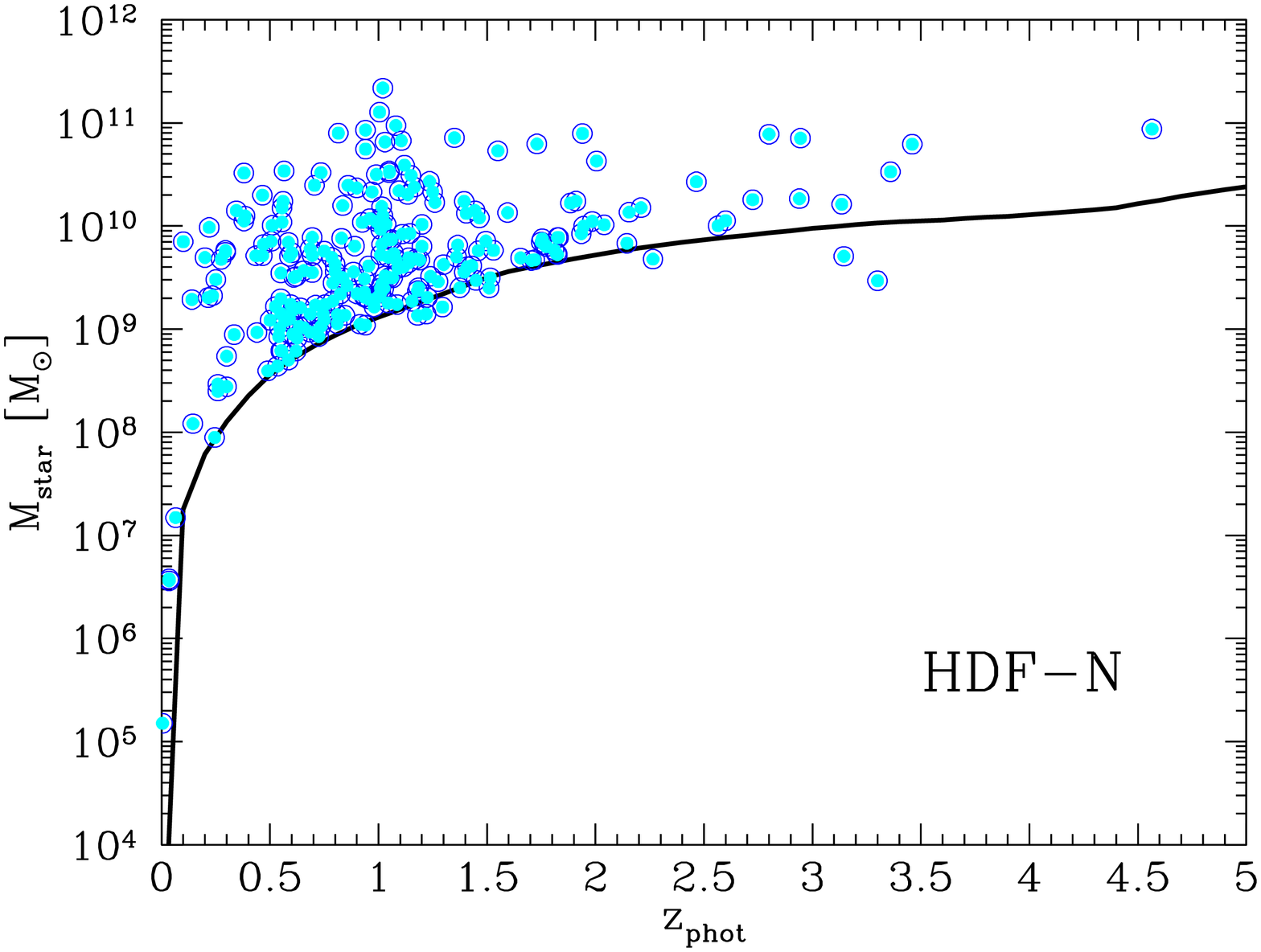,width=.49\textwidth,angle=270} \hfil
\psfig{file=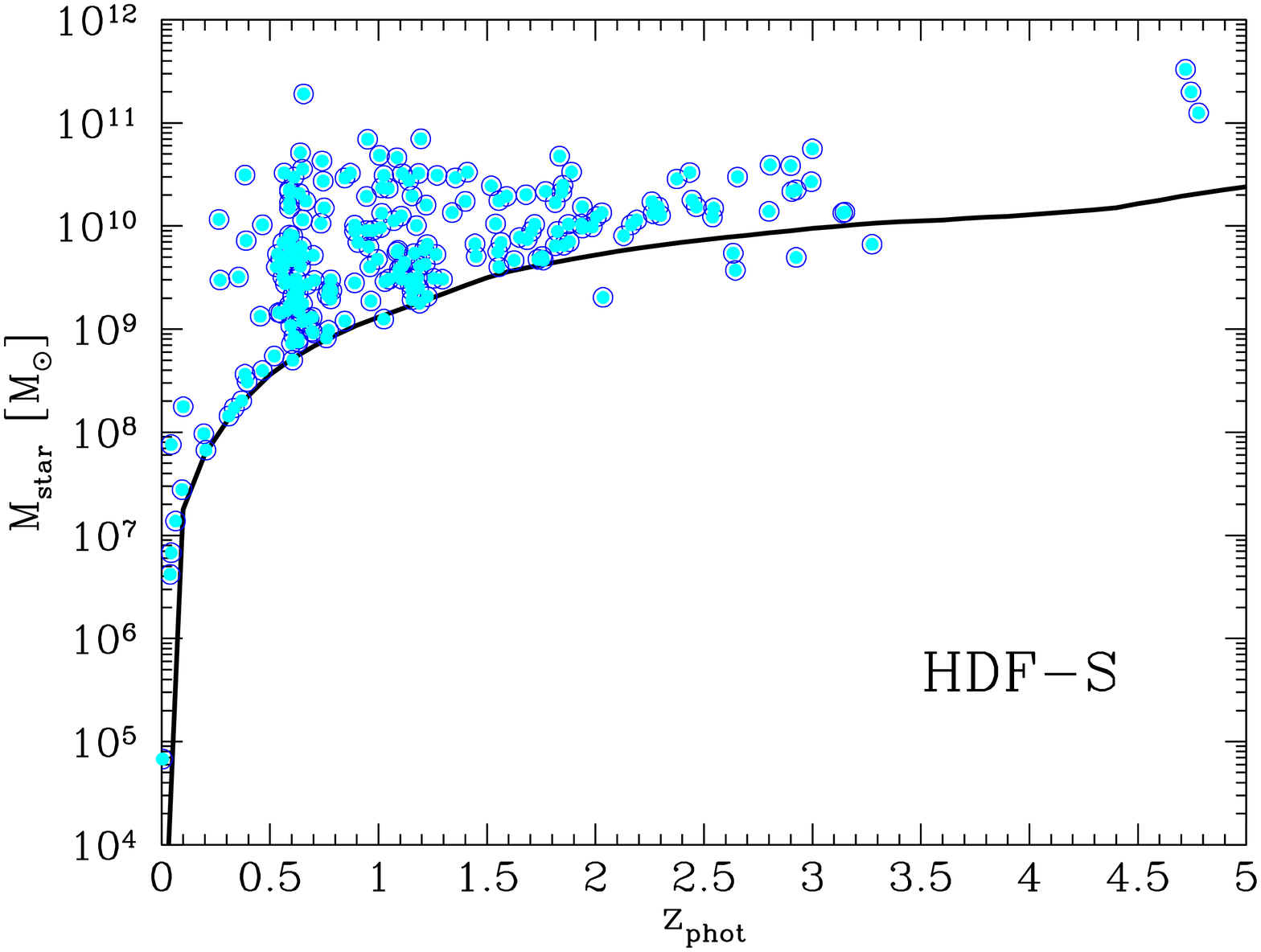,width=.49\textwidth,angle=270} } 
\caption{The masses of the objects in the $K_s$-band selected
subsamples with $K_s \le 24.0$ in the HDF-N and HDF-S, computed
assuming a mass-to-light ration of $0.8$ in solar units.  The solid
line represents the limiting mass inferred from the limiting magnitude
$K_s = 24.0$. }
\label{mass}
\end{figure*}

The $K$-band luminosity is considered as a good tracer of the stellar
mass in galaxies, because the bulk of the emission at these
wavelengths is dominated by solar stars, that constitute most of the
mass in stars.  Moreover, the $K$-band estimator is nearly independent
on the star formation history and then it can be safely used to
compute the baryonic mass contained in galaxies.

We adopted a constant stellar mass-to-light ratio: $M_{\rm star}/L_{K}
= 0.8\, M_{\odot}/L_{K\odot}$, a mean value obtained from the estimate
by Worthey (\cite{worthey}) considering the range of ages spanned by
our objects.  In any case the dependence of $M/L$ with age is not
strong in this filter.  With this assumption, the LF reflects the
stellar mass function: a non evolution in luminosities implies a non
evolution in characteristic masses.  In Fig.~\ref{mass} we show the
masses computed from the $K_s$-band apparent magnitudes of the
$K_s$-band selected subsample used in this study.  We also show the
limiting mass as a function of redshift computed for an elliptical
galaxy, assuming $K_s$-band limiting magnitudes of $24.0$ at which
reach $S/N=3$.  The change using different SEDs is negligible.
According to this figure, the HDFs allow to detect stellar haloes of
$\sim 10^9 M_{\odot}$ up to $z \sim 1$, and around $10^{10} M_{\odot}$
up to $z \ga 3$.


\section{Discussion: Galaxy formation models}
\label{discuss}

LFs are a convenient way to describe the galaxy population and to get
hints about the mechanisms of formation and evolution of galaxies.

Two scenarios are in competition to explain the history of galaxies up
to the present epoch. The formation of elliptical galaxies is
especially intriguing, because despite their old and apparently simple
stellar population, their process of formation is far from being
understood.  The two models in competition are:
\begin{itemize}
\item The hierarchical scenario, which is based on CDM cosmological
models (e.g. Kauffmann et al.\ \cite{kauff1}; Cole et al.\ \cite{cole}).
It assumes that galaxies have formed through merging of sub-galactic
structures. In particular, elliptical galaxies are born from the
merging of two disks of comparable dimensions; during the interaction,
a burst of star formation occurs, consuming the gas present in the
progenitors, and then passive evolution follows.

\item The monolithic scenario, revisited in different ways by many
authors, which basically assumes that galaxies formed approximately at
the same epoch, with particular details depending on the morphological
types (e.g.\ Rocca-Volmerange \& Guiderdoni \cite{rocca}; Pozzetti et
al.\ \cite{pozz}).  Ellipticals are believed to form at $z_{\rm form}
\ga 2-3$ and, at the time of their assembly, a burst of star formation
occurred, followed by passive evolution of the stellar population.
\end{itemize}

The two scenarios foresee different characteristics as a function of
redshift for the progenitors of the blue and red galaxies in the local
universe.  Moreover, since most of the merging would take place at
$z<2$, the redshift range between $1$ and $2$ is fundamental to
discriminate between monolithic and hierarchical scenarios.  Both
mechanisms account for the properties of elliptical galaxies up to
$z\simeq1$, e.g.\ the CM relation (Gladders et al.\ \cite{glad};
Kauffmann \& Charlot \cite{kauff2}), but, between $z=1$ and $z=2$, the
expectations for the photometric properties of all galaxies differ
significantly between the two scenarios.

A powerful test to establish what drives the galaxy evolution, was
proposed by Kauffmann \& Charlot (\cite{kauff}), using the $K$-band
luminosity function.  The luminosity in the $K$ filter is directly
linked to the mass in stars and is barely affected by the presence of
dust extinction, thus making the $K$ photometry a privileged tool to
study galaxy formation. Moreover, galaxies with the same stellar mass
have nearly the same $K$ magnitude, independently on their star
formation history. For these reasons the $K$-band LF can probe if
galaxies were assembled early, according to the monolithic scenario,
or recently from mergers.  Kauffmann \& Charlot (\cite{kauff}) built
two PLE models with density parameters $\Omega_0=1$ and $\Omega_0=0.2$
and two hierarchical models based on the CDM cosmology, with the same
values of $\Omega_0$.  They computed the evolution of the $K$-band LF
at increasing redshifts for the PLE models and for the $\Omega_0=1$
hierarchical model, each one able to reproduce the local LF.  On the
contrary, their low density hierarchical model failed to reproduce the
local $K$-band LF and they did not compute the evolution of the LF in
this framework.

\begin{figure}
\centerline{\psfig{file=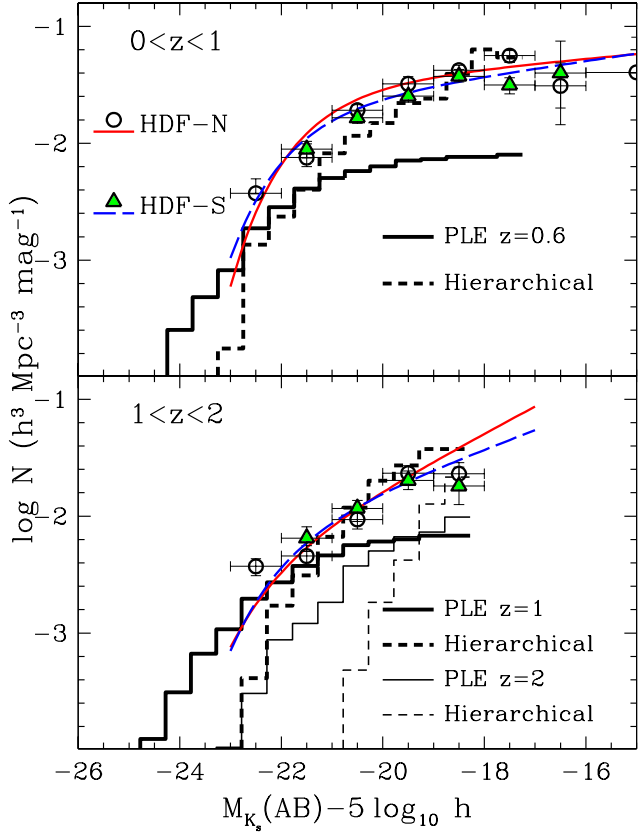,width=0.49\textwidth}}
\caption{Comparison between the theoretical luminosity functions
derived by Kauffmann \& Charlot (\cite{kauff}) and the present results
on the HDFs.  Histograms represent the prediction for the
hierarchical, $\Omega_0=1$ CDM, (dashed line) and PLE scenarios
(solid line).  The solid and long dashed continuous lines display
respectively the STY LFs fitted to the data in the HDF-N and HDF-S.
Circles and triangles show the estimate of the LF with the $C^-$
method in the HDF-N and HDF-S.  In the lower panel we show the model
predictions both at $z=1$ and $z=2$, because we estimated the LF using
galaxies with redshifts between these two extremes.}
\label{lfk_kc}
\end{figure}

At redshifts $z>1$ a sharp difference between the two models is
predicted, as explained in Kauffmann \& Charlot.  The PLE models
foresee a constant bright-end for the LF, with small differences
between the flat and the open cosmology, whereas the hierarchical
model considered by Kauffmann \& Charlot undergoes a shift toward
faint magnitudes (see Fig.~\ref{lfk_kc}).

The development of hierarchical scenarios in open or flat cosmological
models with cosmological constant mitigates the discrepancy between
these two scenarios (monolithic vs. hierarchical), because the epoch
of the major merging moves at higher redshifts (Fontana et al.\
\cite{fontana}, Cole et al.\ \cite{cole}).  In this case, the PLE
model could not be distinguished from a scenario where the galaxy
assembly occurs at early epochs, followed by a passive evolution of
the stellar population. Our results for the cumulative redshift
distribution actually support such scenarios (see Fig.~\ref{nzcum}).

The present results, i.e.\ the lack of significant evolution in the
bright part of the LF from the redshift range $[0,1]$ to $[1,2]$,
provide a stringent clue, supporting the idea that massive
galaxies were already in place at high redshifts, against the old CDM
hierarchical model adopted by Kauffmann \& Charlot.  The comparison
between these theoretical predictions and the observations derived in
the present paper can be found in Fig.~\ref{lfk_kc}.  In the lower
redshift bin, our estimates of the LFs present a faint-end slope in
agreement with the hierarchical model, whereas at bright magnitudes
the small differences between the two models do not support any claim.
In the redshift bin $z=[1,2]$ we have compared our estimate with both
the predictions of the models at redshift $z=1$ and $2$: the model
predictions suitable for our sample should lie between the two.  It is
evident from the lower panel of Fig.~\ref{lfk_kc} that the very bright
part of the LFs remains well above the hierarchical model predictions
at $z=1$ and $2$, being much closer to the predictions of the
monolithic/PLE-like scenario.  On the other hand, the faint end slope
seems to be in better agreement with the hierarchical model, even in
this redshift range.

Other recent studies on the Hubble fields, based on different
selection criteria, seem to indicate that the formation of elliptical
galaxies should be placed at $z>2$ (Ben\'{\i}tez et al.\
\cite{benitez}; Broadhurst \& Bouwens \cite{broad}).  In the present
paper we do not select galaxies according to morphological types, but
the same conclusions apply to the most massive (NIR luminous) galaxies
in our sample.

The lack of evolution in the bright end of the LF is in good agreement
with the results found from spectroscopic surveys.  Glazebrook et al.\
(\cite{glaze}) found no evidence for evolution in the $K$-band LF up
to $z \le 0.5$, and concluded that massive spheroids were in place at
$z \ge 1$ and then evolved passively.  Songaila et al.\
(\cite{songaila}) found a lack of significant evolution in their
$K$-band sample up to a redshift of $\sim 1$.  Cowie et al.\
(\cite{cowie2}) found little evolution in their sample of red (old)
objects to $z \sim 1$.  The present results extend the previous
findings in redshift, up to $z \sim 2$, with the same conclusions with
respect to the evolution of the most massive galaxies.  The comparison
between the present LFs in the near-IR and in the optical-UV bands
(Bolzonella et al.\ in preparation) will provide new insights on the
galaxy formation scenario.


\section{Summary}
\label{sum}

The results of this paper can be summarized as follows:
\begin{enumerate}
\item 
The photometric redshift technique is an indispensable tool to study
the faint-end and the evolution of LFs, pushing the limiting
magnitudes toward fainter limits than the works based on spectroscopic
surveys.

\item 
We have elaborated a method to compute the luminosity function
properly taking into account the characteristic of the uncertainties
of the photometric redshift technique.  This approach can be
conveniently applied to compute LFs from the forthcoming wide and deep
field photometric surveys.  Moreover, this method can be easily
extended to other wavelengths crucial for the study of galaxy
evolution, as optical LFs and the LF at $1500$\,\AA\ (Bolzonella et
al.\ in preparation).

\item We have computed the LF in $K_s$ and $J$ bands for the
HDF-N and HDF-S catalogues with two non-parametric methods, $1/V_{\rm
max}$ and $C^-$, and with the maximum likelihood method STY, assuming
a Schechter functional shape.  We found similar results for the two
fields.  This seems to indicate that clustering does not considerably
affect the estimate of the LFs, even for the $1/V_{\rm max}$ method.

\item The bright-end of the NIR LF remains unchanged within the errors
up to at least $z \sim 2$.  Also the NIR luminosity density is in good
agreement with no-evolution, at least up to a redshift $z \sim 2$.
This result is consistent with the bulk of the stellar mass being in
place and assembled at such redshift.  

\item 
The analysis of the near-IR LF and the cumulative redshift
distributions is a powerful test for galaxy formation models.  The
evidence of an unevolving bright-end of the $K_s$-band LF up to
$z\simeq 2$ supports the assembly of massive galaxies before such
redshift, a result which is rather close to the old monolithic
scenario than to the hierarchical standard CDM cosmological model.
Our results are consistent with hierarchical scenarios in open or flat
cosmological models with cosmological constant (such as Fontana et
al.\ \cite{fontana}, Cole et al.\ \cite{cole}).

\end{enumerate}

\begin{acknowledgements}
We would like to thank G. Bruzual, S. Charlot, G. Mathez, M. Lemoine,
for fruitful discussions and comments.  This work has been done within
the framework of the VIRMOS collaboration.  MB acknowledges support
from CNR/ASI grant I/R/27/00.  Part of this work was supported by the
French {\it Centre National de la Recherche Scientifique}, and by the
TMR Lensnet ERBFMRXCT97-0172.
\end{acknowledgements}


\begin{thebibliography}{}

\bibitem[1999]{arnouts} Arnouts, S., Cristiani, S., Moscardini, L.,
Matarrese, S., Lucchin, F., Fontana, A. \& Giallongo, E. 1999, MNRAS
310, 540

\bibitem[2002]{arnouts2} Arnouts, S., Moscardini, L., Vanzella, E.,
Colombi, S., Cristiani, S., Fontana, A., Giallongo, E., Matarrese,
S. \& Saracco, P. 2002, MNRAS 329, 355

\bibitem[2001]{balogh} Balogh, M., Christlein, D., Zabludoff, A. \&
Zaritsky, D. 2001, ApJ 557, 117

\bibitem[1999]{benitez} Ben\'{\i}tez, N., Broadhurst, T., Bouwens, R.,
Silk, J., \& Rosati, P.  1999, ApJ 515, L65

\bibitem[1996]{bertin} Bertin,  E. \& Arnouts, S. 1996, A\&ASS 117, 393

\bibitem[2001]{blanton} Blanton, M. R., Dalcanton, J., Eisenstein, D.,
Loveday, J., Strauss, M. A., SubbaRao, M., Weinberg, D. H. and the
SDSS collaboration, 2001, AJ 121 2358     

\bibitem[1992]{bower92} Bower, R. G., Lucey, J. R. \& Ellis, R. S. 1992,
MNRAS 254, 601       

\bibitem[1998]{bower} Bower, R. G., Kodama, T.  \& Terlevich, A. 1998,
MNRAS 299, 1193

\bibitem[2000]{hyperz} Bolzonella, M., Miralles, J.-M. \& Pell\'o,
R. 2000, A\&A 363, 476

\bibitem[2000]{broad} Broadhurst, T., Bouwens, R. J.  2000, ApJ 530,
L53

\bibitem[1993]{bruzual} Bruzual, G. \& Charlot, S. 1993, ApJ 405, 538
 
\bibitem[2000]{calzetti} Calzetti, D., Armus, L., Bohlin, R. C.,
Kinney, A. L., Koornneef J. \& Storchi-Bergmann T. 2000, ApJ 533, 682

\bibitem[2000]{casertano} Casertano, S., de Mello, D., Dickinson, M., et
al. 2000, AJ 120, 2747

\bibitem[1987]{cholo} Cho\l oniewski, J. 1987, MNRAS 226, 273

\bibitem[1999]{cimatti} Cimatti, A., Daddi, E., di Serego Alighieri,
S., Pozzetti, L., Mannucci, F., Renzini, A., Oliva, E., Zamorani, G.,
Andreani, P. \& R\"ottgering, H. J. A. 2000, A\&A 532, L45

\bibitem[2000]{cole} Cole, S., Lacey, C., Baugh, C. \& Frenk, C. 2000,
MNRAS 319, 168

\bibitem[2001]{cole1} Cole, S., Norberg, P., Baugh, C. M., Frenk,
C. S. et al. 2001, MNRAS 326, 255

\bibitem[1997]{con} Connolly, A. J., Szalay, A. S., Dickinson, M.,
SubbaRao, M. U., Brunner, R. J. 1997, ApJ 486, L11

\bibitem[1994]{cowie} Cowie, L. L., Gardner, J. P., Hu, E. M.,
Songaila, A., Hodapp, K.-W. \& Wainscoat, R. J. 1994, ApJ 434, 114

\bibitem[1996]{cowie2} Cowie, L. L., Songaila, A., Hu, E. M. \& Cohen,
J. G. 1996, AJ 112, 839

\bibitem[2002]{cross} Cross, N. \& Driver, S. P. 2002, MNRAS 329, 579

\bibitem[1998]{dacosta} da Costa, L. N., Nonino, M., Rengelink, R. et
al., 1998, A\&A submitted (astro-ph/9812105)

\bibitem[1982]{davis} Davis, M. \& Huchra, J. 1982, ApJ 254, 437

\bibitem[2000]{dick} Dickinson, M., Papovich, C., Ferguson, H. C., to
appear in the proceedings of the ESO Symposium, ``Deep Fields'',
ed. S. Cristiani (Berlin: Springer), astro-ph/0105086.

\bibitem[2001]{dye} Dye, S., Taylor, A. N., Thommes, E. M.,
Meisenheimer, K., Wolf, C. \& Peacock, J. A. 2001, MNRAS 321, 685

\bibitem[1976]{felten} Felten, J. E. 1976, ApJ 207, 700

\bibitem[1999]{fontana} Fontana A., Menci N., D'Odorico S., Giallongo
E., Poli F., Cristiani S., Moorwood A.  \& Saracco P. 2001 MNRAS 310,
L27

\bibitem[2000]{fonta1} Fontana, A., D'Odorico, S., Poli, F.,
Giallongo, E., Arnouts, S., Cristiani, S., Moorwood, A. \& Saracco,
P. 2000, AJ 120, 2206

\bibitem[1999]{fsoto} Fern\'andez-Soto, A., Lanzetta, K.M. \& Yahil,
A. 1999, ApJ 513, 34

\bibitem[2000]{furu} Furusawa, H., Shimasaku, K., Doi, M., Okamura, S. 2000,
ApJ 534, 624                                             

\bibitem[1997]{gardner} Gardner, J. P., Sharples, R. M., Frenk,
C. S. \& Carrasco, B. E. 1997, ApJ 480, L99

\bibitem[1998]{gia} Giallongo, E., D'Odorico,, S., Fontana, A.,
Cristiani, S., Egami, E., Hu, E. \& McMahon, R. G. 1998, AJ 115, 2169

\bibitem[1998]{glad} Gladders, Michael D., Lopez-Cruz, Omar, Yee,
H. K. C., Kodama, T. 1998, ApJ 501, 571

\bibitem[1995]{glaze} Glazebrook, K., Peacock, J. A., Miller, L. \&
Collins, C. A. 1995, MNRAS 275, 169

\bibitem[1996]{gwyn} Gwyn, S. D. J. \& Hartwick F. D. A. 1996, ApJ
468, L77

\bibitem[2000]{evanthia} Hatziminaoglou, E., Mathez, G. \& Pell\'o,
R. 2000, A\&A 359, 9

\bibitem[1994]{kauff1} Kauffmann, G., Guiderdoni, B., White,
S. D. M. 1994, MNRAS 267, 981

\bibitem[1998]{kauff} Kauffmann, G. \& Charlot, S. 1998, MNRAS 297,
L23

\bibitem[1998b]{kauff2} Kauffmann, G. \& Charlot, S. 1998, MNRAS 294,
705

\bibitem[2001]{kocha} Kochanek, C. S., Pahre, M. A., Falco, E. E.,
Huchra, J. P., Mader, J., Jarrett, T. H., Chester, T., Cutri, R. \&
Schneider, S. E.  2001, ApJ 560, 566

\bibitem[2002]{leborgne} Le Borgne, D.~\& Rocca-Volmerange, B. 2002,
A\&A 386, 446

\bibitem[1995]{lilly} Lilly, S. J., Tresse, L., Hammer, F., Crampton,
D. \& Le F\`evre, O. 1995, ApJ 455, 108

\bibitem[1998]{liu} Liu, C. T., Green, R. F., Hall, P. B. \& Osmer,
P. S.  1998, AJ 116,  1082

\bibitem[1999]{loveday2} Loveday, J., Tresse, L., Maddox, S. 1999,
MNRAS 310, 281

\bibitem[2000]{loveday} Loveday, J. 2000, MNRAS 312, 557

\bibitem[1971]{lynden} Lynden-Bell, D. 1971, MNRAS 155, 95

\bibitem[1995]{madau} Madau, P. 1995, ApJ 441, 18

\bibitem[1999]{maglio} Magliocchetti, M. \& Maddox, S. J. 1999, MNRAS
306, 988

\bibitem[1986]{moba1} Mobasher, B., Ellis, R. S. \& Sharples,
R. M. 1986, MNRAS 223, 11

\bibitem[1993]{mobasher} Mobasher, B., Sharples, R. M. \& Ellis,
R. S. 1993, MNRAS 263, 560

\bibitem[1996]{moba2} Mobasher, B., Rowan-Robinson, M., Georgakakis,
A. \& Eaton, N.  1996, MNRAS 282, L7

\bibitem[2000]{mori} Moriondo, G., Cimatti, A. \& Daddi, E. 2000,
A\&A 364, 26

\bibitem[1974]{oke} Oke, J. B. 1974, ApJS 27, 21

\bibitem[1998]{pascarelle} Pascarelle, S. M., Lanzetta, K. M. \&
Fern\'andez-Soto, A. 1998, ApJ 508, L1

\bibitem[1998]{pickles} Pickles, A. J. 1998, PASP 110, 863

\bibitem[2001]{poli} Poli, F., Menci, N., Giallongo, E., Fontana, A.,
  Cristiani, S., D'Odorico, S. 2001 , ApJ 551, L 45

\bibitem[1996]{pozz} Pozzetti, L., Bruzual A., G., Zamorani, G. 1996,
MNRAS 281, 953

\bibitem[1999]{renzini} Renzini, A. \& Cimatti, A. 1999, 
ASP Conference Proceedings, Vol. 193, p312 (astro-ph/99010162)

\bibitem[1991]{rocca} Rocca-Volmerange B. \& Guiderdoni, B. 1991, A\&A
252, 435

\bibitem[2001]{rodighiero} Rodighiero, G., Franceschini, A. \& Fasano,
G. 2001, MNRAS 324, 491

\bibitem[2001]{rudnick} Rudnick, G., Franx, M., Rix, H.-W., Moorwood,
A., Kuijken, K., van Starkenburg, L., van der Werf, P., R\"ottgering,
H., van Dokkum, P. \& Labbe, I. 2001, AJ 122, 2205

\bibitem[1979]{sty} Sandage, A., Tammann, G. A. \& Yahil, A. 1979, ApJ
232, 352

\bibitem[1997]{sawicki} Sawicki, M. J., Lin, H. \& Yee, H. K. C. 1997,
AJ 113, 1

\bibitem[1986]{scalo} Scalo  J. M. 1986, Fundam. Cosmic Phys. 11, 1

\bibitem[1968]{schmidt} Schmidt, M. 1968, ApJ 151, 393

\bibitem[2000]{scodeggio} Scodeggio, M. \& Silva, D. R. 2000, A\&A
359, 953

\bibitem[1994]{songaila} Songaila, A., Cowie, L. L. \& Gardner,
J. P. 1994, ApJS 94, 461

\bibitem[1996]{subba} SubbaRao, M. U., Connolly, A. J., Szalay,
A. S. \& Koo, D. C.  1996, AJ 112, 929

\bibitem[1998]{szokoly} Szokoly, G. P., SubbaRao, M. U. \& Connolly,
A. J. 1998, ApJ 492, 452

\bibitem[2000]{take} Takeuchi, T. T., Yoshikawa, K. \& Ishii, T. T.
2000, ApJS 129, 1

\bibitem[2000]{totani} Totani, T. \& Yoshii, Y. 2000, ApJ 540, 81

\bibitem[2001]{vanzella} Vanzella, E., Cristiani, S., Saracco, P.,
Arnouts, S,. Bianchi, S., D'Odorico, S., Fontana, A., Giallongo, E. \&
Grazian, A. 2001, AJ 122, 2190

\bibitem[1998]{wan} Wang, Y., Bahcall, N., Turner, E.L. 1998, AJ 116, 2081 

\bibitem[1996]{williams} Williams, R. E., Blacker, B., Dickinson, M.,
et al. 1996, AJ 112, 1335

\bibitem[1997]{willmer} Willmer, C. N. A. 1997, AJ 114, 898

\bibitem[1994]{worthey} Worthey, G. 1994, ApJS 95, 107

\bibitem[2001]{wright} Wright, E. 2001,  ApJ 556, L17

\bibitem[1993]{yoshii} Yoshii, Y. 1993, ApJ 403, 552

\bibitem[1997]{zucca} Zucca, E., Zamorani, G., Vettolani, G. et
al. 1997, A\&A 326, 477

\end{thebibliography}
\end{document}